\newcommand{\lya}{\mbox{Ly$\alpha$}}

\newcommand{\halpha}{\mbox{H$\alpha$}}
\newcommand{\hbeta}{\mbox{H$\beta$}}
\newcommand{\hgamma}{\mbox{H$\gamma$}}

\newcommand{\hepsilon}{\mbox{H$\epsilon$}}

\newcommand{\arIII}{\mbox{Ar~{\sc iii}}}
\newcommand{\arIV}{\mbox{Ar~{\sc iv}}}
\newcommand{\hI}{\mbox{H~{\sc i}}}
\newcommand{\hII}{\mbox{H~{\sc ii}}}
\newcommand{\heI}{\mbox{He~{\sc i}}}
\newcommand{\heII}{\mbox{He~{\sc ii}}}
\newcommand{\oI}{\mbox{O~{\sc i}}}
\newcommand{\oII}{\mbox{O~{\sc ii}}}
\newcommand{\oIII}{\mbox{O~{\sc iii}}}

\newcommand{\cII}{\mbox{C~{\sc ii}}}

\newcommand{\cIV}{\mbox{C~{\sc iv}}}
\newcommand{\neIII}{\mbox{Ne~{\sc iii}}}

\newcommand{\sII}{\mbox{S~{\sc ii}}}

\newcommand{\siII}{\mbox{Si~{\sc ii}}}

\newcommand{\siIV}{\mbox{Si~{\sc iv}}}
\newcommand{\nII}{\mbox{N~{\sc ii}}}
\newcommand{\nV}{\mbox{N~{\sc v}}}

\newcommand{\mgII}{\mbox{Mg~{\sc ii}}}

\newcommand{\kms}{\mbox{km\,s$^{-1}$}}
\newcommand{\ergsec}{\mbox{erg~s$^{-1}$}}

\newcommand{\msun}{\mbox{M$_\odot$}}

\newcommand{\msunyr}{\mbox{M$_\odot$~yr$^{-1}$}}

\newcommand{\fesclya}{\mbox{$f_\mathrm{esc}^{\mathrm{Ly}\alpha}$}}

\newcommand{\ebv}{\mbox{$E_{B-V}$}}

\newcommand{\Mfuv}{\mbox{$M_\mathrm{FUV}$}}

\newcommand{\Lstar}{\mbox{$L^\star$}}
\newcommand{\Mstar}{\mbox{$M^\star$}}
\newcommand{\Mstell}{\mbox{$M_\mathrm{stell}$}}

\newcommand{\nhi}{\mbox{$N_\mathrm{HI}$}}
\newcommand{\llya}{\mbox{$L_{\mathrm{Ly}\alpha}$}}

\newcommand{\ewlya}{\mbox{$W_{\mathrm{Ly}\alpha}$}}
\newcommand{\ewha}{\mbox{$W_{\mathrm{H}\alpha}$}}
\newcommand{\ewhb}{\mbox{$W_{\mathrm{H}\beta}$}}

\newcommand{\Lbluered}{\mbox{$L_\mathrm{B/R}$}}

\newcommand{\xiion}{\mbox{$\xi_{\mathrm{ion}}$}}

\newcommand{\Lmech}{\mbox{$L_\mathrm{mech}$}}
\newcommand{\Emech}{\mbox{$E_\mathrm{mech}$}}

\documentclass[fleqn,usenatbib]{mnras}
\usepackage{newtxtext,newtxmath}
\usepackage[T1]{fontenc}
\usepackage{xcolor}
\usepackage{longtable}
\usepackage{tablefootnote}

\DeclareRobustCommand{\VAN}[3]{#2}
\let\VANthebibliography\thebibliography
\def\thebibliography{\DeclareRobustCommand{\VAN}[3]{##3}\VANthebibliography}

\usepackage{graphicx}	
\usepackage{amsmath}	

\title[Shaping Ly$\alpha$ Emission in Starburst Galaxies.]{Spectral Shapes of
the Ly$\alpha$ Emission from Galaxies. II. the influence of stellar properties
and nebular conditions on the emergent \lya\ profiles}

\author[M. J. Hayes et al.]{
Matthew J. Hayes,$^{1}$\thanks{E-mail: matthew@astro.su.se (MJH)}
Axel Runnholm,$^{1}$
Claudia Scarlata,$^{2}$
Max Gronke$^{3}$
and T. Emil Rivera-Thorsen$^{1}$
\\
$^{1}$Stockholm University, Department of Astronomy and Oskar Klein
Centre for Cosmoparticle Physics, AlbaNova University Centre, SE-10691,
Stockholm, Sweden.\\
$^{2}$Minnesota Institute for Astrophysics, School of Physics and
Astronomy, University of Minnesota, 316 Church str SE, Minneapolis, MN
55455,USA\\
$^{3}$Max Planck Institut fur Astrophysik, Karl-Schwarzschild-Strasse 1,
D-85748 Garching bei M\"unchen, Germany
}

\date{Accepted XXX. Received YYY; in original form ZZZ}
\pubyear{2023}
\graphicspath{{figs/}}

\begin{document}
\label{firstpage}
\pagerange{\pageref{firstpage}--\pageref{lastpage}}
\maketitle

\begin{abstract} 
We demonstrate how the stellar and nebular conditions in star-forming galaxies
modulate the emission and spectral profile of \hI\ \lya\ emission line.  We
examine the net \lya\ output, kinematics, and in particular emission of
blue-shifted \lya\ radiation, using spectroscopy from with the Cosmic Origins
Spectrograph on HST, giving a sample of 87 galaxies at redshift $z= 0.05-0.44$.
We contrast the \lya\ spectral measurements with properties of the ionized gas
(from optical spectra) and stars (from stellar modeling).  We demonstrate
correlations of unprecedented strength between the \lya\ escape fraction (and
equivalent width) and the ionization parameter ($p\approx 10^{-15}$).  The
relative contribution of blue-shifted emission to the total \lya\ also
increases from $\approx 0$ to $\approx 40$~\% over the range of O$_{32}$ ratios
($p\approx 10^{-6}$).  We also find particularly strong correlations with
estimators of stellar age and nebular abundance, and weaker correlations
regarding thermodynamic variables.  Low ionization stage absorption lines
suggest the \lya\ emission and line profile are predominantly governed by the
column of absorbing gas near zero velocity.  Simultaneous multi-parametric
analysis over many variables shows we can predict 80~\% of the variance on
\lya\ luminosity, and $\sim50$~\% on the EW. We determine the most crucial
predictive variables, finding that for tracers of the ionization state and
\hbeta\ luminosity dominate the luminosity prediction whereas the \lya\ EW is
best predicted by \hbeta\ EW and the \halpha/\hbeta\ ratio.  We discuss our
results with reference to high redshift observations, focussing upon the use of
\lya\ to probe the nebular conditions in high-$z$ galaxies and cosmic
reionization.   
\end{abstract}


\section{Introduction} \label{sec:intro}

This paper draws together several points concerning the Lyman alpha (\lya)
emission line of neutral hydrogen (\hI), when observed from star-forming
galaxies.  The first is that \lya\ has been demonstrated over the last two
decades to be a very efficient observational probe of galaxies at high
redshifts.  The second is that, as a resonance line, the \lya\ spectral profile
is reshaped by properties of the gas in which it scatters.  The important
connection, therefore, is that we may use observations of the \lya\ profile to
infer properties of galaxies in the early universe and the influence of gas that
lies in the foreground of these galaxies -- the circumgalactic and
intergalactic media.

The last decade has seen a great increase in the numbers of high-quality \lya\
spectra obtained from high-redshift galaxies.  This boom has been mainly driven
by large efforts using highly multiplexed, multi-object optical spectrographs
\citep[e.g.][]{Stark.2010,Jiang.2013,Marchi.2019,Hoag.2019} and large-format
integral field spectrographs
\citep[e.g.][]{Drake.2017,Herenz.2019,Claeyssens.2019}.  See \citet{Ouchi.2020}
for a review of the high redshift observations and \citet{Runnholm.2021} for a
compilation of homogeneous spectroscopic measurements.  As a community we have
assembled very large samples of high-$z$ \lya\ spectra, with well understood
selection effects.  In many cases \lya\ is the only well-exposed part of the
ultraviolet spectrum and the only information carrier available from the bulk
of objects at $z\gtrsim 5$.  The time is therefore ripe to extract the encoded
kinematic information from these large samples. 

As a resonance line, the radiative transport of \lya\ (and other optically
thick lines) has been studied for decades \citep{Osterbrock.1962,Adams.1972}.
Both analytical and numerical calculations show that, in the absence of photon
sinks (dust absorption), resonance photons escape from static, optically thick
media with broadened, double-peaked profiles with a separation that scales with
the gas column density \citep[e.g.][]{Neufeld.1990,Verhamme.2006}.  Should the
media show bulk flows in the gas kinematics, the emergent double peaks vary in
relative amplitude.  Since most star-forming galaxies exhibit outflows in their
interstellar media, the blue peak is typically weakened, and frequently
suppressed entirely sometimes giving way to absorption and P\,Cygni like
profiles \citep[][]{Kunth.1998,Mas-Hesse.2003,Shapley.2003}. 

The wide array of observable spectral shapes observed at high-$z$ has
encouraged the community to attempt to recover physical properties from the
resolved profiles
\citep[e.g.][]{Schaerer.2008,Verhamme.2008,Vanzella.2010,Lidman.2012}.
Application of these radiative transfer simulations has shown that blueshifted
\lya\ emission can be used as an informative probe of optically thin gas
falling along the line-of-sight \citep{Verhamme.2015} -- this may be especially
important as the blueshifted \lya\ may then be used as a signpost of galaxies
that emit ionizing radiation.  The theoretical suggestion was almost
immediately back up empirically by low-$z$ by \citep{Henry.2015} who found
strong blueshifted \lya\ peaks in galaxies with signatures of weak metal
absorption.  More recently, \citet{Gronke.2017} used radiative transfer
simulations to directly model the profiles of over 200 \lya-selected galaxies
observed with VLT/MUSE, showing a large fraction of them to have \hI\ column
densities that would imply the gas is optically thin to LyC.  

The connection between blueshifted \lya\ emission and the escape of ionizing
radiation was recently confirmed directly, in small samples of low-redshift
starbursts \citep{Verhamme.2017,Jaskot.2017,Jaskot.2019,Izotov.2021}, and has
been expanded in significance by \citet{Flury.2022pap2}.  The premise is that
low total column densities of \hI\ allow both LyC to escape directly, and
significant fractions of the blueshifted \lya\ to also evade scattering in the
galaxy winds.  This hypothesis is fully consistent with studies that find the
low ionization ultraviolet absorption lines to also be weak in both LyC
emitters and strong \lya-emitting galaxies
\citep{Gazagnes.2020,Saldana-Lopez.2022}.  Finally, resolved 21~cm observations
presented by \citet{LeReste.2022} have shown \lya\ emission on kpc-scales to be
associated with regions of both high- and low column density, supporting the
picture in which scattering on higher density media redistributes photons in
frequency and directs them along low column density channels. 

These advances in \lya\ spectroscopy, and the connection to LyC emission
prompted us to re-examine the spectra of high-$z$ LAEs.  During the development
of the \emph{Lyman alpha Spectral
Database}\footnote{\href{http://lasd.lyman-alpha.com}{http://lasd.lyman-alpha.com}}
(LASD; \citealt{Runnholm.2021}) we obtained and studied large samples of \lya\
spectra, totaling around 150 from $z\lesssim 0.44$ and over 200 from $z>2$
galaxies.  Using 74 of the low-$z$ sample and the full catalog of $z=2.9-6.5$
systems identified with VLT/MUSE \citep{Herenz.2017musewide,Urrutia.2019}, we
showed significant decrease in the relative flux of blue-shifted \lya\ with
increasing redshift \citep[][hereafter Paper I]{Hayes.2021}.  However by
running simulations of the effect of \lya\ absorption by intervening \hI\
clouds, we could attribute this evolution entirely to the evolving average
density of intergalactic hydrogen.  This effort provides support for the idea
that \lya-emitters, and their line profiles, may be used to trace the
emissivity of LyC at all redshifts where \lya\ can be observed. This was more
recently applied by \citet{Matthee.2022b.z2lya} and \citet{Naidu.2022}, who
used a combination of the \lya\ line profiles and observed frequency of \lya\
emission with redshift, to make new inferences of the level of the ionizing
background.

During the experiments of Paper I, we also presented some preliminary analysis
of the blue-shifted emission in the low-$z$ sample: specifically we noticed
that the blue/red flux ratio, \Lbluered, also correlated positively with the
total \lya\ EW.  We speculated that this trend stems from variation in the
interstellar \hI\ column density: because of the outflowing gas that is
ubiquitous in star-forming galaxies, the \hI\ column preferentially attenuates
\lya\ bluewards of line-centre, leading to a scenario in which \nhi\ modulates
the correlation with the total \lya\ output.  In this second paper in the
series, we turn our attention specifically to the question of how this
blue-shifted emission is produced.  We use the great array of mid-resolution,
low-redshift \lya\ spectra available in the HST archive combined with optical
emission line spectroscopy obtained from the SDSS archives to determine many
physical properties of the ionized gas.

The paper proceeds as follows: we describe the observations and data in
Section~\ref{sect:data}, and the processing, measurements, and inference of
physical properties in Section~\ref{sect:meas}.  In
Section~\ref{sect:results_measurement} we characterize the sample and show
distributions of its main properties.  We present the results in three main
sections: Section~\ref{sect:results_basic} shows how the main integrated \lya\
observables scale with basic properties, Section~\ref{sect:results_profiles}
investigates how the line profile is shaped by properties of the stars and
ionized gas, and Section~\ref{sect:mainfinds} introduces the measurements of
the absorbing material and unites the previous Sections into a coherent
physical picture.  In Section~\ref{sect:multipar} we develop this approach and
attempt to quantify how much predictive power can be extracted from the
information we have derived .  We present our concluding remarks and outlook in
Section~\ref{sect:conclusions}.

\section{Data and Processing} \label{sect:data}

\subsection{Ultraviolet Spectra and Lyman alpha}\label{sect:data:uvspec}

We use the same sample of 74 COS galaxies adopted for Paper I with the
inclusion of two further samples for which data have since become public in the
Barbara A. Mikulski Archive for Space Telescopes
(MAST)\footnote{\href{https://archive.stsci.edu/hst/search.php}{https://archive.stsci.edu/hst/search.php}}.
We add nine archival under GO\,15639 (PI: Izotov; published in
\citealt{Izotov.2021}) and eight galaxies from GO\,15865 (PI: Henry, published
in \citealt{Xu.2022.lzlcs}).  These samples were selected in order to study the
\lya\ emission from low mass galaxies ($M_\mathrm{stell} < 10^8$~\msun), and to
study the influence of low optical depths on LyC emission when traced by \mgII.
The galaxies lie at redshifts in the range $z=0.32-0.45$ and hence \lya\ always
fell in the G160M grating of COS. 

Like the data in Paper I, we reprocessed the raw data homogeneously with the
COS pipeline (CALCOS), v.3.3.7.  We correct the spectra for Milky Way
foreground reddening by using the maps of \citet{Schlafly.2011} to look up the
$B-V$ color excess at the coordinates of each object.  We then use the
\citet{Cardelli.1989} reddening law to describe the wavelength-dependent
absorption with standard $V$-band normalization of $R_V=3.1$.

\subsection{Optical Spectra}\label{sect:data:optspec}

We adopt the optical data obtained from the Sloan Digital Sky Survey (SDSS)
Data Release 16 \citep[DR16,][]{Ahumada.2020}.  We obtained all these spectra
from via \texttt{astroquery} for local processing and measurements.  We first
corrected the SDSS spectra for foreground reddening, using exactly the same
method as described above concerning the UV spectra. 

\begin{figure}
\noindent
\begin{center}
\includegraphics[width=0.9\linewidth]{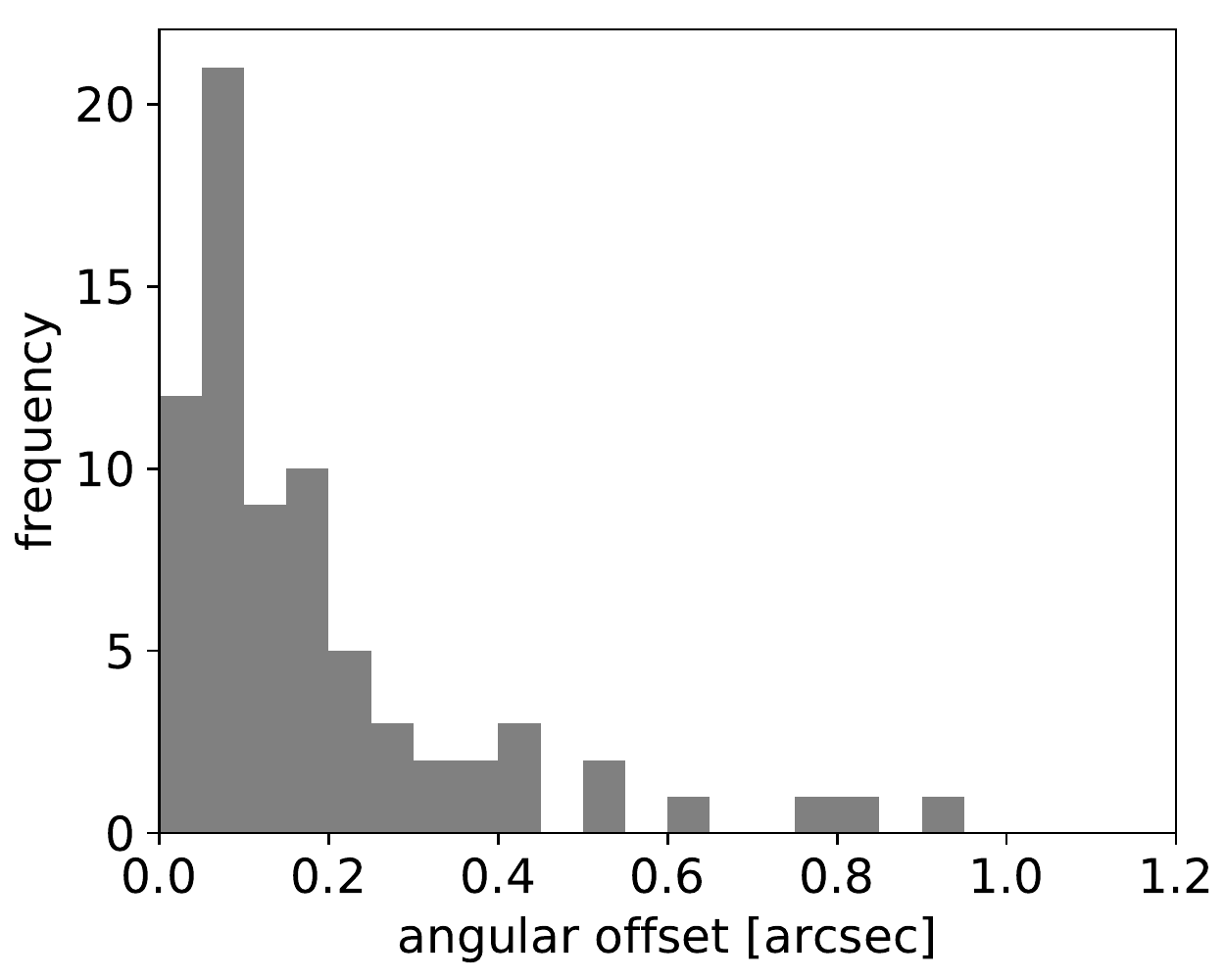}
\end{center}
\caption{Angular offset in arcsec between the \emph{reported} pointings of the
SDSS aperture and COS/PSA.  85\% of the objects show offsets smaller than the
reported $1\sigma$ (68\% of the distribution) accuracy of the absolute WCS of
HST.  The tail to high offsets is confirmed to correspond to WCS uncertainties.
See text for details. } 
\label{fig:angleoff}
\end{figure}

This study requires a quantitative comparison to be made between the UV and
optical spectra, which we note have been obtained in slightly different
apertures.  The COS primary science aperture (PSA) has a diameter of 2\farcs5,
and becomes vignetted at radii greater than $\approx$~0\farcs6.  Comparably to
this, our SDSS spectra are observed in with 3\farcs0 fibers for 10\% of the
sample (with the SDSS spectrograph) and 2\farcs0 fibers for the remainder (9
galaxies, using the BOSS fibers). Concerning the positional alignment, we
calculate the distribution of angular offsets, which we show in
Figure~\ref{fig:angleoff}.  In the interpretation of this we first note that
both telescopes performed acquisition centered upon the brightest local source,
and we have verified in the COS acquisition images that the target was always
acquired and centered to within a precision better than 3 NUV pixels
(0\farcs066).  However we also note that the absolute astrometric solution of
the HST focal plane is $\approx$~0\farcs25, while typical ground-based
astrometric solutions are usually good to an rms of $\approx$~0\farcs3.
Assuming wavelength invariance of the morphologies, we would expect differences
of $\approx$~0\farcs4 (1--$\sigma$) between reported COS and SDSS coordinates
in case of perfect aperture matching, purely as a result of differing WCS
solutions.  In light of this, it is surprising that the average offset in
Figure~\ref{fig:angleoff} is as small as it is.  We anyway inspect the SDSS
fibre-positions of the nine objects with reported offsets more than ~0\farcs4,
and in fact find no evidence that the aperture centers actually differ by the
reported amounts -- we have verified that the tail of this histogram is due to
the reported HST world coordinate system.

We next examine photometric issues: the size of the apertures with respect to
the seeing.  The median seeing of the SDSS survey is 1\farcs43, from which we
calculate at 4.7\% flux loss when feeding a fibre of 3" if the objects are
point-sources.  While there is no need for the underlying galaxy to be
point-like, the overwhelming majority of COS acquisition images show the
morphology of UV light would not be resolved from the ground.  The sources are
so compact in the UV continuum that the spectrophotometric effects of the COS
vignetting function are negligible, and our spectral modeling (described in
Section~\ref{sect:meas:stell}) will not be affected.  However the same may not
be true for \lya, which may be spatially extended with respect to the
continuum.  We examine the COS vignetting function\footnote{COS Instrument
Science Report 2010-10}, and calculate that for a flat surface of \lya\
emission, 70\% of the light would be recovered, although we estimate this loss
to be somewhat conservative because the \lya\ should also be centrally
concentrated.  We conclude that this effect remains as a systematic source of
uncertainty that would enter at the $\approx 20$~\% level on average, but
without explicit information on the \lya\ light profile we are unable to
quantify the effect further.  

Finally we remark that while we study a large number of properties in this
paper, only one -- the \lya\ escape fraction, \fesclya\ -- is actually
sensitive to the relative flux calibration of COS and SDSS.  Other UV
properties are either equivalent widths or kinematic properties, while all the
optical properties are line ratios, and aperture effects will mostly cancel out
when the ratios are taken. 

\begin{figure}
\noindent
\begin{center}
\includegraphics[width=0.9\linewidth]{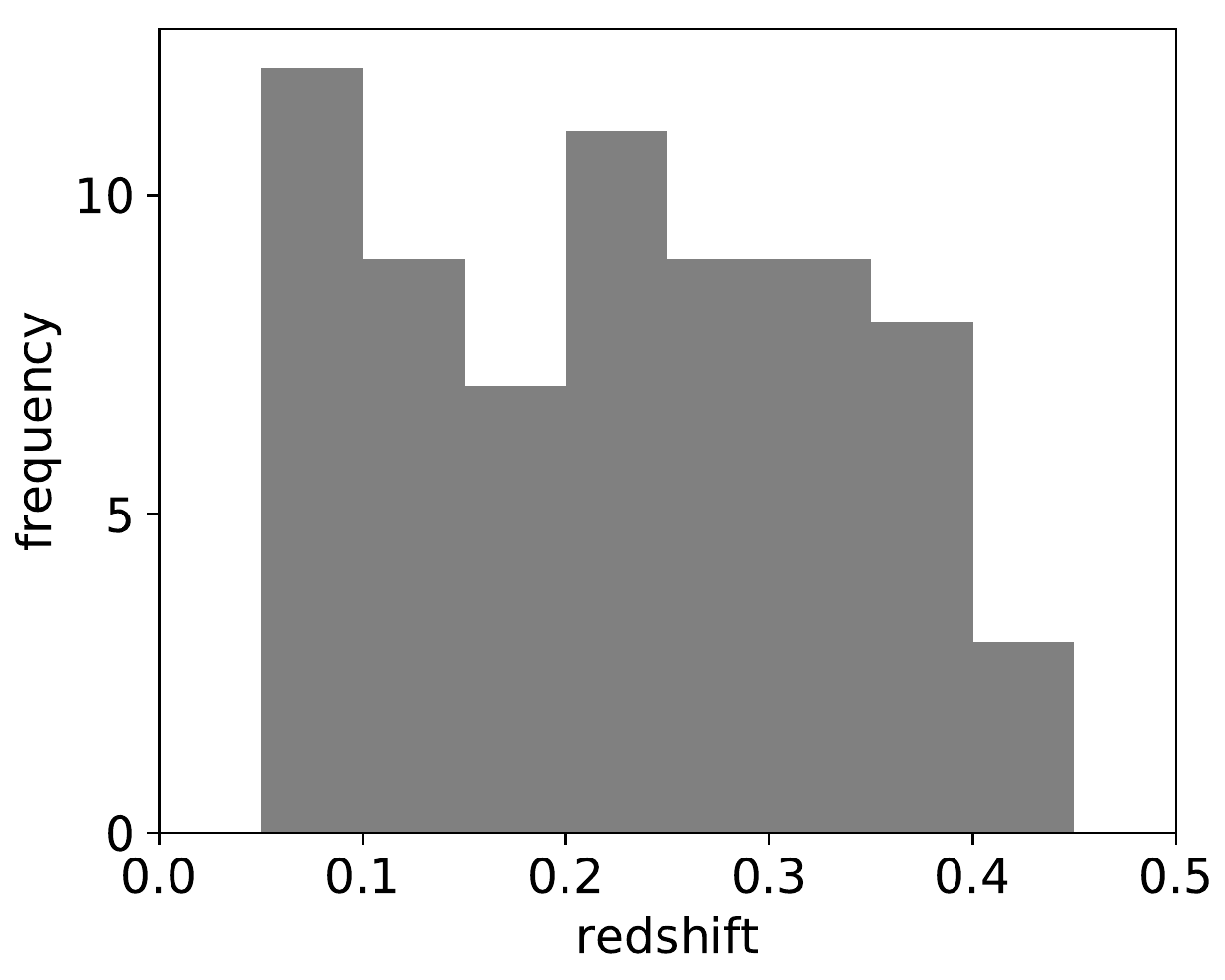}
\end{center}
\caption{Redshift distribution of the galaxies.  The upper limit is imposed by
the necessity of having \lya\ fall in the G160M grating of COS ($z<0.44$),
while the lower limit is imposed by us to ensure that the bulk of the galaxy is
encompassed within the aperture.}
\label{fig:zdist}
\end{figure}

\section{Data Processing and Measurements} \label{sect:meas}

\subsection{Stellar Continuum Modeling}\label{sect:meas:stell}

\subsubsection{Motivation}

The first step in our data processing is to model the continuum of the galaxies
and estimate the properties of the stellar population.  This serves a number of
purposes.  Firstly the modeling estimates and corrects for the stellar
absorption features in the optical spectra, especially the Balmer and Paschen
series of \hI, and optical \heI\ lines.  This naturally improves the
measurements of these emission lines.  Secondly, we derive a model of the
ultraviolet continuum that is free from interstellar absorption and emission
lines.  This is especially important near the high-ionization UV lines such as
\nV\ ($\lambda\approx1240$~\AA),  \siIV\ ($\lambda\approx1400$~\AA), and \cIV\
($\lambda\approx 1550$~\AA), which may mix contributions of stellar wind
features, interstellar absorption, and nebular emission.  The narrow
interstellar and nebular lines are masked during the fitting process, leaving
the much broader (several thousand \kms) P\,Cygni features to be modeled
without contamination. 

Thirdly, these models directly estimate the recent star formation history and
stellar metal abundances, both of which are key quantities to investigate.
Moreover, the direct estimate of the starburst age provides us with an
independent route to obtain the instantaneous ionizing photon production rates
(e.g. $Q_0$ for \hI) and the mechanical energy returned to the interstellar
medium by both winds from massive stars and supernova explosions.  Inferences
from these quantities were the subject of \citet{Hayes.2023} and the software
formed the basis for calculations in \citet{Sirressi.2022}. Finally we also
obtain a measurement of the internal dust reddening experienced by the stellar
continuum, in addition to the nebular reddening from \hI\ Balmer emission.  As
well as estimating physical properties, we also use these accurate models of
the UV and optical continuum to normalize the observed spectra to measure
interstellar absorption lines (see Section~\ref{sect:meth:uvcont}), and
subtract the stellar continuum to measure nebular emission line fluxes (see
Section~\ref{sect:meth:optical}).  The process therefore corrects for P\,Cygni
stellar features in the UV and stellar Balmer absorption in the optical.

\subsubsection{Stellar Fitting Method}\label{sect:method_stellarfit}

Our software fits multiple generations of stellar population, their ages,
metallicity ($Z$), and dust obscuration.  We adopt the high resolution
libraries of \emph{Starburst99}
\citep{Leitherer.1999,Vazquez.2005,Leitherer.2014}, computed using the
evolutionary tracks of the Geneva group for high mass-loss rates.  We adopt a
Salpeter initial mass function (IMF) between mass limits of 0.1 and 100~\msun,
and permit metallicities across the full range of model libraries: $Z \in
\{0.001, 0.004, 0.008, 0.020, 0.040\}$.  We remind the reader that the
evolutionary tracks from the Geneva group are computed using the outdated solar
abundance patterns, and there may be a factor of $\approx 0.2$~dex offset
\citep{Asplund.2009} when scaling between metallicity expressed as mass
fraction ($Z$) and the oxygen abundance.  The optical spectra include nebular
continuum by default, under the assumption that 100\% of the ionizing photons
are reprocessed into nebular light that is emitted cospatially with the
starlight. 

We first mask the regions of the observed spectrum that are contaminated by
known strong nebular emission and interstellar absorption lines. Milky Way
absorption lines are also masked in the UV spectra.  Stellar modeling then
proceeds by first building a 3-dimensional data structure holding the
luminosity density ($L_\lambda$) as a function of stellar age, metallicity, and
wavelength ($\lambda$).  We fit a 3-dimensional cubic spline field to this cube
to build a interpolation function that produces $L_\lambda$ as a function of
these three main quantities (age, $Z$, $\lambda$).  This numerically
differentiable function can then be used in standard minimizer algorithms, for
which we adopt \texttt{lmfit} in \texttt{Python}.

The next step is to redshift the stellar models into the observed frame of the
galaxies, and scale the luminosity density by a factor of $4\pi
d_\mathrm{L}^2(1+z)$ (where $d_\mathrm{L}$ is the luminosity distance) to
obtain spectra in flux densities ($f_\lambda$).  We model the UV and optical
spectra simultaneously so as to take advantage of the long spectral baseline
afforded by the combined data ($\sim 1200-7000$~\AA).  This is firstly designed
to estimate (or place limits upon) the presence of a more evolved underlying
stellar population with higher mass-to-light ratio, and secondly to improve
estimates of the stellar obscuration with the longer lever-arm in wavelength
than using the UV data alone. 

It is important to account for the fact that the COS and SDSS spectrographs
have different spectral resolutions (by a factor of $\sim 5$ depending upon the
light distribution of the source).  Our fitting algorithm includes the
intrinsic velocity dispersion of the stellar population as a free parameter,
which is implemented by a two-step convolution.  Within the function that
generates the spectrum, we first convolve with a single Gaussian (to account
for intrinsic stellar velocity dispersion), and then by the resolutions of the
independent spectrographs: the UV models are convolved to $R\approx$\,12,000
(approximately that of HST/COS), while the optical spectra are convolved to
$R\approx$\,2,000, which holds for the SDSS spectrograph at the central
wavelength of interest ($\lambda \approx 5000$~\AA).

Star formation is an inherently clustered process, with starbursts typically
forming stars in a number of discrete star clusters.  This is especially
obvious at ultraviolet wavelengths \citep[e.g.][]{Meurer.1995}, and we may
expect multiple populations of stars to exist within the COS aperture.  The
youngest stars dominate the light output, but more evolved massive stars (down
to B-type, with ages to 40~Myr) may be more important in determining the
mechanical energy release if supernova contribute significantly.  Thus the more
evolved population cannot be ignored \citep[see][]{Sirressi.2022}.  When
fitting just one population of stars, the stellar ages are almost always found
to be 3--4~Myr, and are drawn to this value by the strong P\,Cygni lines (e.g.
\nV, \cIV) that form in the winds of the most massive stars.  However these
wind lines represent only a tiny minority of the wavelength coverage: a lot of
information is also present from narrower photospheric lines that are much
weaker but far more numerous, as well as the overall shape of the continuum
\citep[see][]{Chisholm.2019,Senchyna.2022}.  As we expect more protracted
episodes of star formation (free-fall timescales are $\gtrsim 10$~Myr), we fit
the additive combination of one, two, and three stellar populations to the
data, selecting the final model with the lowest $\chi^2$ per degree of freedom
as the final model.  In order to better estimate the star formation history
over timescales relevant for mechanical energy return and the development of
galaxy winds, we allow the ages to range freely over the limits of $10^6$ to
$10^{10}$~yr. 

\begin{figure*}
\noindent
\begin{center}
\includegraphics[width=0.9\linewidth]{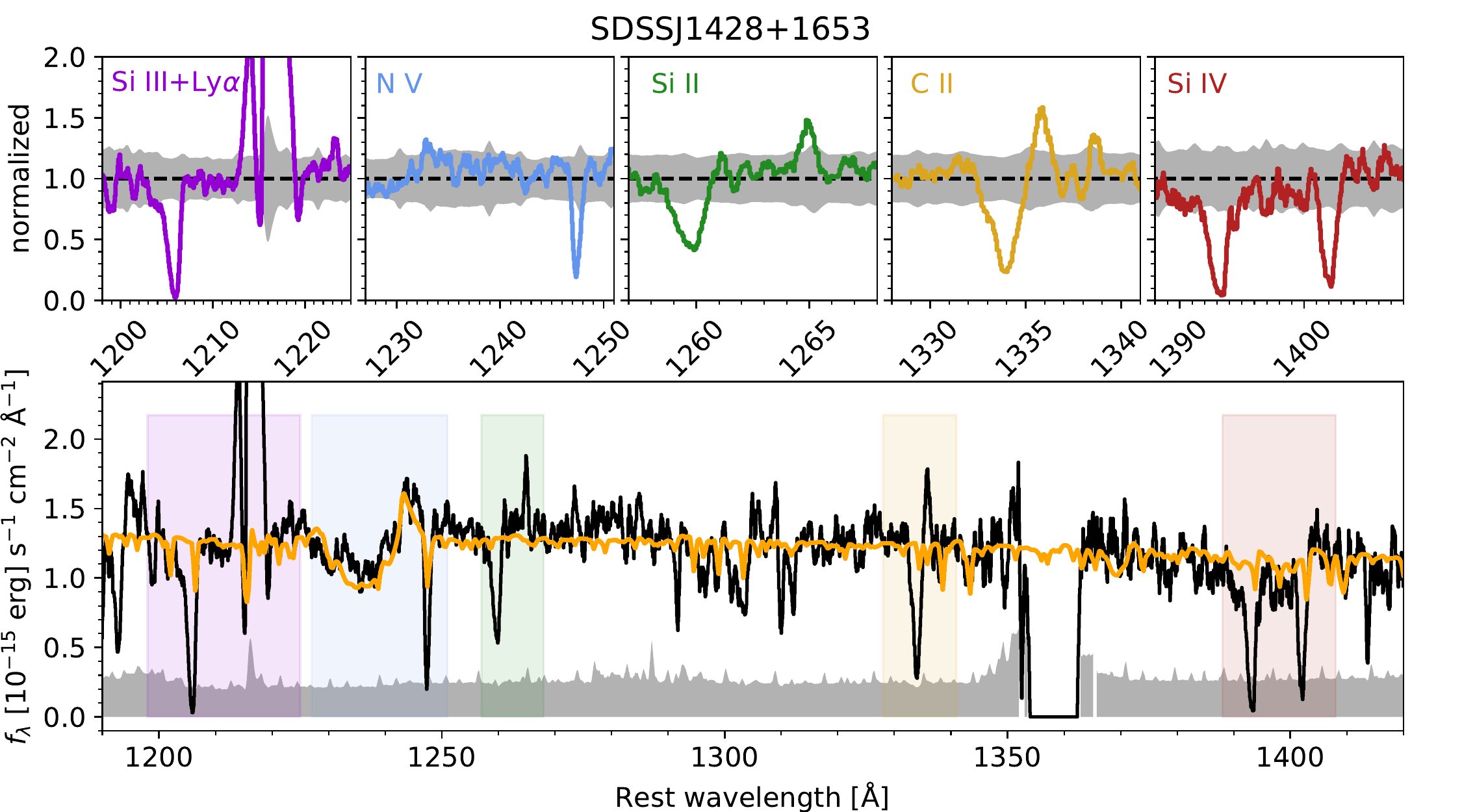}
\end{center}
\caption{Spectral modeling of the stellar continuum, using well-exposed object
SDSS\,J1428.  The lower panel shows the observed UV spectrum in black, shifted
into the restframe. The model fit is in orange, and the uncertainty level is
indicated by the grey shaded region.  Large deviations from the fit are mostly
caused by ISM absorption features, fluorescent emission lines, and \lya.  Upper
panels show zooms around some relevant features, each of which is labeled, and
colors correspond to the shaded regions in the lower panel.}
\label{fig:contmodels}
\end{figure*}

We redden the spectra using the dust attenuation law derived for starburst
galaxies by \citet{Calzetti.2000}; we have experimented with other
parameterizations, such as \citet{Charlot.2000} and clumpy dust distributions
\citep[][revisited for \lya\ emitting galaxies by
\citealt{Scarlata.2009}]{Natta.1984} without a noticeable increase in the
quality of our fits.  We apply the same dust attenuation to each stellar
population.  We acknowledge this as a limitation of method, but large
degeneracies are introduced when fitting multiple extinctions that effectively
decouple the optical and UV spectra.  This leads to very different extinctions
at fixed age, and very different stellar metallicities that are also hard to
motivate.  We show an example of the ultraviolet spectral modeling in
Figure~\ref{fig:contmodels}.

\subsubsection{Summary of Estimated Properties}\label{sect:meas:uvprop}

In total we measure and record: 

\textbf{Normalization of each stellar population.}  This is recovered directly
from the modeling.  Since we fit simple stellar populations from
\emph{Starburst99}, this corresponds directly to a stellar mass, which we have
for the total population (all ages) and `starburst event' (defined as ages
below 20~Myr). 

\textbf{Stellar Age.}  This is recovered directly from the modeling.  With the
normalizations computed above and relative contributions of light at different
wavelengths, we can then describe the total population with mass or
light-weighted ages.

\textbf{Stellar Metallicity.}  As with ages, this is a basic fit parameter and
can be assembled into mass- and light-weighted descriptors of the population.

\textbf{Total stellar attenuation.}  This is a basic result of the fitting that
is recovered directly assuming the \citet{Calzetti.2000} prescription.

\textbf{Ionizing photon production rates.}  \emph{Starburst99} tabulates the
production rates of H-, He-, and H$^+$-ionizing photons as a function of age
and metallicity.  We compute these quantities, which we refer to as $Q_0$,
$Q_1$, and $Q_2$, respectively, using the quantities estimated above; as with
all properties we have an estimate for each population and the sum over all
stars.

\textbf{Mechanical energy return.}  Similarly, \emph{Starburst99} also
tabulates the instantaneous mechanical luminosity, which we refer to as \Lmech.
We treat \Lmech\ in the same way as ionizing photon rates above, and sum to
obtain the current value.  However mechanical luminosity is the time derivative
of the energy, and by integrating \Lmech\ along the evolution
\emph{Starburst99} tables, we can compute the total energy deposited by each
stellar population at its estimated age.  We refer to this as the integrated
mechanical energy, \Emech. 

\subsection{Optical Emission Lines}\label{sect:meth:optical}

\subsubsection{Measurements}

After subtraction of the stellar continuum, we measure many relevant nebular
emission lines.  These range in wavelength between the [\oII]$\lambda\lambda
3726,3729$~\AA\ doublet in the blue, and the Pa~8 line at 9229~\AA, but what
lines are actually measured for a given galaxy depends on redshift.  There are
42 lines in total, the highest ionization potential of which is that of \heII\
$\lambda 4686$~\AA\ with IP=54~eV, although also includes other lines of
Ar$^{3+}$ and Ne$^{2+}$ that also probe particularly highly ionized gas phases.
The stellar modeling described in Section~\ref{sect:method_stellarfit}
automatically corrects for underlying absorption in the Balmer lines.

We perform a simultaneous kinematic fit to all the emission lines. We fit one
single recession velocity (redshift) and Gaussian FWHM to represent the
velocity dispersion of the gas.  We account for the wavelength-dependence of
the spectral resolution by interpolating the estimated resolving powers
(provided in the instrument manuals) to the observed wavelength of each line,
and convolve each model line with the instrumental resolution.  For each line,
the only free parameter is the normalization, which then gives the flux.
Fitting the sum of these lines to the data we automatically deblend nearby
lines (e.g. \halpha\ and [\nII]) and get a better estimate of the total [\oII]
flux by partially deblending the doublet.  We estimate errors using a Monte
Carlo simulation, where we use the error spectrum to weight a regeneration of
each pixel in the spectrum using randomized Gaussian deviates.

Some of the strongest lines are clipped by the SDSS data reduction pipeline,
where they are mis-identified as cosmic rays.  These are relatively easy to
identify as either absent entries in the error vector, or unphysical ratios for
the strongest lines. We fix clipped 5007~\AA\ lines using  2.98 times the flux
of the 4959~\AA\ line, and clipped \halpha\ lines in a handful of cases are
repaired by estimating the nebular reddening from the \hgamma/\hbeta\ ratio and
scaling dust-corrected \hbeta\ flux and re-applying the reddening suitable for
6563~\AA. 

\subsubsection{Estimated Nebular Conditions}\label{sect:indvar}

We examine a large number of commonly used line ratios, and are specifically
interested in: 

\textbf{Specific star formation rate/stellar evolutionary phase.} Formally we
have these quantities estimated from the stellar modeling above, but we also
compute equivalent values from the emission lines.  We estimate the SFR from
dust-correct \halpha\ (see below) using the calibration of
\citet{Kennicutt.2012}. We record the EW of all emission lines: the \halpha\ EW
strongly correlates with the sSFR and \hbeta\ EW strongly with the evolutionary
stage.  We also examine the EW of the strong [\oIII]5007 line, as it is so
complicit in the selection of galaxies at both low- and high-redshifts.

\textbf{Abundance of dust and metals.}  We use the \halpha/\hbeta\ ratio that
traces dust reddening, and adopt the extinction curve of \citet{Cardelli.1989}.
We also record ratios that predominately trace ionic abundance:
[\nII]6484/\halpha\ ($N_2$), [\sII]6717,31/\halpha\ ($S_2$), and combinations
of [\oIII]/\hbeta\ ($R_3$) and ([\oIII]+[\oIII])/\hbeta\ ($R_{23}$).

\textbf{Ionization state.}  We include basic ratios of lines that have very
different ionization potentials, and may encode information on the ionization
levels of the gas.  We use recombination lines (RLs) such as \heI/\hbeta, and
\heII/\hbeta, and commonly used ratios of collisionally excited lines (CEL),
such as [\oIII]/[\oI] ($=$O$_{31}$), [\oIII]/[\oII] ($=$O$_{32}$),
[\neIII]/[\oII] (Ne3O2), and [\arIV]/[\arIII] (hereafter Ar$_{43}$).  

\textbf{Optical depth tracers.} We record the `\sII\ deficit'
\citep{Wang.2019,Wang.2021}, which is defined as the logarithmic distance of a
galaxy in [\sII]/\halpha\ from the sequence of galaxies in the plane of
[\oIII]/\hbeta\ vs.  [\sII]/\halpha.  As such, it approximately encodes the
lack of [\sII] emission at fixed metallicity and ionization state, which could
imply low column densities as the partially neutral medium is truncated.  We
also add the ratios of \heI\ lines 3888/6678 and 7065/6678, which have both
been shown by \citet{Izotov.2017} to be subject to transfer effects.  

\textbf{Thermodynamic variables.} We calculate electron density
($n_\mathrm{e}$) from [\sII] 6717/31 ratio and electron temperature
($T_\mathrm{e}$) from [\oIII]4363/5007 ratio using the iterative method in
\texttt{PyNeb} \citep{Luridiana.2015}.  We calculate pressure ($P_\mathrm{e}$)
from the product of the two.

\subsection {Lyman alpha Measurements -- the Dependent Variables} 

We make most of our measurements concerning the \lya\ emission (fluxes,
equivalent widths, and kinematic properties) using \emph{Lyman alpha Spectral
Database} \citep[LASD,][]{Runnholm.2021} and refer the reader to that paper for
a detailed description.  In summary we rely upon the following quantities: 

\textbf{\lya\ luminosity.} This is  computed by numerical integration of the
continuum-subtracted spectrum over a 2500~\kms\ window centered around the
systemic velocity.  This quantity, and all that follow from it, also includes
any interstellar \lya\ absorption -- I.e. \lya\ absorbed out of the stellar
continuum (most frequent at $v<0$) subtracts from the net \lya\ flux.  

\textbf{\lya\ equivalent width.} This follows from luminosity computed above.
Continuum placement is handled using clipping algorithms inside of the LASD,
and interpolated to 1216~\AA.  In this paper we adopt the `nebular definition'
of EW, that emission is positive.

\textbf{\lya\ escape fraction, \fesclya.} This is computed using dust-corrected
Balmer line emission.  For this we use the prescription defined in
\citet{Hayes.2005} as $f_\mathrm{esc}^{\mathrm{Ly}\alpha} =
L_\mathrm{obs}^{\mathrm{Ly}\alpha} / (8.7\times
L_\mathrm{int}^{\mathrm{H}\alpha})$, where $L_\mathrm{obs}^{\mathrm{Ly}\alpha}$
is the observed \lya\ luminosity, $L_\mathrm{int}^{\mathrm{H}\alpha})$ is the
intrinsic (dust corrected; see Section~\ref{sect:meth:optical}) \halpha\
luminosity. The factor 8.7 stems from the intrinsic \lya/\halpha\ ratio
\citep[see arguments in][]{Hayes.2015}.

\textbf{The first moment.} This characterizes the velocity shift relative to
line centre and is calculated over the same wavelength range as above.

\textbf{The second moment.}  This characterizes the velocity width of the line
and is calculated over the same wavelength range as above.

\textbf{The third moment.}  This quantity, commonly known as the skewness,
characterizes the asymmetry of the emission line. It is is calculated over the
same wavelength range as above.

As well as integrated over the entire spectral region (full width of 2500~\kms),
we also calculate all these quantities for the negative-only and positive-only
velocities, which we refer to as blue and red components.  We acknowledge that
it may not be clear how to interpret some of these quantities when divided into
the blue and red parts, as frequency redistribution can shift wave packets
across line centre.  We therefore emphasize that these flux-related quantities
should be thought of as the \emph{contribution of the blue/red-shifted emission
to the total output \lya.}

\subsection {UV Continuum Measurements -- `Explanatory' Variables}
\label{sect:meth:uvcont}

We use ultraviolet absorption lines to explain the phenomena we see when
contrasting nebular measurements with the \lya, and focus on three species:
\siII$\lambda 1260$~\AA, \cII$\lambda 1334$\AA, and \siIV$\lambda 1403$~\AA.
The former two ions have relatively low ionization potentials: Si$^+$ requires
8.2~eV to be produced, and requires 16.3~eV to photoionize to Si$^{2+}$; the
corresponding range for C$^+$ is 11.3--24.4~eV.  The Si$^{+}$ and C$^+$ zones
therefore overlap with the neutral zone of hydrogen (0--13.6~eV).  \siIV, on
the other hand, requires at least 33.5~eV to be produced and probes gas ionized
by higher energy radiation, potentially with a significant contribution from
collisional processes.  The stellar modeling described in
Section~\ref{sect:method_stellarfit} automatically corrects for contamination
of narrow interstellar features by broad P\,Cygni wind lines. 

Unfortunately, the individual UV spectra do not have sufficient signal-to-noise
ratio in the continuum to well measure these absorption lines target-by-target.
We therefore rely upon a stacking analysis, similar to that performed for \lya\
(Section~\ref{sect:res:lyaprof}).  Because of the required stacking we cannot
treat these as independent variables upon which to base a differential
comparison, but use them to further explain the trends seen in \lya; we
therefore refer to these measurements as \emph{explanatory variables}, and
investigate them in a post-hoc fashion.

The absorption line studies are concerned with the relative strength compared
to the continuum itself, so we perform continuum normalization prior to
analysis (in contrast to the continuum \emph{subtraction} performed for the
\lya\ measurements).  This is done by dividing out the best fitting
\emph{Starburst99} model described in Section~\ref{sect:meas:stell}, which
leaves a normalized spectrum that is free from stellar features and contains
only interstellar lines (see the insets of Figure~\ref{fig:contmodels}).  In
the normalized spectra we measure: 

\textbf{Equivalent width of absorption lines.}  The flux is computed by direct
numerical integration, and divided by the continuum level (in normalized
spectra this is anyway set to 1).  We conduct the integral over a velocity
range of $-800$ to $+200$~\kms.  This asymmetric window may appear as though it
would bias our results, however (1.) no absorption is seen in our spectra at
velocities higher than 200~\kms, as the overwhelming majority of the absorption
is blue-shifted.  Moreover (2.) there is a fluorescent emission line from \cII\
280~\kms\ red-ward of the resonance absorption line, that would contaminate our
measurement.  Note that for our definition, the absorption lines have negative
EW.

\textbf{Velocity offset of absorption lines.}  We take the first moment of the
absorption line, computed over the same window as described above.  This
provides an estimate of the characteristic velocity at which the bulk of the
foreground gas is outflowing.

\textbf{Absorbed fraction at zero velocity.}  In normalized spectra, this is
just the flux density measured at $\Delta v=0$, and encodes the amount of
foreground gas that is static (recall that the systemic redshift is derived
from optical line emission).  We take the direct average of the four resolution
elements closest to $\Delta v=0$ (two at positive and two at negative
velocities).

\textbf{Equivalent width of fluorescent emission lines.}  For the \siII\ and
\cII\ lines we also compute the EW of their fluorescent counterparts, in
exactly the same way as for the absorption lines above.  More details can be
found in Section~\ref{sect:find:fluor}.

\section{Results I -- Properties of the Sample from Stellar and Nebular
Spectroscopy}\label{sect:results_measurement}

\subsection{Ionization Conditions and Nebular Excitation } 

\begin{figure*}
\noindent
\begin{center}
\includegraphics[width=0.99\linewidth]{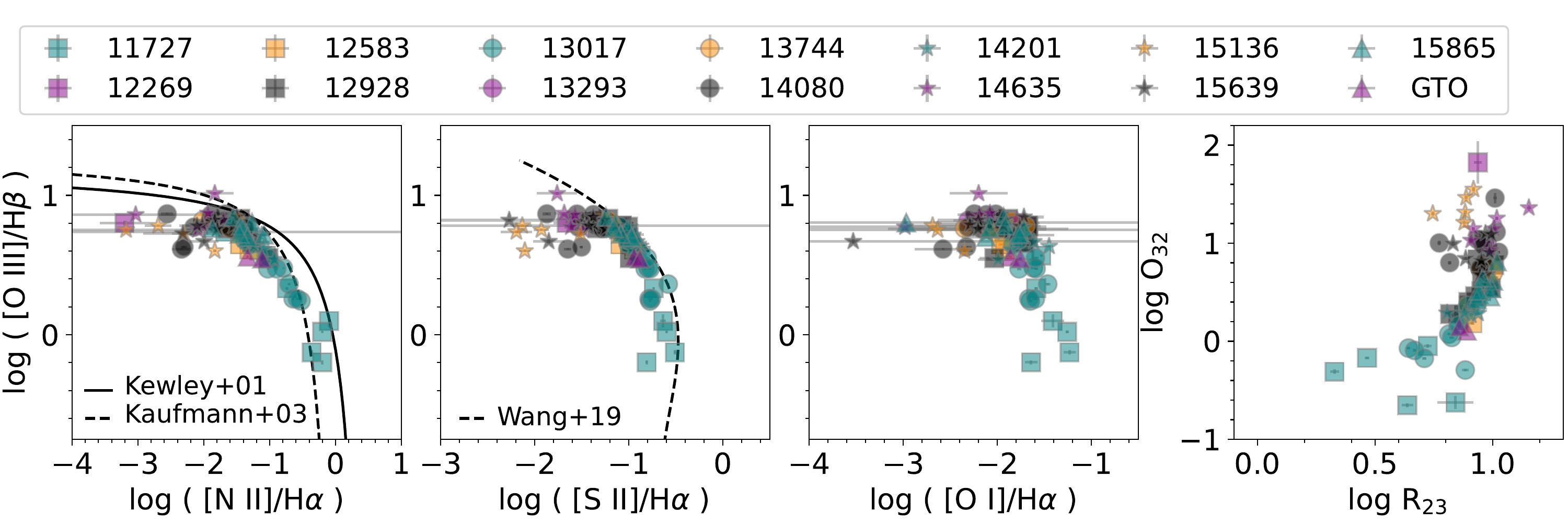}
\end{center}
\caption{BPT and excitation diagrams.  The three common diagnostic diagrams of
\citet{Baldwin1981} are shown, along with the excitation diagram of
\citet{Nakajima.2014} to the far right.  The BPT diagrams all show the log of
the [\oIII]/\hbeta\ ratio on the ordinate axis.  Abscissae show the
[\nII]/\halpha, [\sII]/\halpha, and [\oI]/\halpha\ ratios.  Demarcation lines
of \citet{Kewley.2001} and \citet{Kauffmann.2003agnhost} are shown that
segregate starburst galaxies from AGN in the upper left figure, while the line
defining the [\sII]-deficit \citep{Wang.2019} is shown in the second diagram.
The far right figure shows the logarithm of the [\oIII]5007/[\oII] ratio
\emph{vs.} the ([\oII]+[\oIII]4959,5007)/\hbeta\ ratio, specifically used to
show the most highly ionized ISM conditions towards the upper right.  Program
identifiers for the individual COS programs are indicated using various colors
and symbols (labelled in the legend).  } 
\label{fig:bpt}
\end{figure*}

We begin by characterizing the sample, using the stellar and nebular properties
derived in Section~\ref{sect:meas}.  We show the commonly-used sequence of
diagnostic diagrams \citep[][]{Baldwin1981}, hereafter `BPT-diagrams', in
Figure~\ref{fig:bpt}.  These figures encode the ionization conditions,
excitation, and metallicity along various loci.  Upper left shows the
[\oIII]/\hbeta\ \emph{vs} [\nII]/\halpha\ plane that is typically used to
identify star-forming galaxies from AGN, demarcated by theoretically and
observationally determined functions
\citep[][respectively]{Kewley.2001,Kauffmann.2003agnhost}.  Galaxies are spread
along the star-forming sequence.  A handful of galaxies do cross into the
AGN-dominated regions, but none exceed the higher curve by more than $1\sigma$,
and there is little indication of AGN contamination in the sample.  Program
identifiers from the HST observational campaigns are coded by various symbols
and colors: 11727 and 13017 spread to more metallic and low excitation end of
the diagram, where cooling is more efficient.  In contrast, programs targeting
low-mass and metallicity, highly excited starbursts fall at the upper left end
of the diagram.

The upper right plot shows a similar pattern, with galaxies falling along a
narrow sequence. The dashed line was defined in \citet{Wang.2019}, and
represents a fit to all the star-forming galaxies in the SDSS spectroscopic
samples.  All galaxies fall close to, or somewhat to the left
(`[\sII]-deficient') side of the line.  This deficit is most clearly visible in
GO\,15136 and 15639, which were both selected to be LyC- and \lya-emitting
candidates with the most highly ionized interstellar media; they were not,
however, selected by $\Delta$[\sII].  

The lower right panel shows where the galaxies fall in the plane of
[\oIII]/[\oII] (O$_{32})$ \emph{vs} ([\oII]+[\oIII])/\hbeta\ (R$_{23}$) ratios.
Similar to the BPT diagrams, this `excitation diagram' encodes the level of
excitation on both axes, while the ordinate includes an additional factor of
nebular metallicity.  It therefore appears as a rotated version of the BPT
locus.  The diagnostic diagram has been used extensively as an indicator for
potential LyC leakage as it identifies highly ionized interstellar media
\citep[e.g.][]{Nakajima.2014}.

\begin{figure}
\noindent
\begin{center}
\includegraphics[width=0.49\linewidth]{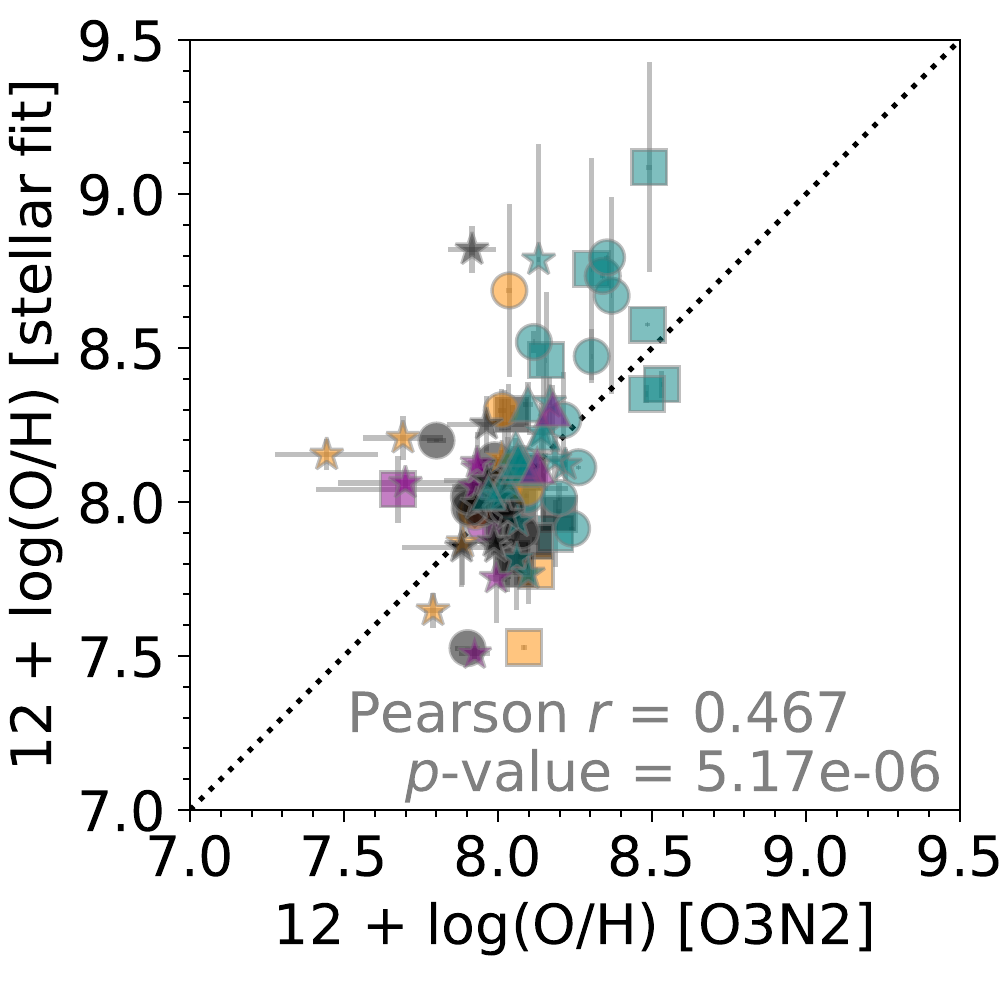}
\includegraphics[width=0.48\linewidth]{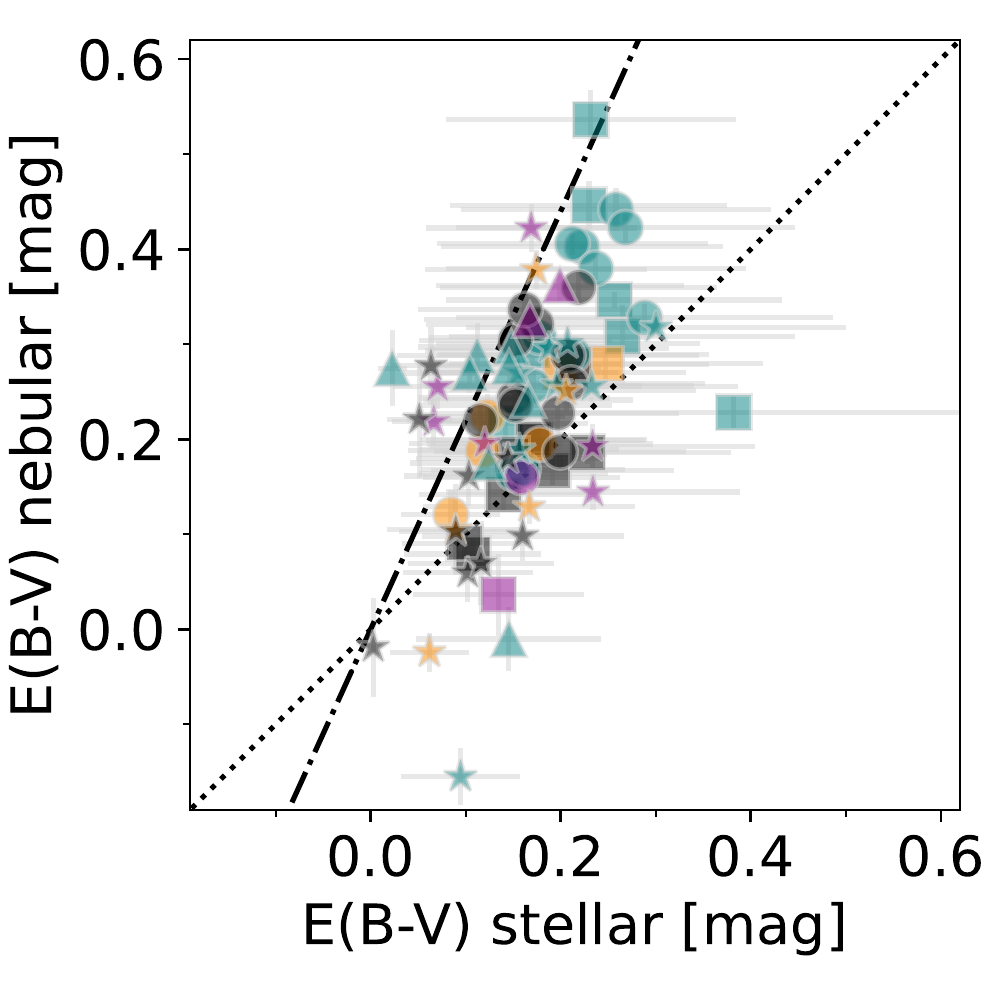}
\end{center}
\caption{Left shows a comparison of stellar and nebular metallicity estimates.
This example shows O3N2 strong-line calibration \emph{vs} the estimates made by
fitting the stellar continuum.  Note that stellar metallicities were estimated
using models calculated with the outdated solar oxygen abundance of
$12+\log(\mathrm{O/H})=8.9$.  Right shows a comparison of stellar and nebular
dust attenuation estimates.  This example shows \ebv\ derived from the
\halpha/\hbeta\ ratio against that corresponding value derived by modeling the
stellar spectra.  Dotted lines show equivalence in both cases, while the
dot-dashed line in the right panel shows nebular nebular extinction at 2.2
times the stellar value (see text for details). The color and shape coding of
the markers are the same as in Figure~\ref{fig:bpt}. }
\label{fig:metallicity_comparison}
\end{figure}

\subsection{Heavy Element Abundances and Dust
Obscuration}\label{sect:res:meas:metaldust}

In the left panel of Figure~\ref{fig:metallicity_comparison} we show how the
metallicities compare when derived using stellar and nebular methods.  For the
nebular gas we adopt the strong-line method of the O3N2 index
([\oIII]5007/\hbeta~$\times$~\halpha/[\nII]6584), using the calibration of
\citet{Marino.2013}.  While we have made estimates using temperature sensitive
methods, the faintness of the [\oIII]4363 feature in many of our galaxies means
this method is of little use for comparing galaxies at very different
distances.  Overall there is reasonable agreement around 12+log(O/H)\,$\approx
8.0$, but the scatter is large at all abundances.  Pearson's $r$ coefficient,
which measures the degree of linearity between the two measurements recovers a
$p$-value of $5.2\times10^{-6}$, but it is noteworthy that at higher
metallicities the stellar methods tend to estimate higher metallicities.
Moreover by eye it appears that the relation between the two measurements is
steeper than the 1--1 slope.  

It has been noted at high-redshifts that metallicities derived from stellar
continua tend to be lower than those derived in the gas.  \citet{Steidel.2016}
show that stellar metallicities fall below those of the nebular phase by a
factor of 3--5 for $z=2-3$ Lyman Break Galaxies (LBG). They  attribute this
phenomenon to enhancement of alpha-based products of nucleosynthesis in young
star forming events, suggesting that the recent explosion of many core-collapse
supernovae have enriched the ISM with oxygen (which dominates nebular
measurements in the optical). In contrast the lack of thermonuclear supernovae
results in a relative absence of iron, which is the dominant source of opacity
in stellar atmospheres and the main source of metallicity information in
ultraviolet stellar fitting.  In similar lower redshift studies that use
similar techniques, \citet{Chisholm.2019} do not find this lack in stellar
metals: they instead report very tight, linear agreement between stellar and
nebular metallicities, suggesting that the enrichment processes occur on longer
timescales.

The right panel of Figure~\ref{fig:metallicity_comparison} shows how the
stellar and nebular estimates of the dust attenuation compare. There is
generally a strong trend between the two estimators, although they clearly do
not align along the one-to-one relation (dotted line), and nebular attenuation
is typically higher than that of the stars. In fact, the locus of points is
closely aligned with a factor of 2.2 scaling between the nebular and stellar
values. This is consistent with results from local starburst galaxies
\citep[][see also discussion in
\citealt{Reddy.2015}]{Calzetti.2000,Kreckel.2013}.  We note, however, that
there is a slight offset between the locus of points and the 2.2-scaling line,
of about $\Delta E_\mathrm{B-V}\approx +0.05$ mag in the nebular value. Both
the general slope and offset are confirmed when replacing the \halpha/\hbeta\
ratio with that of \hgamma/\hbeta.  This offset remains unexplained, but we
speculate it could be due to the differing relative geometries of stars and gas
that may be implicit in our sample compared to those of more local starbursts
where these comparisons have more-often been performed.

\begin{figure}
\noindent
\begin{center}
\includegraphics[width=0.49\textwidth]{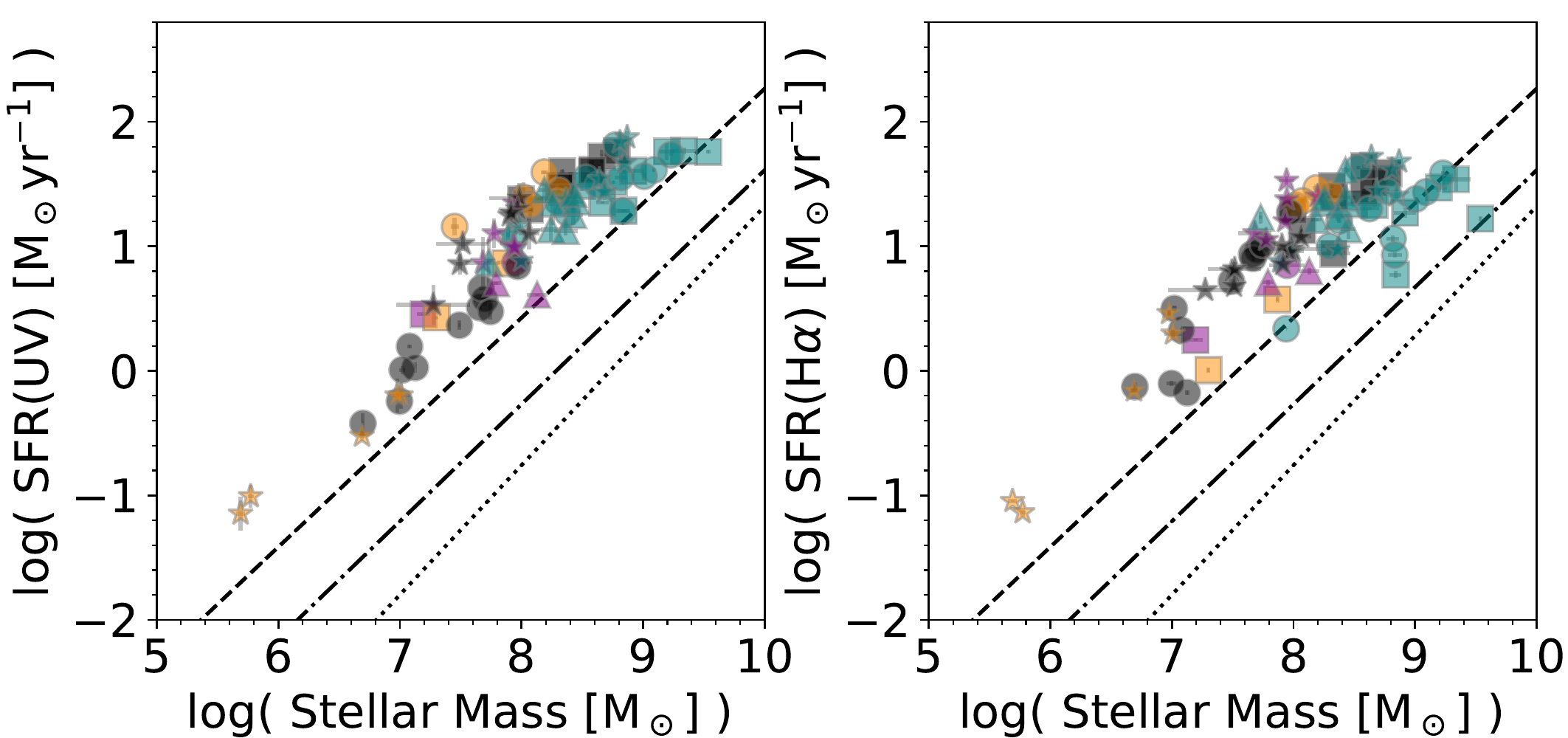}
\end{center}
\caption{The relationship between stellar mass and star-formation rate.
Stellar mass is always derived from spectral modeling.  \emph{Left}: the
UV-based SFR, corrected for stellar reddening derived from stellar population
fitting.  \emph{Right}: the \halpha-based SFR, corrected for nebular reddening
using the \halpha/\hbeta\  ratio (or \hgamma/\hbeta\ on occasion).  Linear fits
to the same relation  at redshifts 1,4, and 5 are shown by increasing lines,
derived by \citet{Santini.2017}.  The color and shape coding of the markers is
the same as in Figure~\ref{fig:bpt}.  }
\label{fig:mainsequence}
\end{figure}

\subsection{Mass and SFR Relations} 

Figure~\ref{fig:mainsequence} shows the relation between star formation rate
and the stellar mass, sometimes referred to as the star-forming main sequence.
Stellar masses are derived routinely from the continuum measured in the COS and
SDSS spectroscopic apertures (Section~\ref{sect:method_stellarfit}), and must
On the left we show the SFR derived from the UV continuum flux, corrected for
dust attenuation on the continuum, and using the calibration of
\citep{Kennicutt.2012}.  The right figure shows the SFR estimated using the
\halpha\ luminosity, corrected for dust using the \halpha/\hbeta\ (or on
occasion the \hgamma/\hbeta) ratio.  Overlaid are relations derived at
redshifts 1, 4, and 5 by \citet{Santini.2017}, which increase in their
normalization with redshift, but otherwise follow a very similar slope. 

\begin{figure}
\noindent
\begin{center}
\includegraphics[width=0.49\textwidth]{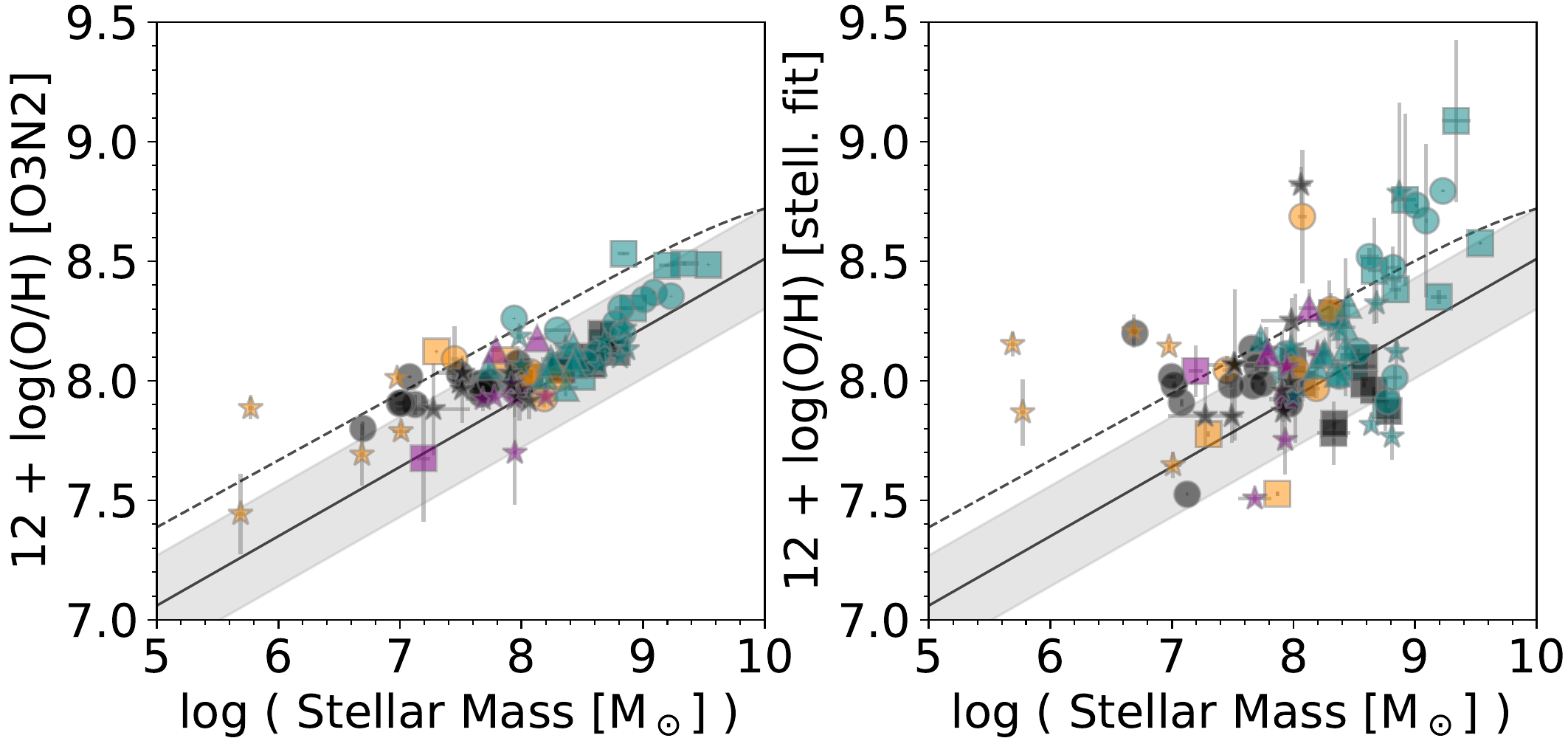}
\end{center}
\caption{The mass-metallicity relation.  \emph{Left} shows the metallicity
derived from the O3N2 index, and \emph{right} shows the metallicity derived
from the stellar fitting (see also Figure~\ref{fig:metallicity_comparison}).
Note that stellar metallicities were estimated using models calculated with the
outdated solar oxygen abundance of $12+\log(\mathrm{O/H})=8.9$.  The solid line
and shaded region shows the low-$z$ relationship derived from the Local Volume
Legacy Survey \citep{Berg.2012}, while the single dashed line is the
relationship of \citet{Curti.2020}.  The color and shape coding of the markers
are the same as in Figure~\ref{fig:bpt}.}
\label{fig:mzr}
\end{figure}

\begin{figure*}
\noindent
\begin{center}
\includegraphics[width=0.95\textwidth]{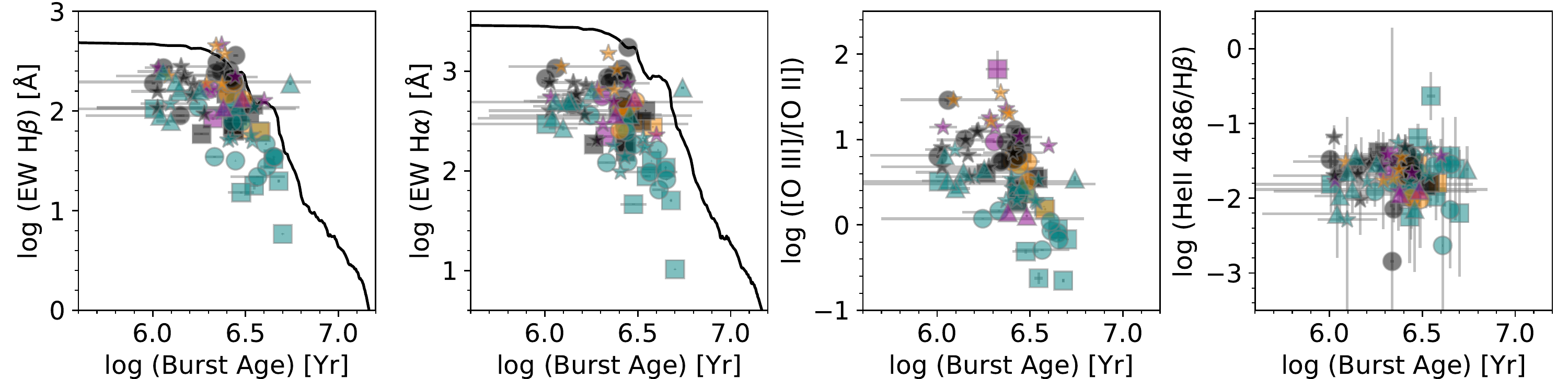}
\end{center}
\caption{The evolution of some characteristic optical observables with derived
age of the starburst component (younger). \emph{Far left} shows the equivalent
width of \hbeta\ and  \emph{centre left} the EW of \halpha.  The EWs have been
corrected for the differing levels of dust attenuation experienced by the stars
and gas, which are estimated independently by stellar modeling and the Balmer
decrement.  The tracks correspond to instantaneous burst Starburst99 models for
$Z=0.008$ and a Salpeter initial mass function.  \emph{Center right} shows the
O$_{32}$ ratio and \emph{far right} the ratio of \heII4686/\hbeta.  The color
and shape coding of the markers are the same as in Figure~\ref{fig:bpt}.}
\label{fig:age_vs_optlines}
\end{figure*}

Our sample is offset from the SFR-stellar mass relations in the direction of
higher SFR at fixed mass.  Here we caution that our methods and selection are
both very different from those applied in other studies.  Firstly in selection,
some of these objects are identified as being among the most extreme starburst
events present in the local universe. They have the highest equivalent widths
in the optical emission lines (\halpha\ and [\oIII]), which would naturally
shift the locus of points to higher values, especially compared to the lower
redshift sequences.  Secondly the stellar masses are measured in small
apertures corresponding to 1\farcs 25 in radius.  Our modeling is designed to
recover the recent star formation history by fitting multiple populations of
stars; this should add more leverage to capture underlying stellar populations,
but may still remain biased towards the youngest population with the lowest
mass-to-light ratio.  This should accurately reproduce the burst mass, but is
likely to leave the total stellar mass underestimated, shifting galaxies
towards the left on this Figure.

Figure~\ref{fig:mzr} shows the mass metallicity relation (MZR), including
metallicities derived by both strong line calibrations (left) and stellar
continuum modeling (right).  The overlaid relations are derived from SDSS
galaxies \citep[][dashed black line]{Curti.2020} and for dwarf galaxies in the
Local Volume Legacy Survey \citep[][black line]{Berg.2012}. The agreement with
the MZR for local galaxies is surprisingly good, given the lack of agreement
with the SDSS-derived SFR-\Mstell\ relation (Figure~\ref{fig:mainsequence}).
Together these figures suggest that the disagreement in
Figure~\ref{fig:mainsequence} is actually not because the mass has been
underestimated because of aperture effects, but because the SFR is elevated
compared to typical local galaxies of equivalent stellar mass.

Based upon the SFR main sequence (Figure~\ref{fig:mainsequence}) we could
expect our sample to lie above the $z=0$ MZR, since they are currently
producing more metals than the average galaxy at fixed mass. This is true, but
only marginally so: points appear to occupy roughly the upper half of the grey
band describing local dwarfs.  At higher redshifts, however, the MZR evolves
towards \emph{lower} metallicity at fixed mass \citep{Erb.2006,Maiolino.2008}.
Thus, while a agreeing reasonably with the local MZR, our sample diverges from
the mass-metallicity relation seen at higher redshifts.

Figure~\ref{fig:age_vs_optlines} shows how some nebular quantities evolve with
starburst age.  For age, we adopt that of the starburst (youngest) population
that provides the majority of the hydrogen-ionizing photons.  We first show the
the equivalent widths of the Balmer lines, \hbeta\ and \halpha, which should
drop rapidly when the most massive and ionizing stars explode.  In this figure
only we take the unconventional step of correcting the EWs for dust, using the
independently-derived stellar attenuation (from modeling) and nebular reddening
(calculated from the \halpha/\hbeta\ ratio).  It is often noted that inferred
absorption due to dust is higher when measured from the nebular gas when
compared with stellar light.  This is believed to be the result of geometrical
effects \citep[e.g.][]{Calzetti.2000}, and is also verified in this sample (see
Section~\ref{sect:res:meas:metaldust}). 

We overplot \emph{Starburst99} evolutionary tracks for the same quantities
\citep{Leitherer.1999}.  It is clear that the theoretical track always
over-predicts the EW, in an effect that is stronger for \halpha\ than for
\hbeta.  We believe this is mostly because the redder wavelengths include
larger contributions from more evolved stars: the multi-component modeling
described in Section~\ref{sect:method_stellarfit} does account for a more
evolved population, and we confirm that strong absorption is seen (typically
\hepsilon\ to about H12) in the high-order Balmer lines in these cases.  This
is consistent with dilution of the EWs typical of single-population starbursts.
Moreover, the escape of ionizing photons can also reduce the EW of
recombination lines while leaving all colors basically unaffected
\citep[e.g.][]{Bergvall.2013}.  While this scenario is possible, we do not
expect escape fractions to typically be high enough to cause significant
reductions to the observed EWs.  It is encouraging to see such agreement for
the \hbeta\ line where the contamination is lower than for \halpha, which would
be expected in the case of contamination from cooler stars.  Overall, we
interpret these findings as support for the spectral modeling, reinforcing the
belief that it has produced acceptable ages for the hard-to-model starburst
component, and reasonable production rates for hydrogen-ionizing photons. 

The centre right plot shows the time-evolution of the O$_{32}$ ratio, which
also decreases substantially with time.  Like Balmer line fluxes, this ratio
peaks at very early times when stars produce more $h\nu\ge 34.4$~eV photons
that can doubly ionize oxygen.  In this case it may be tempting to attribute
the evolutionary effect directly to an excess of photons at $h\nu\ge 34.4$~eV
compared to 13.6~eV.  The \emph{far right} plot, however, shows this
interpretation unlikely to be correct as the same phenomenon is not noticed
when comparing recombination lines of \heII\ (54.4~eV) with hydrogen.  This
result mirrors that of \citet{Marques-Chaves.2022}, who find that the
\heII/\hbeta\ ratio does not correlate with either the ionizing escape
fraction, while the O$_{32}$ ratio in the same sample does
\citep{Flury.2022pap2}.  We expect, therefore, that the evolution of O$_{32}$
seen here is instead related to a truncated O$^+$ zone, as the O$^{2+}$ region
is elevated due to large scale photoionization in younger galaxies.

\begin{figure}
\noindent
\begin{center}
\includegraphics[width=0.48\textwidth]{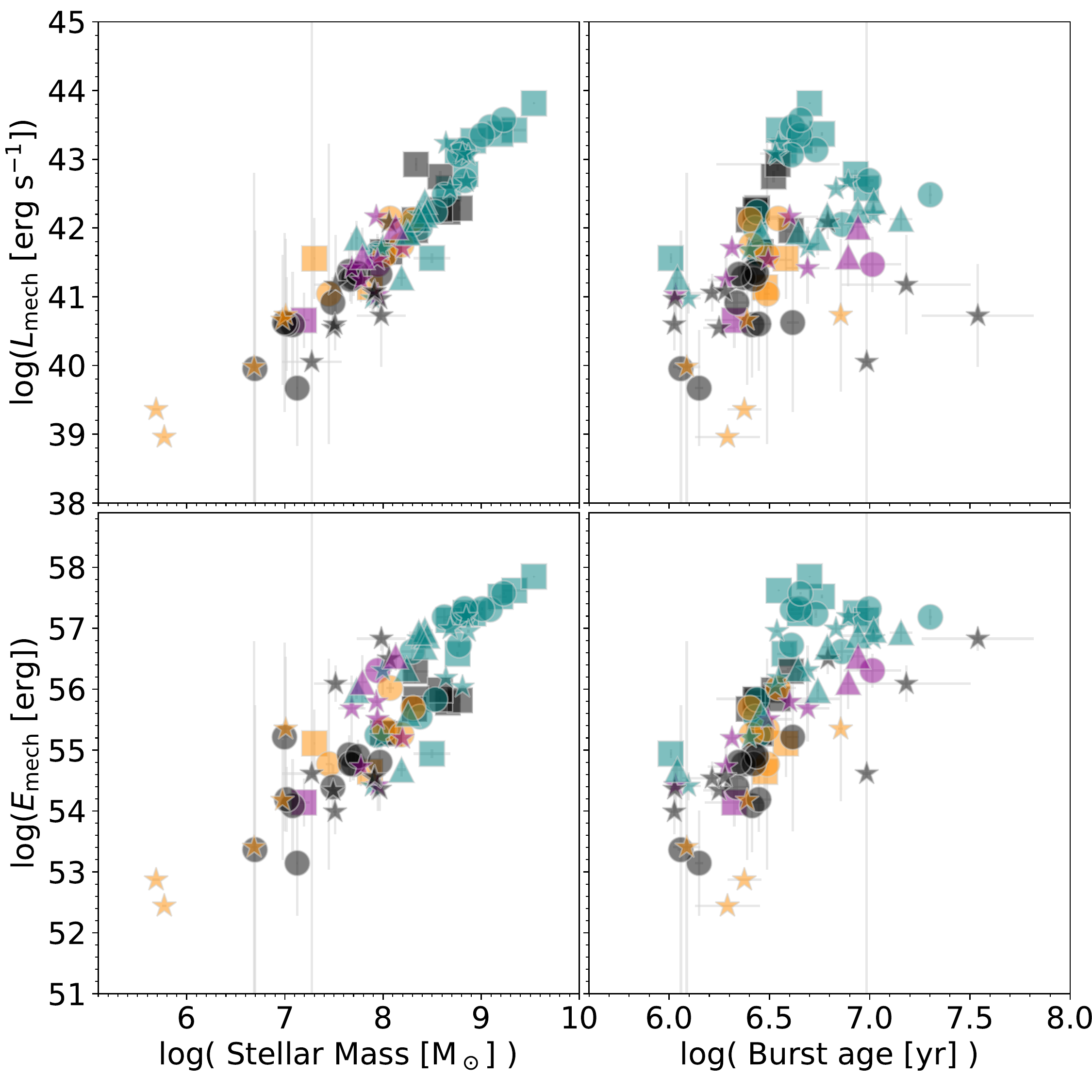}
\end{center}
\caption{The mechanical energy return from stellar feedback.  The upper row
shows the instantaneous rate of energy injection (mechanical luminosity) from
stellar winds and supernovae, projected against the mass formed in the
starburst (left) and age of that population (right).  The lower row shows the
integral of the mechanical luminosity over time since the onset of the burst
(the total mechanical energy injected).  The color and shape coding of the
markers are the same as in Figure~\ref{fig:bpt}.}
\label{fig:energy_return}
\end{figure}

\subsection{Mechanical Luminosity and Kinetic Energy Return}

With reasonable estimates of the mass and evolutionary point of the starburst
episode, we estimate the rate of mechanical energy injection into the ISM as a
result of stellar feedback processes (the mechanical luminosity, \Lmech). The
total amount of mechanical energy returned since the onset of the starburst
(\Emech; see Section~\ref{sect:meas:uvprop}) then follows by integrating along
the star formation history.  We show the relations between \{\Lmech, \Emech\}
and \{mass, age\} in Figure~\ref{fig:energy_return}.   We recover a large range
in \Lmech\ between $10^{39}$ and $10^{44}$~\ergsec, and as shown in the upper
left panel, the bulk of this is explained almost linearly by the large range of
burst masses.  The dispersion on this relation is quite small and comes from
the time-dependent instantaneous mechanical luminosity injection, which varies
with dominant mode (stellar winds or supernova) and position along the
(current) stellar mass function \citep[see, for example, Figure~111 of
][]{Leitherer.1999}\footnote{\href{https://www.stsci.edu/science/starburst99/figs/lmech_inst_c.html}{https://www.stsci.edu/science/starburst99/figs/lmech\_inst\_c.html}}.
We plot \Lmech\ as a function of burst age in the upper right panel, which also
shows a broadly positive correlation, in this case because the supernova
explosion rate increases with time until about 40~Myr. The very large
dispersion results from the large dynamic range in burst masses already
discussed.

The lower panels in Figure~\ref{fig:energy_return} show the total mechanical
energy injected since the onset of star-formation, again plotted against mass
formed and burst age. The range of \Emech\ runs from $10^{52}$ and
$10^{58}$~erg, which for comparison corresponds to the energy returned by $\sim
10$ to $10^7$ supernovae.  The relation between total \Emech\ and mass is
somewhat more dispersed than the corresponding relation with \Lmech, spanning a
range of $\approx 2.5$~dex at \Mstell~$=10^8$~\msun.  This expected, since
\Emech\ is the integral of \Lmech\ from $t=0$ to now, and depends more
sensitively upon the time since the burst ignited -- see the corresponding
points in the lower right panel, which show this variation of \Emech\ with
stellar age.

\subsection{Summary of Estimated Properties}

This section demonstrates that our combined sample of almost 90 starburst
galaxies spans a large and useful range of the multi-dimensional parameter
manifold.  Galaxies span a broad range of positions on the BPT diagrams from
the very low metallicity and highly excited end, to more `normal', metal rich
end. They trace the star-forming sequence throughout, with only a handful of
partial exceptions. The sample spans around three orders of magnitude in
stellar mass, from $10^6$~\msun\ to more than $10^9$~\msun; it is true that
these are aperture-based values and will be somewhat lower than the total mass
in stars, but the burst mass in which we are most interested is likely
accurately recovered.  The SFR spans a similar dynamic range from 0.1~\msunyr\
to around 100~\msunyr.  

Metal abundances span a range of 12+log(O/H)\,$= 7.5-8.5$, and the
correspondence between gas phase metallicity is close to the trend that would
be expected for dwarf galaxies at low-redshift.  There is reasonable agreement
between the gas-phase and stellar metallicities; while this need not
necessarily be the case, the correlation is encouraging from the perspective of
stellar modeling. It suggests that our efforts to break the age-metallicity
degeneracy are successful, and that there is meaningful information to be
extracted from the star-formation histories/stellar ages. This is important
because it later feeds into the estimates of ionizing photon production rates
and mechanical energy return.  The conclusion is further backed up by the
correspondence between inferred starburst age and independent estimates based
upon the equivalent width of optical recombination lines.

\section{Results II -- Lyman Alpha Output and Fundamental Variables}
\label{sect:results_basic}

\begin{figure*}
\noindent
\begin{center}
\includegraphics[width=0.9\textwidth]{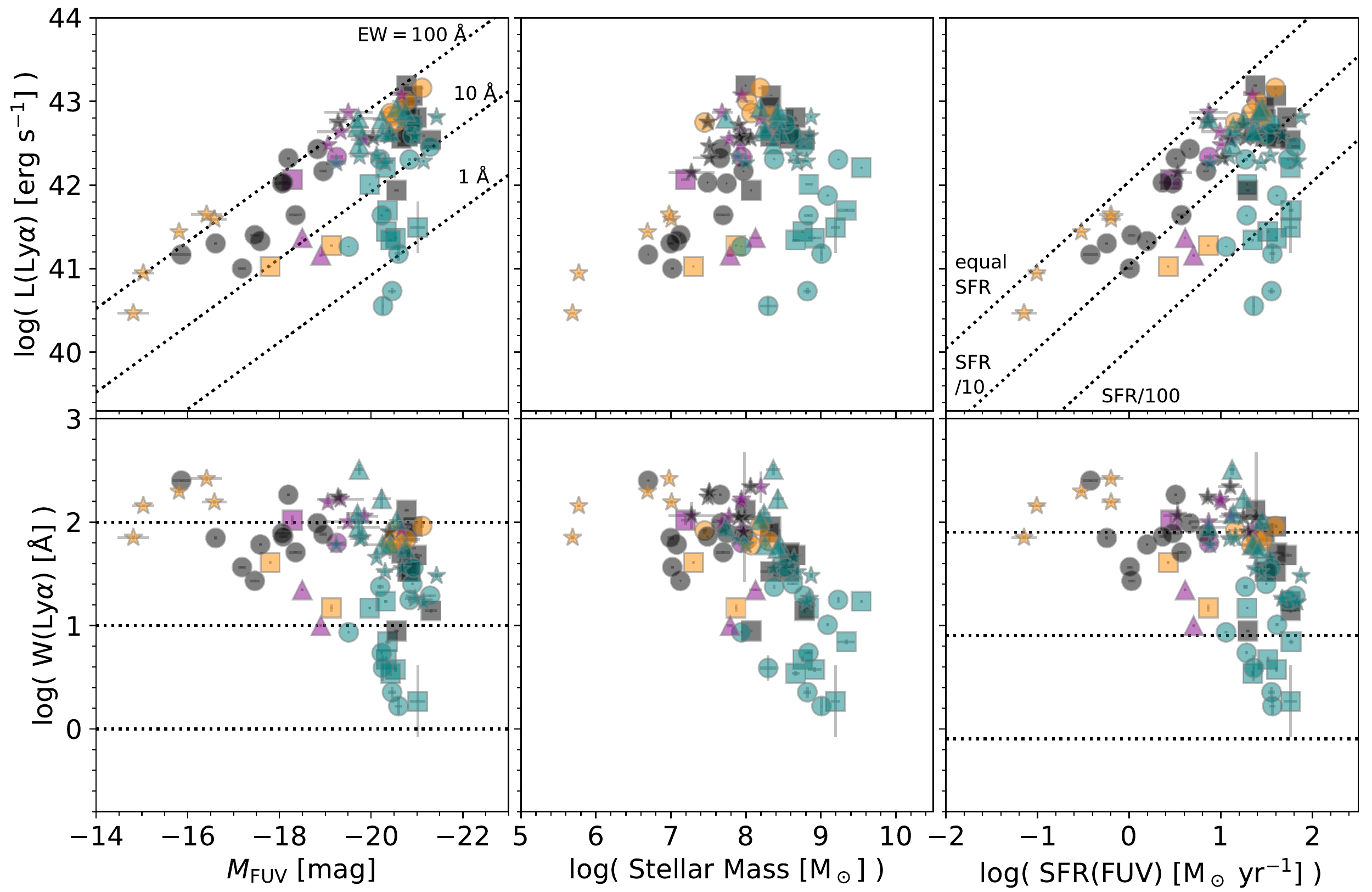}
\caption{Variation of the total \lya\ observables of luminosity (\emph{upper})
and equivalent width (\emph{lower}) with $M_\mathrm{FUV}$ (\emph{left}),
stellar mass (\emph{center}), and SFR (\emph{right}).  Dashed lines on the left
figures are of constant \lya\ EW showing 1, 10, and 100~\AA; comparable lines
on the right plots show equivalent SFR in \lya\ and FUV, and values for
SFR(\lya) scaled down by factors of 10 and 100.  Mass is derived from spectral
modeling, UV magnitude is the observed quantity and has not been corrected for
dust obscuration, while in the SFR calculation dust has been accounted for.
See Section~\ref{sect:results_basic} for details.  Color-coding shows the
original proposal from which each FUV observation was drawn, and is consistent
with Figure~\ref{fig:bpt}. } 
\label{fig:global_lya}
\end{center}
\end{figure*}

In this section we characterize the sample in terms of the bulk \lya\
properties, where we focus on the observables of \lya\ luminosity and EW.  We
show this in Figure~\ref{fig:global_lya} to address the question of where the
strongest \lya\ emitting galaxies may be found, and to facilitate more direct
comparison with high-$z$ observations.

The upper left figure shows \llya\ vs. \Mfuv, where dotted lines show constant
EW, and the lower panel shows this EW directly.  Note that the dashed lines do
not precisely agree with the EWs measured in individual galaxies; for example
the two galaxies with EW\,$<1$~\AA\ in the upper figure have
log(EW/\AA)\,$\approx 0.2$ in the lower plot.  The reason for the difference is
that \Mfuv\ is measured in GALEX data to provide the total magnitude, and
therefore samples the UV continuum at observed wavelength of 1527~\AA,
regardless of galaxy redshift.  The EWs, on the other hand, are measured in the
COS spectra to cancel out any aperture effects, and estimate the restframe
continuum flux density at restframe 1216~\AA\ directly.  While neither of these
numbers is incorrect, the inconsistency arises because of the different
restframe wavelengths and apertures sampled by COS and GALEX.

\llya\ does not scale directly with \Mfuv, but an upper envelope is clear,
where a maximum value of \llya\ appears to be imposed at EW\,$\approx 100$~\AA.
Points typically scatter below this line, with EWs of several tens of \AA, but
scatter is larger towards higher luminosities.  At \Mfuv\,$\approx -21$
($\sim$\,\Mstar\ for $z=3$ LBGs, and 3 magnitudes brighter than \Mstar\ at
$z=0$), \lya\ luminosities reach their largest values of $\approx
10^{43}$~\ergsec, which is also close to \Lstar\ for LAEs at $z\sim 3$
($10^{42.7}$~\ergsec; \citealt{Herenz.2019}). However the lowest \lya\
luminosities of $\approx  10^{40.3}$~\ergsec\ are also found at the same \Mfuv,
corresponding to EWs as low as 1~\AA.  This dispersion, in which \lya\ EWs
range between $\sim 1$ and $\gtrsim 100$~\AA\ is more visible in the lower left
plot, and is driven largely by the inclusion of cyan points, corresponding to
earlier COS programs (11727 and 13017) that targeted more massive LBG-like
galaxies.  We see less dispersion in \llya\ and EW towards the fainter end of
the \Mfuv\ distribution, where the sample is dominated by galaxies selected to
have more highly ionized ISM (e.g.  15136 from \citealt[][]{Izotov.2020}, and
14080 from \citealt[][]{Jaskot.2017}).  We do not claim that comparably low EW
\lya-emitters do not exist among such faint galaxies: this is probably a result
of sample selection where these objects, while faint, have been identified by
having the highest production efficiencies of ionizing radiation among all the
known samples.  We confirm this directly in
Sections~\ref{sect:results_profiles} and \ref{sect:mainfinds}.

A similar distribution of points is visible in the central two panels of
Figure~\ref{fig:global_lya}, which shows the variation of \lya\ with respect to
mass.  The \llya\ vs. \Mstell\ figure, however, shows a less clear upper
envelope, and the upper-right corner of the EW distribution is no longer
populated.  Both these differences can be attributed to the cyan points (again
11727 and 13017), which exhibit higher stellar masses than other samples of
comparable UV luminosity.  This higher mass-to-light ratio is the result of a
more evolved stellar populations in these sub-samples, as demonstrated in
Figure~\ref{fig:age_vs_optlines} -- it is likely that part of the reason these
galaxies less luminous in \lya\ and showing lower EWs is because less \lya\ is
produced intrinsically by the more evolved population.

The right panels of Figure~\ref{fig:global_lya} show the \llya\ and EW as a
function of SFR.  The dashed lines now illustrate equivalent SFRs in the UV
continuum and \lya, assuming star-formation proceeds at a constant rate; we
also scale down the \lya-inferred SFRs by factors of 10 and 100.  These plots
naturally resemble the left-most figures very closely: at the high SFR end
there is a lot of dispersion, again driven by the more massive and luminous
galaxies that exhibit a wider range of evolutionary stages.  There are a number
of galaxies for which the \lya\ EW exceeds expectations based upon star
formation that proceeds at a constant rate, as evidenced by targets with \lya\
EW greater than $\approx 100$~\AA. The explanation for this is most likely a a
younger star-formation episode, which causes the emission lines to be stronger
with respect to the continuum. 

The conclusion of this section must be that, while more massive, luminous, and
rapidly star-forming galaxies must by construction produce more \lya\
intrinsically, this is only sometimes visible in their \lya\ output.  At higher
luminosities and SFRs, a range of additional processes must be available to, in
some cases, suppress the \lya. This is consistent with the notion that larger
galaxies possess not only, on average, a larger \ion{H}{1} column density which
increases the path length of \lya\ photons and thus their attenuation -- but
also a wider distribution of  column densities.

\section{Results III -- Shaping Lyman alpha and Ultraviolet Absorption Lines}
\label{sect:results_profiles}

In Section~\ref{sect:meas} we presented the many quantities we investigate in
this article, and we now explore how these variables shape the \lya\ spectral
shape.  We explore \lya\ observables concerning the luminosity, EW, escape
fraction, and higher order moments, examining how these depend upon quantities
that encode dust attenuation, abundances, ionization parameter, age, ionizing
photon production efficiency, etc.   We first present a detailed discussion of
a single independent variable, the O$_{32}$ ratio, in
Section~\ref{sect:res:o32}.  We then expand this discussion to other quantities
in Section~\ref{sect:res:morevar}.

\subsection{Lyman alpha Profiles and the [\oIII]/[\oII] Ratio}\label{sect:res:o32}

\subsubsection{The Lyman alpha Profiles}\label{sect:res:lyaprof}

\begin{figure*}
\noindent
\begin{center}
\includegraphics[width=0.98\textwidth]{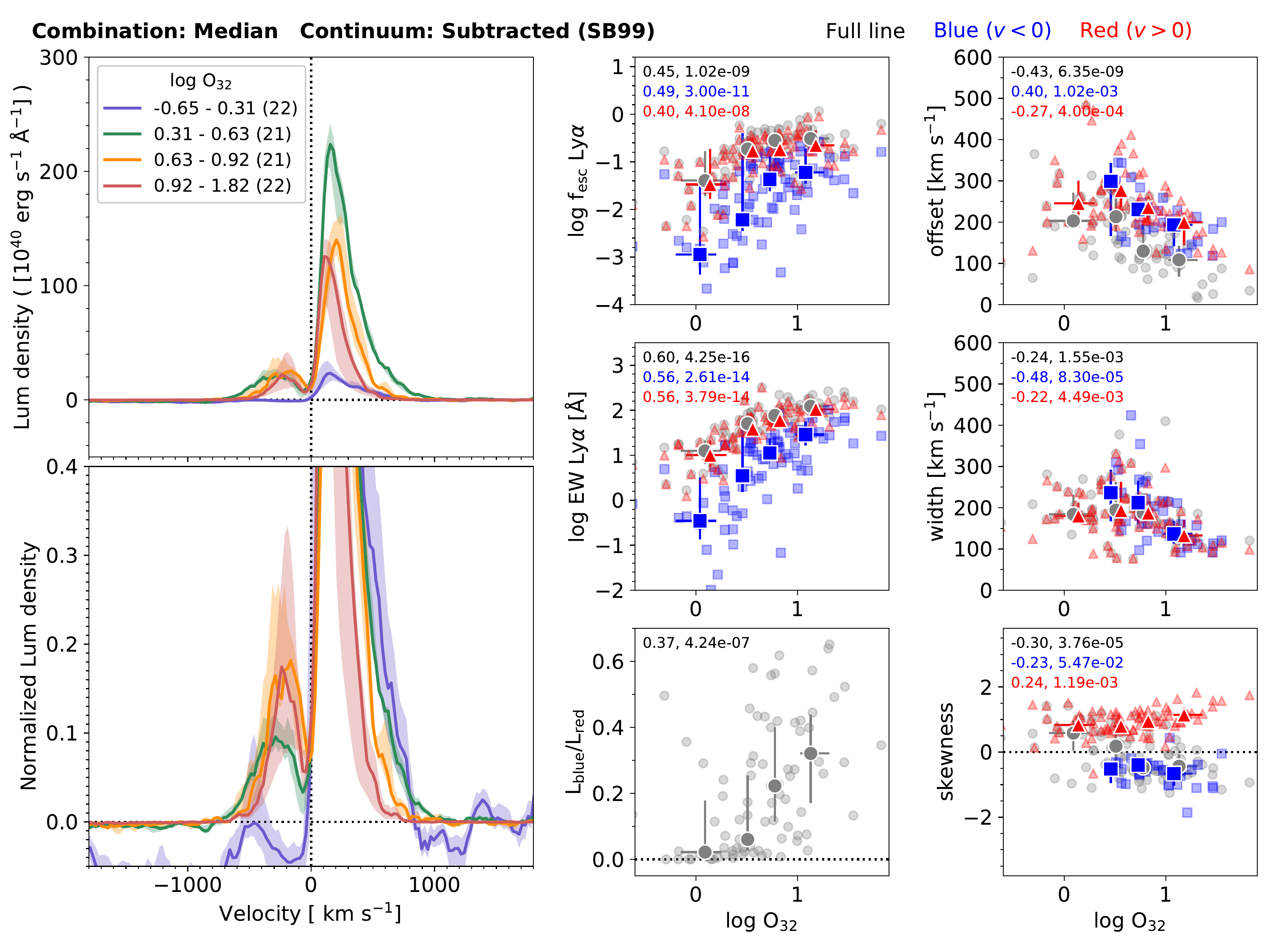}
\end{center}
\caption{The influence of the [\oIII]/[\oII] (O$_{32}$) ratio on the emergent
\lya\ line profiles.  The left two panels show combined spectra in velocity
space, stacked into four equally sized bins sorted by O$_{32}$ and
median-combined.  The \emph{upper} panel shows the absolute luminosity density,
and the \emph{lower} shows the spectra when normalized by the luminosity
red-wards of line centre to highlight variations in blueshifted emission.
Values of O$_{32}$ that define the bin edges are given in the caption.  The
$2\times 3$ panels to the right show correlation diagrams.  Quantities derived
for the whole line profile are shown in grey, and the blue (red) points show
the contribution from emission at negative (positive) velocities.  In the six
figures, we show the quantities of \fesclya\ and \ewlya\ at upper-left and
centre-left respectively, and the lowest left plot shows the \Lbluered\ ratio.
The right-most column shows the higher order moments: velocity offset (moment
1), width (moment 2), and the skewness (moment 3).  Kendall's $\tau$
coefficient is written into each figure, together with the corresponding
$p$-value.  Solid points with errorbars correspond to the stacked profiles
shown in the spectra to the far left.}
\label{fig:corrLyaO32}
\end{figure*}

In Figure~\ref{fig:corrLyaO32} we show an example of the main correlation
studies presented in this paper.  This illustration shows the O$_{32}$ ratio
([\oIII]/[\oII]), which is shown to be a strong correlate of the ionization
parameter.  The figure is divided into eight sub-figures: the left half
illustrates the average line profiles as a function of the independent
variable, derived by stacking analysis.  Solid lines show the median stacked
spectra, and shaded regions show the interquartile range (25--75 percentiles)
calculated by bootstrap analysis.  Many similar figures concerning other
quantities may be found in the supplementary online-only materials.

Several trends become clear from Figure~\ref{fig:corrLyaO32}.  Attending to the
normalized stacked spectrum (lower left), the blueshifted emission provides a
successively greater contribution to the total emergent \lya\ at higher values
of O$_{32}$.  At the lowest values of O$_{32}$\,$\approx 0.3$, \lya\ absorption
is visible at negative velocities, as more continuum radiation is absorbed in
the \lya\ resonance than is emitted by the nebular gas.  Despite absorbing
\lya, these galaxies remain in the sample because they are net \lya-emitters.
The contribution of blueshifted emission then increases to $\gtrsim 30$~\% at
O$_{32}$\,$\approx 10$. Across the same range of O$_{32}$, the redshifted
component of the emission becomes narrower, which is most easily seen at a
normalized flux density of about 0.1, where the red edge of the line profile
recedes from $\sim 600$ to $\sim 400$~\kms.

The $2 \times 3$ grid of plots to the right of Figure~\ref{fig:corrLyaO32}
shows how the O$_{32}$ ratio influences the global \lya\ output.  Grey points
show the total \fesclya\ and EW, derived over the full 2500~\kms\ window, and
include both redshifted and blue-shifted emission.  Blue (red) points show the
same quantities, but calculated only for emission at $\Delta v<0$ ($\Delta
v>0$), respectively, and can be thought of as the contribution to these
frequencies to the total output.  Kendall's $\tau$ coefficient -- a
non-parametric rank statistic testing dependence between two variables -- is
calculated for the each of the figures, and written into the upper left corner
of each subplot.  The corresponding $p$-value is written immediately below.
Firstly, and most obviously, there is a strong trend for the total escape
fraction (\fesclya) and EW to increase with the O$_{32}$ ratio, which spans
more than an order of magnitude on both axes.  In this example, the total
\fesclya\ (black) relation is significant at $p\approx 10^{-9}$, and the EW
relation is significant at $p\approx 10^{-15}$.  

Notably in this example, there are equally strong relationships in place that
describe the respective contribution of blue- and red-shifted emission.  The
red points closely trace the grey ones, which is natural because the red peak
almost invariably dominates the total \lya.  The less expected result is that
not only does the blue-shifted emission correlate as strongly, but the slope of
these correlations is steeper than those for the redshifted emission, and spans
a factor of five greater dynamic range: \emph{the blue part of the \lya\
profile is more sensitive to ionization conditions than the red part.} 

The lowest panel of the left column addresses the relative contribution of the
blue and red peaks directly, by showing the dependence of the \Lbluered\ ratio
on O$_{32}$.  In line with the more rapidly increasing EW of the blue-shifted
emission, the \Lbluered\ ratio increases sharply from effectively zero at an
average O$_{32}$ of $\approx 1$, to \Lbluered\,$\approx 0.35$ at
O$_{32}$\,$\approx 10$.

The right-most column shows various kinematic properties measured on the \lya\
profiles, by the \emph{Lyman Alpha Spectral Database} \citep{Runnholm.2021}.
In descending order we show the variation on higher order moments, 1, 2, and 3:
velocity offset, line width, and skewness.  Again they are computed for the
full line profile (black), emission bluewards of line centre (blue) and
red-wards (red).  We note here that for the left column, where \lya\ variables
are derived from moment 0 (i.e. by numerical integrations), that it is not
necessary to identify a clear peak: the integral is meaningful if the emission
is not isolated to a peak.  Higher order moments (offset, width, and skewness),
however, demand the emission be peaked.  The right-hand column, therefore, only
shows data-points where peaks have been clearly detected -- grey points and red
points will be defined for the overwhelming majority of galaxies, but blue
peaks are not ubiquitous, and there are fewer blue points than than red or
grey.  We use the peak-finding algorithm of \citet{Runnholm.2021}, which
requires a (usually second) flux maximum be found at velocities in the range
$-1000 < \Delta v < 0$ ~\kms, and that this be identified in $>95$~\% of the
Monte Carlo realizations.  We also show the absolute value of velocity offset
of the blue peak, in order to contrast the blue and points on the same axes;
this can still be thought of as the distance between the peak and zero
velocity.

The trends concerning the higher order moments are not as strong as those
related to fluxes, but a number of interesting features still become apparent.
There is a trend for the \lya\ centroid shift to decrease with increasing
O$_{32}$ (upper right).  This trend is identified for both the total and red
components, but is equally strong for the offset of the blue-shifted emission:
for the highest three bins in O$_{32}$ the blue and red-peak offsets mirror
each other.  A very similar result is seen for the second moment, where both
blue and red peaks become narrower with increasing O$_{32}$ (centre right
subpanel).  The lower right plot shows that only weak trends in skewness can be
identified for the individual peaks, and their overall symmetry is not strongly
dependent on the ionization conditions.  The skewness of the total \lya,
however, becomes systematically more negative as O$_{32}$ increases, because of
the increasing contribution of the weaker blueshifted emission.

\begin{figure*}
\noindent
\begin{center}
\includegraphics[width=0.98\textwidth]{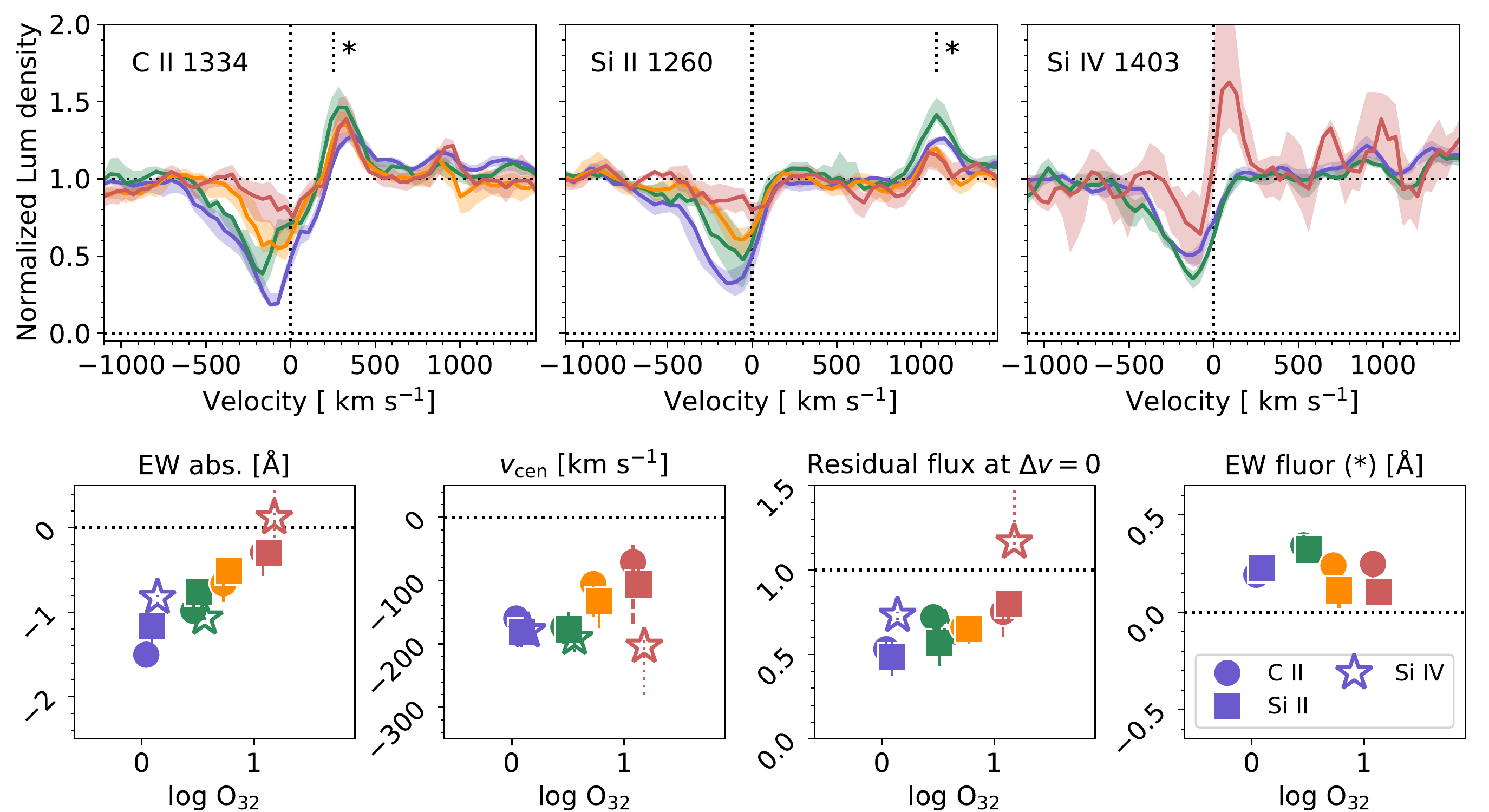}
\end{center}
\caption{The influence of the [\oIII]/[\oII] (O$_{32}$) ratio on the UV
absorption lines.  We show the \cII$\lambda 1334$\AA, \siII$\lambda 1260$, and
\siIV$\lambda 1403$~\AA\ lines in the upper row from left to right.  Spectra
are normalized to 1 in the continuum, and binning and color-coding are the same
as Figure~\ref{fig:corrLyaO32}.  Dotted vertical lines show systemic velocity
of each line, and for the \cII\ and \siII\ features we also indicate the
velocity of fluorescent fine-structure transition with a star.  The lower
panels show measurements of various sample-averaged properties measured in the
above stacks, and contrast them against O$_{32}$ -- from left to right we show:
the equivalent width of the absorption line; the centroid velocity of the
absorption line (first moment); the residual intensity at zero velocity (the
relative intensity at which an absorption feature crosses $v=0$); and the
equivalent width of the fluorescent emission line (\cII\ and \siII\ only).  The
color coding is the same in both rows, and also matches stacks in
Figure~\ref{fig:corrLyaO32}.  In the lower panels the absorbing species is
encoded by different symbols (caption in the lower right panel).}
\label{fig:corrLisO32}
\end{figure*}

\subsubsection{Lessons from Ultraviolet Absorption Lines}\label{sect:res:lis}

The results presented in Section~\ref{sect:res:lyaprof} show how various
tracers of the \lya\ output vary with a flux ratio in optical emission lines.
Whatever the underlying physics is, it influences both total \lya\ output
(luminosity), relative output (EW and \fesclya), and kinematics of the line
profile (velocity offset and width).  However this investigation itself does
not point towards a causal mechanism; to bring our analysis one step closer to
this we also examine the UV absorption lines, which we show in
Figure~\ref{fig:corrLisO32}. 

The upper row is comparable to the left panels of Figure~\ref{fig:corrLyaO32},
and the galaxies entering each sub-stack are the same in both figures.  Again,
solid lines show the median stacked spectra, and shaded regions show the
interquartile range.   Attending first to the \cII\ absorption (left-most), the
color sequence makes the opposite pattern that seen in
Figure~\ref{fig:corrLyaO32}: the absorption becomes weaker as the O$_{32}$
ratio increases.  The same effect can be seen in the \siII\ absorption line,
shown in the central panel of the upper row.  In the rightmost plot we present
the same analysis for the highly ionized gas using the \siIV$\lambda 1403$~\AA\
line.  This feature does not display such a clear trend with O$_{32}$, but it
is noteworthy that the absorption almost vanishes for the highest O$_{32}$ bin.
In fact, the net EW becomes slightly positive in this bin.

To better quantify these trends, we also compute the absorption EW, velocity
centroid, and zero-velocity flux of each of the sub-stacks, and plot them
against the independent variable in the lower panel.  The decreasing absorption
with O$_{32}$ can clearly be visualized, as the low ionization absorption lines
exhibit EWs of $\approx 1.3$~\AA\ for O$_{32}$~$\approx 1$, but fading to
0.4~\AA\ at the highest O$_{32}$.

The centre left panel shows the first moment of the absorption line, which
should trace the velocity of the (mostly outflowing) gas in front of the
hottest (UV-brightest) stars.  Average outflow velocities range from $\approx
-250$~\kms, which is quite normal for compact starburst galaxies
\citep[e.g.][some of the galaxies in whose studies overlap with
ours]{Rivera-Thorsen.2015,Henry.2015,Heckman.2015,Heckman.2016,Chisholm.2017}.
In this example there is a modest trend for galaxies with more gas covering the
stars to show faster outflows.  Concerning \lya\ emission, the faster outflows
would accelerate more absorbing material away from the \lya\ resonance and
enhance the \lya\ emission.  However, it is also clear that despite the
outflows being slower in the more highly ionized galaxies, there is also less
gas covering overall.

To address this last point, the third lower panel shows the normalized residual
flux density at zero velocity.  The \cII\ and \siII\ species in this figure
very clearly correlate with O$_{32}$: higher ionization states in the ISM
clearly imply less absorbing material close to line centre.  It is impossible
for \lya\ to avoid scattering in this static material, and the amount
(especially the column density) is responsible for splitting the intrinsic
\lya\ feature into double peaks.

Finally we draw attention to the fluorescent transitions associated with the
\cII\ and \siII\ absorption features.  These result from the split ground-state
($^2P^0_J$, where $J=1/2$ or $3/2$), and manifest as an emission line red-wards
(slightly lower energy) of the main (resonant) transition.  These are
illustrated by the dotted vertical line at $\Delta v\approx 250$~\kms\ for
\cII\ and $\approx 1000$~\kms\ for \siII\ in Figure~\ref{fig:corrLisO32}.
These fluorescent transitions (which have been studied in detail by, e.g.
\citealt{Jaskot.2014,Scarlata.2015,Carr.2018}) are detected at high
significance in all of the sub-stacks presented for O$_{32}$.  This C$^+$ and
Si$^+$ exists within the COS aperture and is excited in the same transition by
which the main lines are absorbed.  However, as the strength of these emission
components does not differ substantially with the strongly varying absorption,
it appears the emitting and absorbing gas is not the same.  We will return to
this point in Section~\ref{sect:find:fluor}.

\subsection{More Variables}\label{sect:res:morevar}

\begin{figure*}
\noindent
\begin{center}
\includegraphics[height=8.63cm,trim={0 0 0.9cm 0},clip]{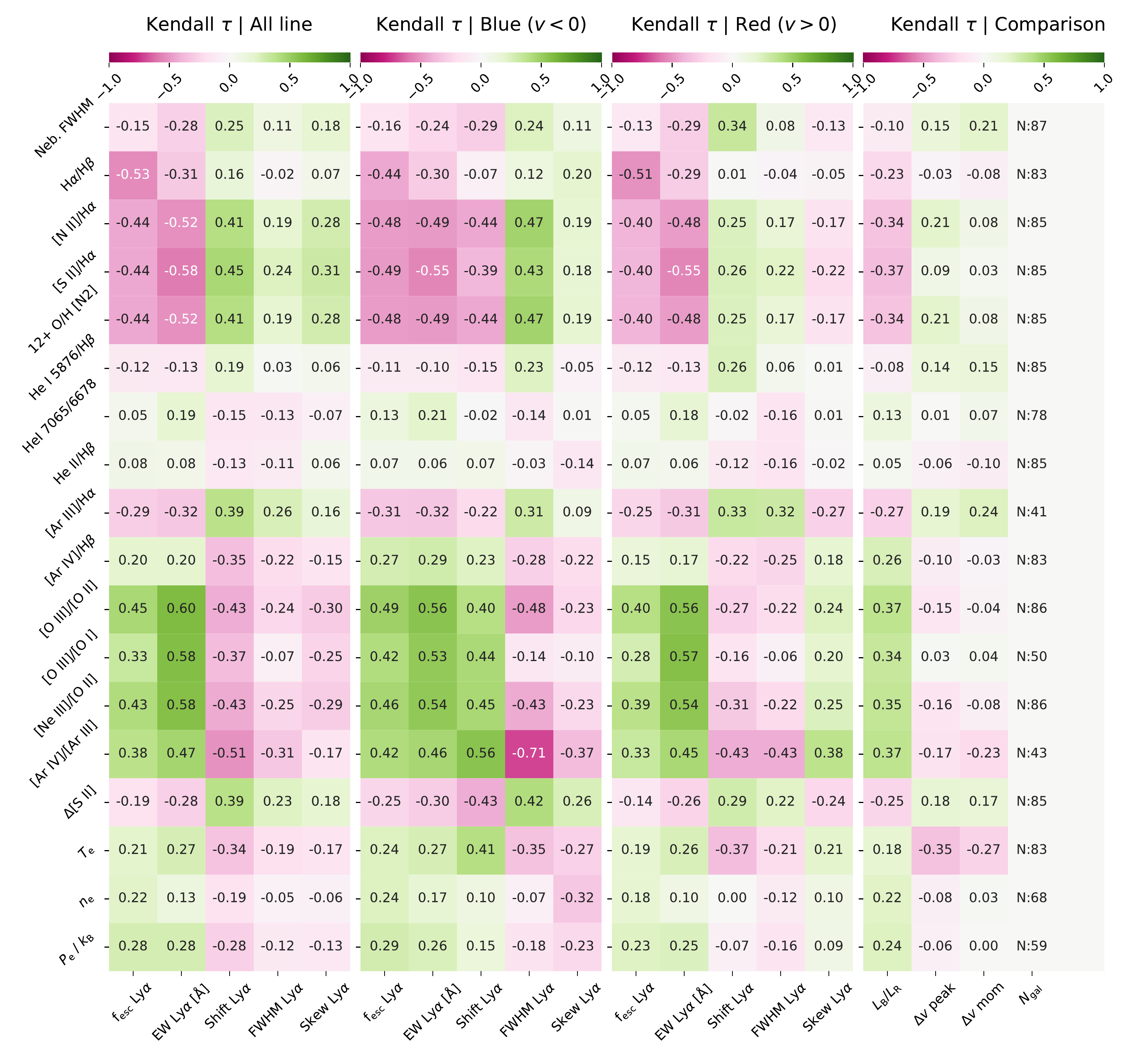}
\includegraphics[height=8.63cm,trim={2.99cm 0 0 0},clip]{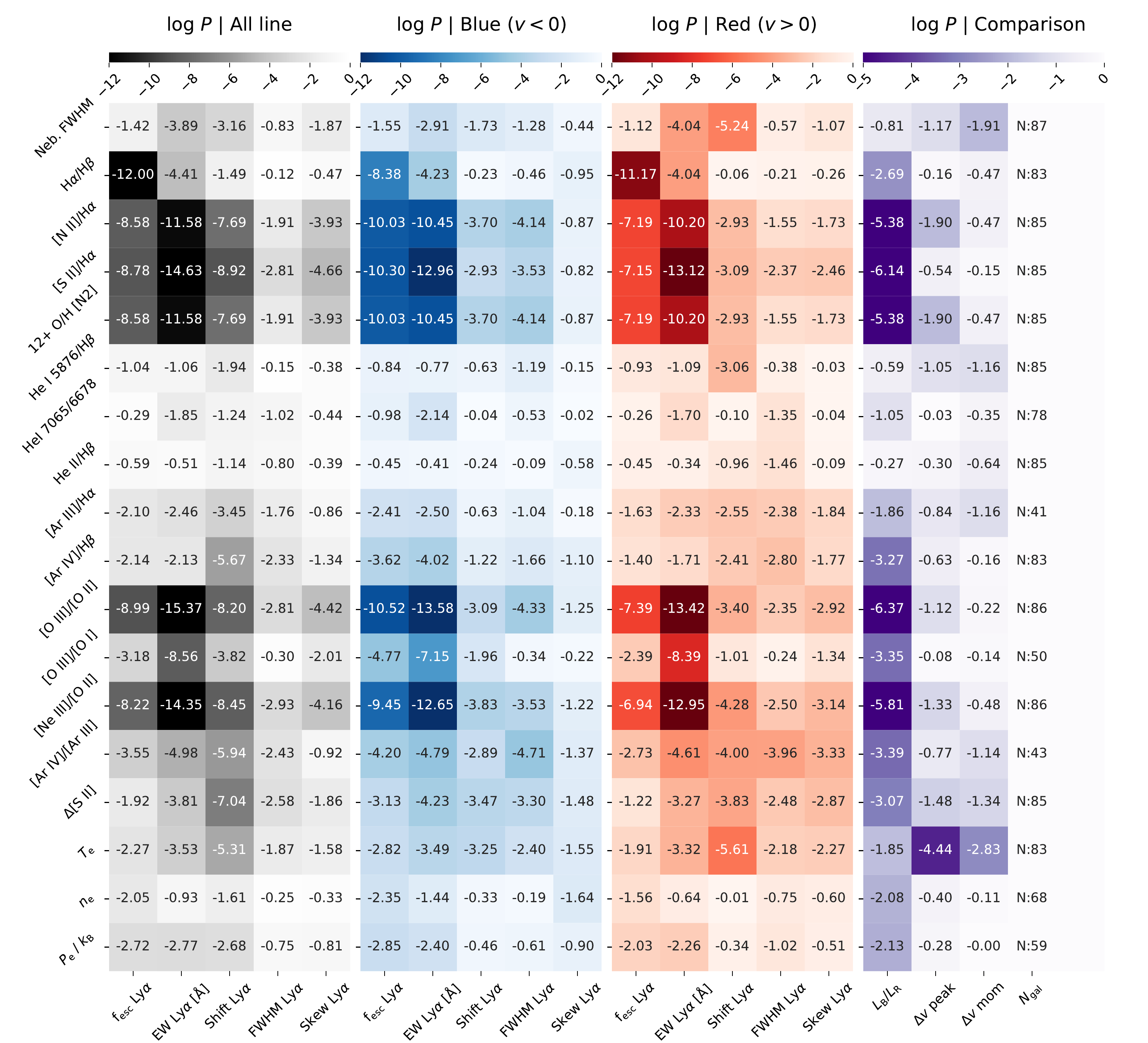}
\end{center}
\caption{Heatmap showing strength and significance of relationships calculated
between \lya\ quantities ($x$-direction) and properties of the nebular gas
($y$-direction).  The figure is divided in half, with Kendall's $\tau$ shown to
the left, and the associated $p$-value to the right.  $\tau$ and $p$ are
written into each cell.  For $\tau$, the diverging colorbar is symmetric around
zero and spans the range from --1 to +1, while $p$ is shown in logarithmic
scale with the range showed by the colorbar.  Each heatmap is divided into four
groups in the $x$-direction, which are color-coded for the $p$-values: the
first block of five shows coefficients for the quantities measured over the
full line, including both blue-shifted and redshifted emission (i.e. derived
from grey data points in Figure~\ref{fig:corrLyaO32}).  The next two blocks
show \lya\ quantities computed only for the blue- and red-shifted emission,
respectively (i.e. derived from the blue and red data points in
Figure~\ref{fig:corrLyaO32}). The final block shows the same correlation
coefficients for relative quantities computed between the blue-shifted and
redshifted emission. In order: the \Lbluered\ ratio, the velocity shift between
the peaks, and the shift between the moments.  The number of galaxies entering
the correlation are shown in the final column.}
\label{fig:heatmap_neb}
\end{figure*}

\begin{figure*}
\noindent
\begin{center}
\includegraphics[height=8.27cm,trim={0 0 0.9cm 0},clip]{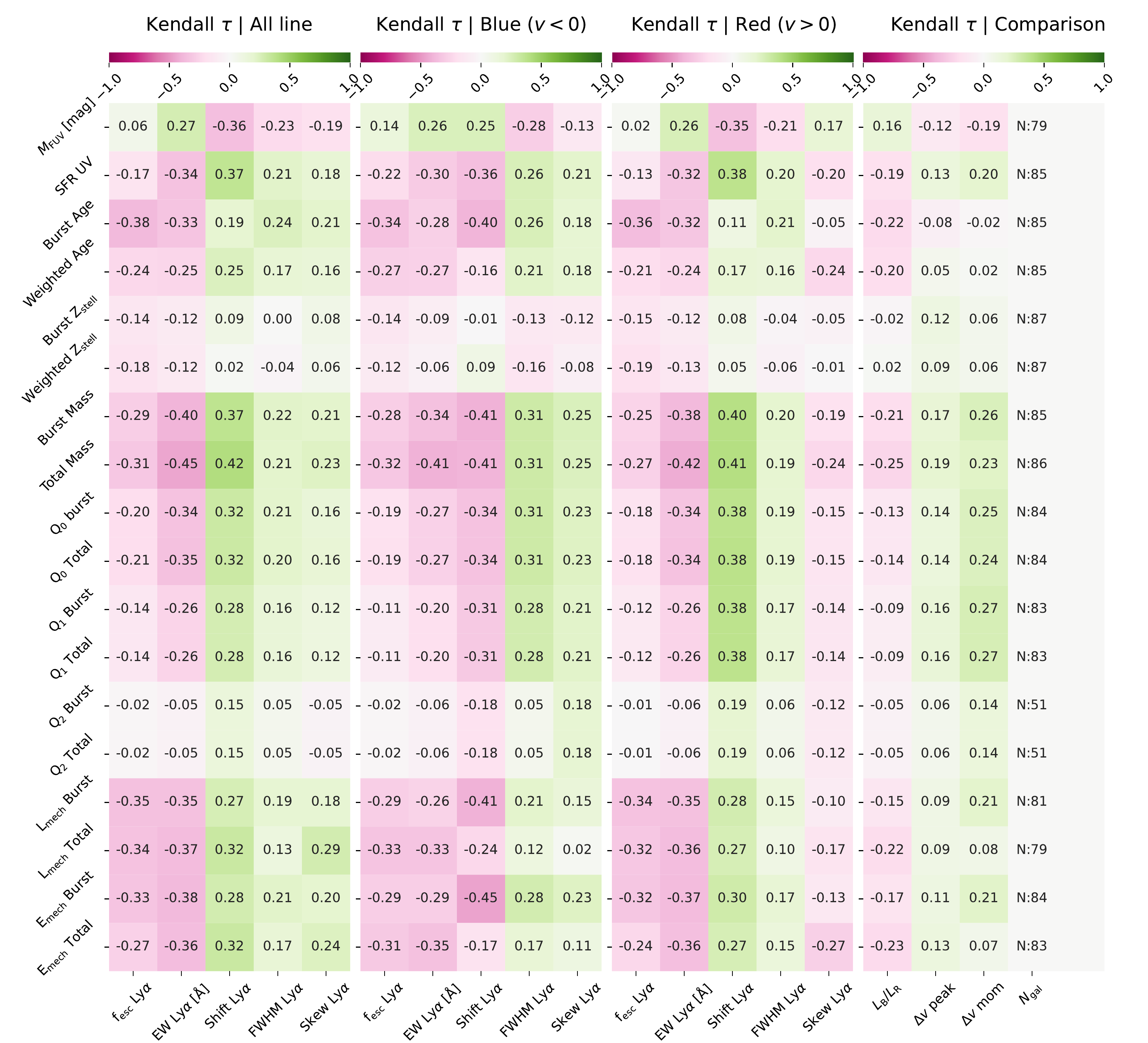}
\includegraphics[height=8.27cm,trim={2.7cm 0 0 0},clip]{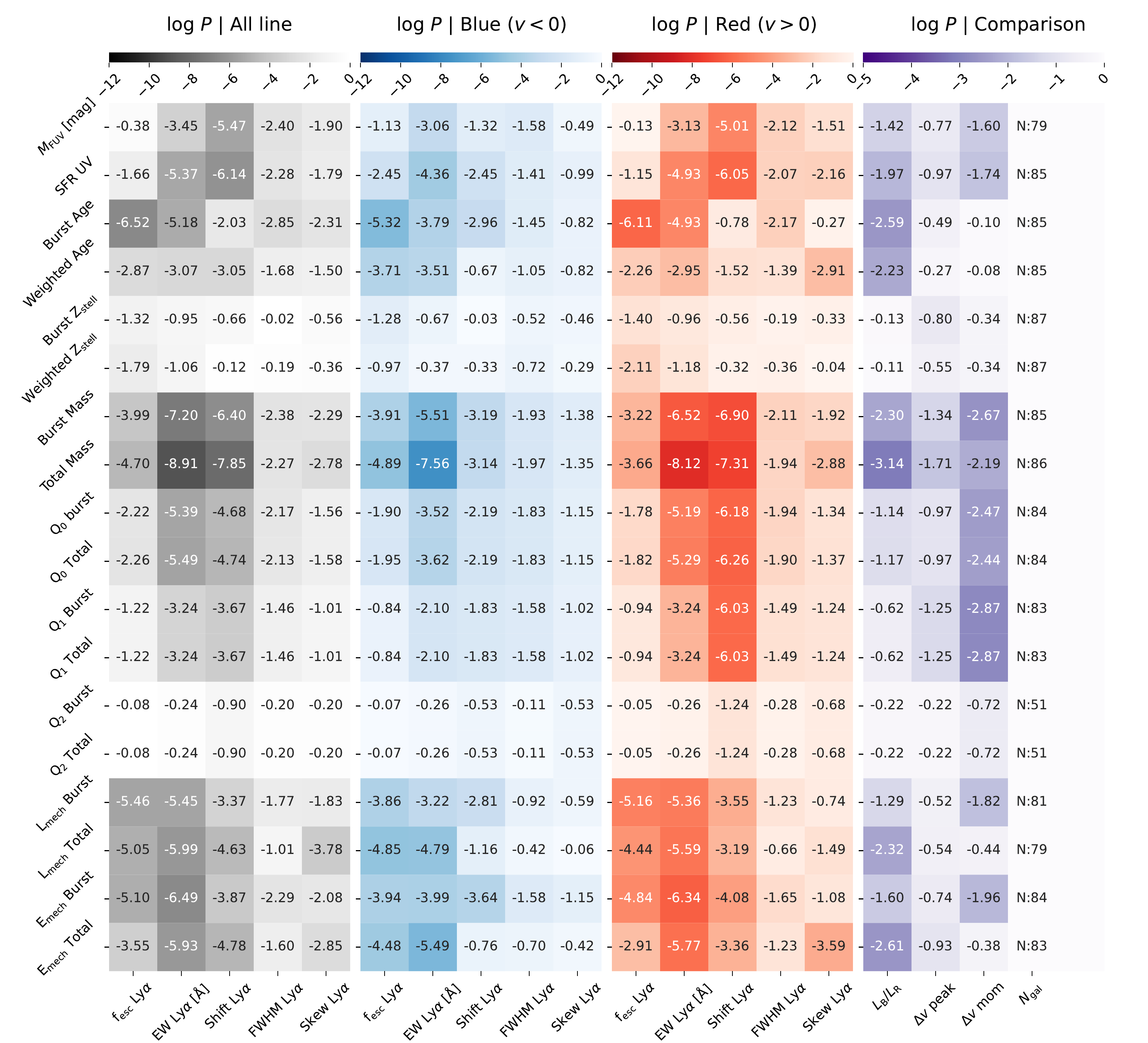}
\end{center}
\caption{Heatmaps for Kendall's $\tau$ and $p$ for properties derived from
stellar population modeling.  Caption the same as for
Figure~\ref{fig:heatmap_neb}.}
\label{fig:heatmap_stell}
\end{figure*}

\begin{figure*}
\noindent
\begin{center}
\includegraphics[height=4.33cm,trim={0 0 0.95cm 0},clip]{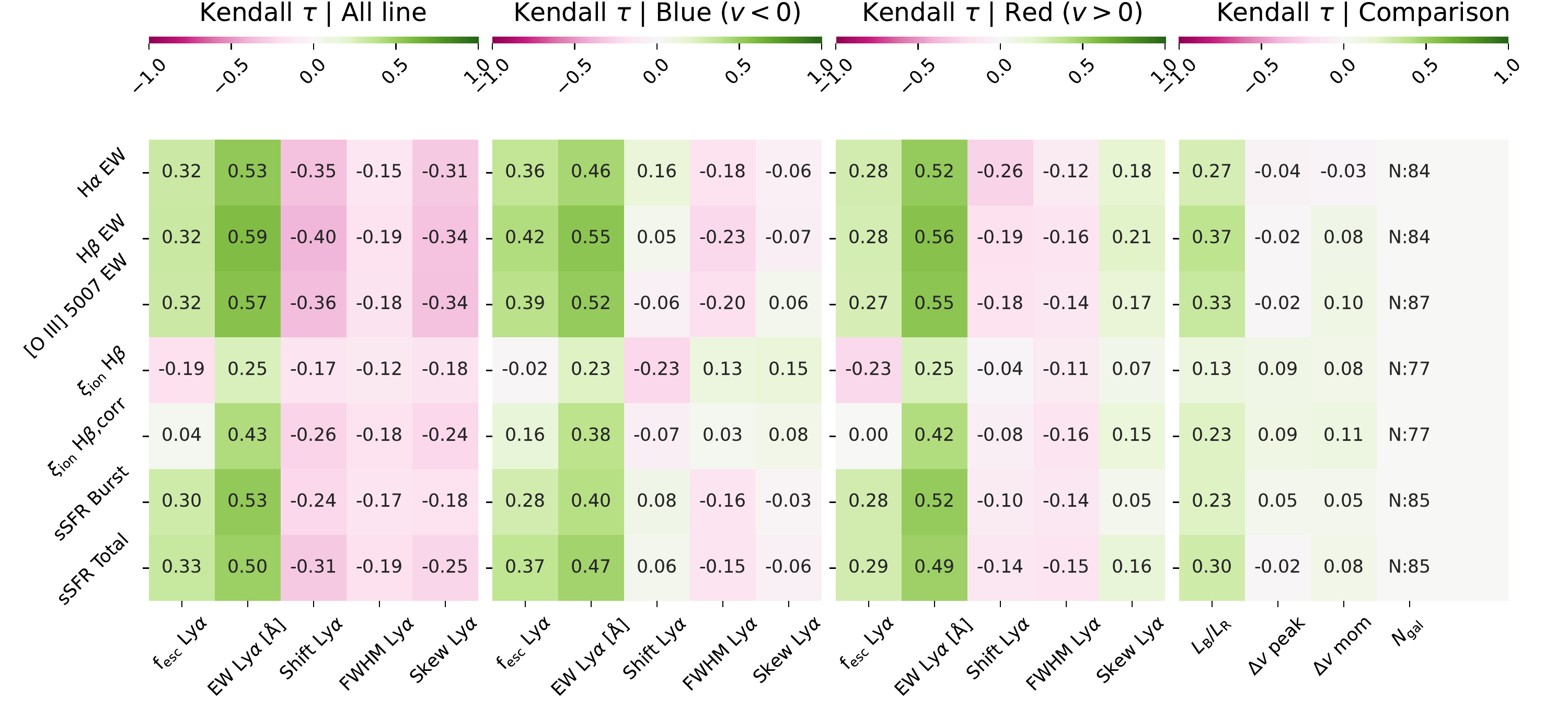}
\includegraphics[height=4.33cm,trim={2.65cm 0 0  0},clip]{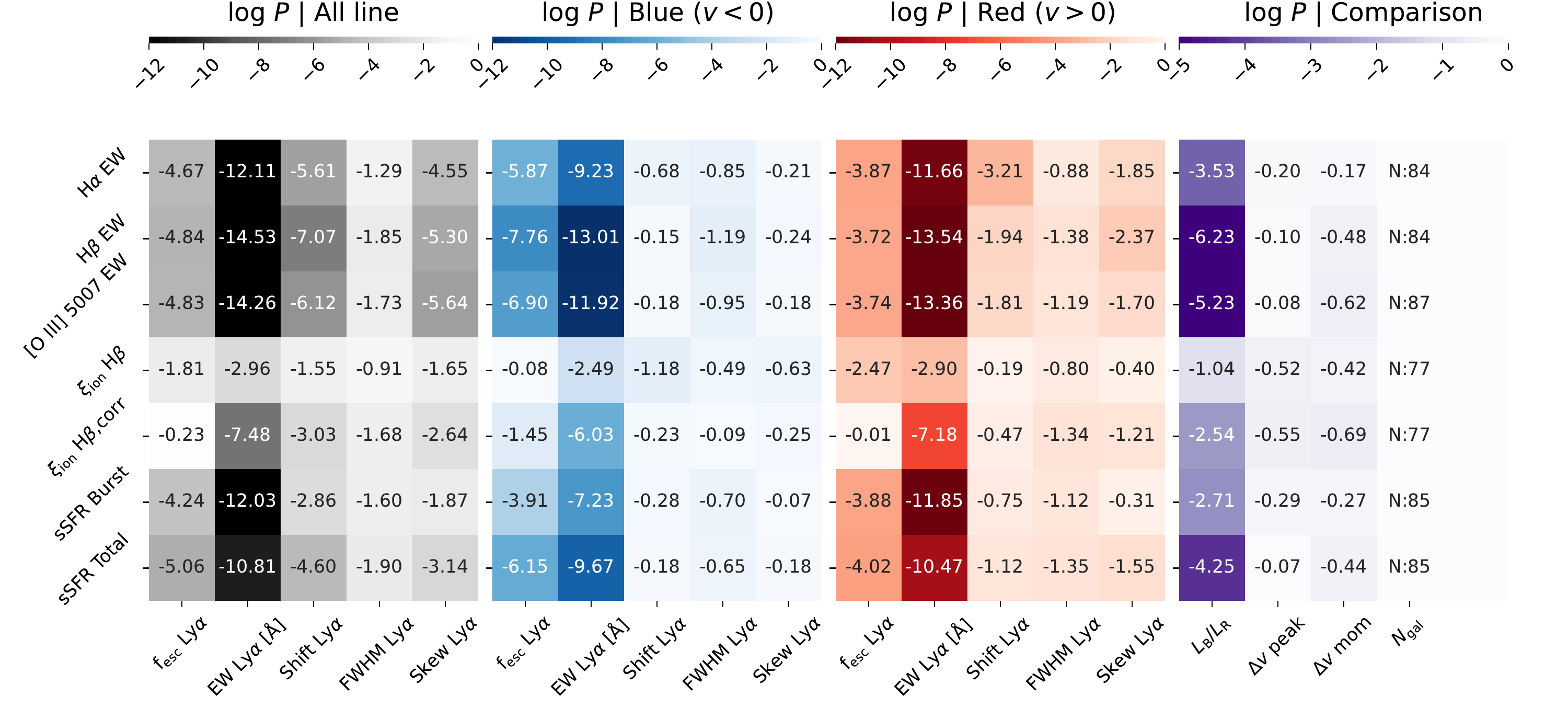}
\end{center}
\caption{Heatmaps for Kendall's $\tau$ and $p$ for properties derived from
information that combines stellar and nebular modeling.  Caption the same as
for Figure~\ref{fig:heatmap_neb}.}
\label{fig:heatmap_mix}
\end{figure*}

Given the number of quantities we wish to investigate, and the wealth of
information in Figures~\ref{fig:corrLyaO32} and \ref{fig:corrLisO32}, we
distribute the equivalent figures as online-only material.  For other
independent variables, we present the results as a series of heatmaps that
record the Kendall $\tau$ coefficient and $p$-value.
Figure~\ref{fig:heatmap_neb} shows these heatmaps for the purely nebular
properties, which we discuss first.  In each cell we color code $\tau$ on a
diverging color-scale between $-1$ and $+1$, where strong positive correlations
are dark green, and strong anti-correlations are dark pink.  The $p$-values are
encoded on black, blue, and red colormaps to correspond to total, blueshifted,
and redshifted \lya\ for consistency with Figure~\ref{fig:corrLyaO32}.  In each
cell we also record $\tau$ (left group) and $p$ (right group) for quantitative
use by the reader. Figure~\ref{fig:heatmap_stell} shows the corresponding
heatmaps for properties derived from stellar modeling, and
Figure~\ref{fig:heatmap_mix} shows quantities that are derived from
combinations of stellar and nebular measurements. 

Rows in the heatmaps (Figure~\ref{fig:heatmap_neb}) are grouped in approximate
order by independent variables.  First come variables associated with
abundance: we show the \halpha/\hbeta\ ratio, which scales with dust reddening,
and the N2 and S2 indices that roughly encode the abundance of nitrogen and
sulphur ions relative to hydrogen.  Then come a number of variables that should
scale with the ionization state: these include recombination line variables at
the top (e.g.  \heI/\hbeta\ and \heII/\hbeta), followed by metal ions relative
to hydrogen (e.g.  [\arIII]/\halpha), followed by metal line ratios such as
[\oIII]/[\oII] (already discussed in Section~\ref{sect:res:lyaprof} and
Figures~\ref{fig:corrLyaO32} and \ref{fig:corrLisO32}), [\neIII]/[\oII], etc.
Finally towards the bottom of the figure we show the influence of thermodynamic
variables: temperature, density, and pressure. 

In terms of the dependent variables, we include the majority of the \lya\
variables discussed in Figure~\ref{fig:corrLyaO32}.  The x-direction of the
heatmap is divided into four groups: the first shows \lya\ quantities computed
over the entire line profile, comprising both negative and positive velocities
(corresponding to grey/black points in Figure~\ref{fig:corrLyaO32}).  We show
\fesclya, the \lya\ EW, and the first, second, and third moments.  The second
block shows these same quantities computed over negative velocities only (blue
points in Figure~\ref{fig:corrLyaO32}), and the third for positive velocities
(red points in Figure~\ref{fig:corrLyaO32}).  The final block shows
correlations regarding direct comparisons between blue and red emission: the
first is the blue/red flux ratio, the second is the distance between the peaks
in velocity space, given by peak identification (first) and moment calculation
(second).  Finally, in the last column, we show the number of data-points from
which each correlation is calculated. 

Figure~\ref{fig:heatmap_stell} proceeds in the same fashion.  From top to
bottom we show stellar properties of age, which is further divided into
starburst age and mass-weighted total age, followed by abundance (similarly
divided), and mass (divided into burst and total).  Then follow ionizing photon
production rates (for H, He$^0$, and He$^+$), and mechanical luminosities and
integrated energies, all of which are shown for both starburst and total
components. 

Figure~\ref{fig:heatmap_mix} shows quantities combining nebular and stellar
estimates.  We first show equivalent widths, which encode the intensity of line
radiation in comparison to the underlying stellar radiation at the same
wavelength.  This most strongly scales with the specific star formation rate or
recent evolutionary history, depending upon wavelength.  To tease out these
contributions we also examine the ionizing photon production efficiency
(\xiion), which we derive from the dust-corrected \hbeta\ luminosity, and the
specific SFR for which we adopt the stellar masses of both the burst and total
population.

\section{Main empirical findings}\label{sect:mainfinds}

We begin by presenting a summary of our main findings, which concern how
observables that trace differing physical processes influence the \lya\ output,
its shape, and the behaviour of the ISM absorption lines.  We then proceed to
discuss the \lya\ kinematic properties and inferences made from the fluorescent
lines of \cII\ and \siII.

\subsection{Lyman alpha Output and LIS Absorption with Respect to Conditions}

\subsubsection{Evolutionary Phase of the Starburst} \label{sect:finds:evol}

Given the dependence of \lya\ production on the photoionization rate, and the
implication of stellar feedback in ionizing and clearing the ISM, it is
unavoidable that stellar age must have a significant effect on the \lya\
output.  We have several tracers of age, that stem directly from spectral
modeling (Sect~\ref{sect:meas:stell}), and from the nebular response in terms
of the EW of hydrogen and helium recombination lines.

Stellar ages show strong anti-correlations with \ewlya, which decreases by a
factor of $\sim 10$ (roughly from 100~\AA\ to 10~\AA\ over the 1--10~Myr
timescale.  The trend is more significant when the starburst age is considered,
removing the contamination from more evolved stars that no longer contribute to
photoionization: $\tau$ increases from $\sim 0.25$ to $\sim 0.4$ and $p$ drops
from $\sim 10^{-3}$ to $\sim 10^{-6}$.  This is entirely expected from the
perspective of \lya\ production, and similar results are shown for the EW of
\hbeta\ in Figure~\ref{fig:age_vs_optlines}.  It is less obvious that \fesclya\
should behave similarly, since \fesclya\ is not directly causally related to
the intrinsic \lya\ luminosity. However, \fesclya\ also anti-correlates with
the stellar age and with similar dynamic range and significance, which implies
two things.  Firstly, \fesclya\ must be connected to the age by a hidden third
variable, such as increased dust production or the loading of the winds with
cool material.  Secondly, the trends of \ewlya\ and age are not purely related
to intrinsic \lya\ production but must also be modulated by these transfer
effects.  It is interesting that when we study the ionizing photon production
efficiency (\xiion) directly that the trends with \ewlya\ remain, but \fesclya\
almost vanish. 

An almost identical picture is revealed by the Balmer line EWs, but the
significance is improved and $\tau$ reaches 0.6 with $p\approx10^{-15}$ for the
relation between \ewlya\ and \ewhb.  This is naturally expected since both the
EWs intrinsically reflect the number of ionizing photons compared to the
underlying stellar light.  However the almost-as-strong correlations concerning
\fesclya\ ($p\approx 10^{-5}$ for \ewhb) clearly demonstrate that the escape of
\lya\ is also heavily modulated.  This correlation is stronger for \ewhb\ than
for \ewha\ because of the smaller wavelength difference between \lya\ and
\hbeta, which is less contaminated by underlying, older stellar generations
than \halpha. 

A partial explanation for the escape fraction behaviour is demonstrated by the
evolution of the LIS absorption in the stacked UV continuum spectra: with
\ewhb\ increasing from $\sim 30$~\AA\ to 300~\AA, the EW of absorbing gas
decreases by a factor of four in the \cII~$\lambda 1334$\AA\ absorption line
($-1.6$ to $-0.4$~\AA).  This traces a decrease in the combined effects of
covering fraction and column density of cool gas, which absorb the \lya.  

An interesting observation is how the absorption EW of \cII\ and \siII\
increase with age (and decreasing Balmer EW).  We argue this results from an
increasing loading of the wind with time as it is accelerated. It is clear also
that the velocity offsets are smallest ($\lesssim 100$~\kms) when the starburst
is youngest, but then increase by a factor of about two.  The individual UV
spectra are not sufficiently deep for us to robustly derive outflow masses,
mass-loading factors, etc. for each galaxy. Indeed, it is for this reason that
we resort to stacking analyses for UV absorption line measurements.  However,
we studied these quantities in detail in \citet{Hayes.2023} in exactly the same
dataset, using the same sub-bins for stacking.  In that paper we showed that
the covering of cool gas, and consequently the outflow rate and mass loading
factors increase over the duration of the starburst episode.  Wind masses grow
from $\sim 10^4-10^8$~\msun\ over the 1--10~Myr duration, which could either be
because it takes time to accelerate cool gas or advect it into the flow, or for
cooler absorbing material to condense out of the warmer outflowing gas.  In
either scenario, the column of \lya-absorbing column increases with time, which
would contribute to the negative relationship between both \ewlya\ and
\fesclya\ with evolutionary independent variables.

We previously hypothesized that the \lya\ output should be modulated by the
amount of mechanical energy returned by feedback.  Both \fesclya\ and \ewlya\
are strongly anti-correlated with the mechanical luminosity (\Lmech) and its
total integral since the onset of the starburst (\Emech).  This relationship
runs contrary to our hypothesis, in that more \lya\ is emitted when less
mechanical energy is available (or has been deposited).  This relationship is
attributed to second order effects, and may indicate that while this feedback
must be responsible for accelerating large scale winds, it is sub-dominant to
other processes when considering the emission of \lya\ (see
Section~\ref{sect:finds:ionstate}).

As discussed in Section~\ref{sect:res:lyaprof} the total \lya\ output is
dominated by the redshifted component, as shown by the alignment of grey and
red data-points in the \fesclya\ and \ewlya\ figures.  However the slope of the
points for the blue-shifted emission rises more steeply, and the blue-shifted
emission contributes more at larger \ewhb.  The fraction of blue-shifted
emission, \Lbluered, is shown directly, which rises from effectively zero at
\ewhb~$\approx 30$~\AA\ to 0.3 at \ewhb~$\approx 300$~\AA\ ($p\approx
10^{-6}$).  This rapid increase requires a decreasing column densities of gas
at negative velocities, which is supported once more by the absorption lines:
not only is there less absorption in total at higher \ewhb, but the level of
absorption at zero velocity also falls by a factor of $\approx 2$.

The basic interpretation for the above is that evolutionary stage modulates the
\lya\ output, both by affecting the intrinsic \lya\ production and its
transfer.  A working scenario is one in which the abundance of ionizing photons
at smaller ages also leads to higher ionization gas, and less cool absorbing
material -- this would also be consistent with the decreasing \cII\ and \siII\
at higher \ewhb.  In other words, that in terms of \lya\ the `rich get richer',
i.e., since ionizing photons not only increase the intrinsic emissivity is
increased but also the escape is facilitated \citep{Kakiichi.2021}. The same
effect on the LIS lines can also be explained by the changing the ISM
abundances of carbon and silicon, which is the subject of the next subsection.

\subsubsection{Abundances of Dust and Metals} \label{sect:finds:abundances} 

We use the \halpha/\hbeta\ ratio (or \hgamma/\hbeta\ where \halpha\ is clipped)
to estimate the dust obscuration, and the ratios of [\nII]6584/\halpha\ (N2),
[\sII](6717+6731)/$\halpha$ (S2) indices as proxies for the nebular metallicity
\citep[e.g.][]{Pettini.2004,Marino.2013,Kewley.2002,Yin.2007}.

The absolute \lya\ output (both \fesclya\ and \ewlya) strongly anti-correlate
with the dust obscuration in way that are both intuitive and have been shown
before
\citep{Scarlata.2009,Hayes.2010,Hayes.2014,Atek.2009galex,Atek.2014galex}.
This is also very clear when comparing the \halpha/\hbeta\ ratios with the LIS
absorption lines, which shows that the covering gas also reddens the light
\citep[e.g.][]{Shapley.2003,Gazagnes.2018}.  It is apparent that sorting the
sample by dust obscuration also sorts by the amount of absorbing gas at $\Delta
v=0$: the least dusty galaxies show almost no absorption at zero velocity,
which is probably responsible for the downwards trend in \Lbluered\ with
\halpha/\hbeta.

While the above relations are quite intuitive, it is not clear whether
\halpha/\hbeta\ stacking sorts the sample along an age sequence (both dust
production and destruction/clearing scenarios could be envisaged).  Outflow
velocities are smaller in more obscured galaxies, firstly suggesting that
radiation pressure on dust grains is not a significant accelerator of winds (or
that the trend is otherwise obscured).  A plausible explanation is that dustier
winds are more massive and more energy is required to accelerate the gas. 

The effects with nebular metal abundances (traced by N2 and S2 indices) assist
in the interpretation: the anti-correlation of these quantities with \lya\
output are stronger than those for \halpha/\hbeta, with $p\approx 10^{-15}$ for
S2 \emph{vs.} \ewlya\ and $p\approx 10^{-9}$ for \fesclya.  Abundances can
explain this in two ways: by lower metallicity stars having higher ionizing
photon production, and by more metals producing more dust to obscure the
emitted \lya.  The EW of LIS absorption lines is also strongly correlated with
the S2 index, where more nebular [\sII] emission aligns with more interstellar
absorption.  This could be the result of ISM abundances and it is curious that,
for example, the \siII\ EW varies by a factor of $\approx 8$ while the S2 index
changes by a similar amount, which would imply close-to constant gas hydrogen
if metallicity were the only action. (Note, however, that metallicity
calibrations using strong lines are seldom linear with the index.)  Almost
constant hydrogen columns would not cause major differences to the \lya\
output, which changes significantly, and by larger factors than the LIS
absorption.  This would argue that the ISM lines are saturated and the trend
with \hI\ covering is in fact significant. 

Here we may use the expected correlation between abundances in the ISM and the
stars.  Lower metallicity stars experience less photospheric opacity from metal
absorption, and produce more ionizing photons at fixed mass; this increases the
\lya\ production (intrinsic EW) and also produces a more highly ionized ISM and
increases the \lya\ transmission.  However we see no noticeable influence of
stellar abundance on the observable \lya\ properties ($p\gtrsim 0.05$),
implying that this effect is likely not an important driver of either \lya\
production or, by secondary effect, emission.

A further interesting point is that when testing nebular metallicity as an
independent variable, we again recover the strong evolution of the \Lbluered\
ratio, which peaks at at $\approx 0.4$ for the least metallic galaxies.  There
is no case for this unless the amount of absorbing material close to zero
velocity is not strongly correlated with the metallicity, which would further
argue for ionization levels to be the main driver. 

A final point of interest is that the S2 index correlates more tightly with
\lya\ variables than the N2 index.  N2, however, is the preferred measure of
metallicity, which suggests there may be confounding factors.  Because S$^+$
forms in the partially neutral medium (I.P. range: 10.4--23.3~eV) but N2 does
not (14.5--29.6~eV), an extra deficiency factor for S2 may result when the ISM
becomes highly ionized \citep{Wang.2019}.  The [\sII] deficit
(Section~\ref{sect:indvar}) encodes the lack of [\sII] emission at fixed
metallicity, and has been argued as a tracer of low optical depths to ionizing
photons.  When we contrast the \lya\ output against $\Delta$[\sII] we see
weaker but significant trends ($p\approx 10^{-4}$) for \lya\ to be stronger
when the deficit is larger, lending weight to the use of this variable to
segregate between galaxies where the hydrogen optical depth is low
\citep[see][]{Wang.2021}. 

\subsubsection{Ionization State} \label{sect:finds:ionstate}

Without a detailed photoionization model, we do not have direct estimates of
the ionization state of the gas.  However many line ratios are available, and
the nebular lines measured in this paper trace species ionized by photons with
energies between $\approx 11$~eV at the low ionization end, up to 54~eV at the
high end.   Ratios of some of these are often invoked as indicators of the
ionization level of the nebular gas, although like with abundances there are
many caveats to single line ratios.

This has already been discussed in Section~\ref{sect:results_profiles}, where
we focussed upon the O$_{32}$ ratio, but Figure~\ref{fig:heatmap_neb} shows
many line ratios. We sort these by ratios of pure recombination lines (RL; e.g.
\heI/\hbeta\ and \heII/\hbeta), followed by ratios of collisionally excited
lines (CEL) to RL, (e.g. [\arIII]/\halpha, [\arIV]/\hbeta), and then ratios
involving collisional lines only (e.g.  [\neIII]/[\oII], [\oIII]/[\oI], and
[\arIV]/[\arIII]).

Beginning with the RL ratios, effectively no trends are seen when using the
\heI~5876/\hbeta\ ratio, implying the ratio does not separate out physical
properties that influence \lya\ emission.  (There is a trend for higher
\heI/\hbeta\ ratios to show both thicker and faster outflows, but currently we
do not have an explanation for this.) The lack of correlation with \lya\
probably indicates that in all these galaxies, the helium-ionizing photon
production rate ($Q_1$) is a sufficient fraction of $Q_0$ ($Q_1/Q_0 \gtrsim
0.15$; \citealt{Draine.2011}) that the H$^+$ and He$^+$ Str\"omgren spheres are
bound together and the line ratio contains little diagnostic information.  

One may expect the \heII\ line to provide more diagnostic power than \heI, as
the requirement of a significant photoionization rate at energies above 54~eV
must identify the hottest stars.  In a sample of $z\approx 2$ galaxies
identified by \citet{Matthee.2021}, it was later shown that the higher EW LAEs,
and galaxies with more blueshifted emission, did indeed show higher
\heII/\hbeta\ ratios \citep{Naidu.2022}.  In this context, our result is
surprising: the \lya\ output is not significantly correlated with \heII/\hbeta.
A similar result has already been obtained for the LyC output of starburst
galaxies \citep{Marques-Chaves.2022}, who argue variations in this line ratio
are driven mainly by metallicity; having addressed abundances in
Section~\ref{sect:finds:abundances} it is surprising we do not see the
secondary correlation here.  A possibility is that because the He$^{2+}$ zone
is small, it does not contain much information regarding the environments where
\lya\ is absorbed, but the same could be argued for [\arIV], [\neIII] and even
[\oIII].  Another alternative is that \heII\ is produced by processes other
than photoionization, such as fast radiative shocks.  This could also explain a
lot of the difficulty experienced in reproducing the \heII\ fluxes with pure
photoionization models
\citep{Kehrig.2018,Senchyna.2020,Olivier.2021,Wofford.2021}.

Moving to the CEL/RL ratios, [\arIII]/\halpha\ and [\arIV]/\hbeta\ both show
weak but significant trends with \lya\ output behave in opposite directions.
[\arIII]/\halpha\ anti-correlates with \ewlya\ and \fesclya, while
[\arIV]/\hbeta\ positively correlates ($p\approx 0.005$ typically).  It appears
that the [\arIII]/\halpha\ ratio is more closely aligned with the metallicity,
which produces the trend on this variable, and indeed the same [\arIII] line
has been proposed as an abundance diagnostic, albeit with a different
denominator \citep{Stasinska.2006}.  \lya\ output increase for higher fluxes in
the higher ionization potential [\arIV] lines (40--60~eV).  \lya\ also becomes
bluer at high [\arIV]/\hbeta, and the LIS absorption becomes weaker, with
behaviour similar to the O$_{32}$ diagnostic discussed previously.  The
faintness of the [\arIV] doublet, however, reduces our sample, and we cannot
afford the same number of bins as for stronger lines. 

Pure CEL ratios far surpass ratios involving recombination lines in predicting
\lya\ output, as demonstrated with O$_{32}$ in Section~\ref{sect:res:o32}.
Abundance variations cannot be the cause, because ratios of comparable
ionization potential to hydrogen (e.g. N2 and S2;
Section~\ref{sect:finds:abundances}) also out-perform higher potential ratios
like those of [\arIII] and [\arIV] above.  The commonplace O$_{32}$ and Ne3O2
indices \citep{Levesque.2014} are the best-predictors in our study, with
$\tau\gtrsim 0.6$ ($p\sim 10^{-15}$) against global \lya\ metrics in both
cases.  This is partly intuitive as abundance variations can have zero
influence in the case of O$_{32}$, and very little influence for Ne3O2 since
neon is produced in the same way as most oxygen, with the addition of one extra
alpha particle.  Similar results are obtained for the [\oIII]/[\oI] ratio
(O$_{31}$) ratio although the trend is weaker. 

It is important to point out that the EW of LIS absorption features strongly
correlates with O$_{32}$ and Ne3O2 indices.  We discussed a similar trend with
abundance in Section~\ref{sect:finds:abundances}, but noted that if abundance
scaled directly with N2 and S2 ratios, then variations in the absorption EWs of
\siII\ and \cII\ could actually point to fixed hydrogen column.  That the same
LIS absorption EWs also vary strongly with O$_{32}$ show that this cannot be
the case as this intensity must also modulate the ionization correction
factors, and amount of H traced by C$^+$.  The ionization state must be
increased for the extreme O$_{32}$ galaxies, pushing more of the carbon into
the C$^{2+}$ state. 

Further support for this comes from the argon lines, whose predictive power is
increased when combined with each-other.  This is observationally remarkable
given the relatively shallow nature of SDSS spectra, and the fact that these
lines are $\approx 100\times$ weaker than the strong oxygen lines.  The
[\arIV]/[\arIII] ratio (Ar43) also strongly correlates with \Lbluered\ and with
the UV absorption lines, revealing similar traits to O$_{32}$ at somewhat
higher potential.  Both the \cII\ and \siII\ lines almost vanish in the highest
Ar43 bin, and even at $\Delta v=0$ the stack shows little absorption (residual
intensity of $\approx 80$~\% in \siII\ and \cII.  It appears that galaxies with
the strongest [\arIV] emission are fully ionized, explaining the low column
densities of absorbing material, and the large contribution of blueshifted
\lya\ emission. 

\subsubsection{Thermodynamic Variables} \label{sect:finds:thermodyn}

Finally we draw attention to some quantifiable physical conditions in the ISM:
those of temperature, density, and pressure.  These are inferred from
variations in the ratios of (often) weak emission lines: the 6717/6731~\AA\
doublet of [\sII] for the electron density ($n_\mathrm{e}$), and the
5007/4363~\AA\ ratio of [\oIII] for temperature ($T_\mathrm{e}$).  The product
of these two quantities is the pressure ($P_\mathrm{e}$), although note that
$P_\mathrm{e}$ and $n_\mathrm{e}$ are almost equivalent as variables because
the dynamic range of $n_\mathrm{e}$ far exceeds that of $T_\mathrm{e}$.  One
may intuitively expect these variables to rank as important correlates in this
study, as they are direct assessments of physical conditions.   

Weak correlations exist between all of these variables and \ewlya\ and
\fesclya, in the direction that \lya\ is stronger at higher temperature and
density.  However these are weaker than for many other variables, with
$\tau\sim 0.2$ and $p\sim 10^{-3}$ for the sample.  Some suggestive
relationships can be seen when examining the LIS absorption lines.
Specifically, the absorption EW is correlated with ISM temperature, in the
direction that hotter galaxies show less absorption -- the most likely
explanation is that the galaxies with hotter ISM are also more highly ionized,
lowering column densities in the lower ionization C$^+$ and Si$^+$ stages.
Also the outflow velocities of the gas increase dramatically --by a factor of
3-- with ISM density and pressure (regardless of $T_\mathrm{e}$).  In itself
this is also intuitive because more pressurized environments should accelerate
the surrounding gas more rapidly; it is not intuitive, however, why these
effects of $T_\mathrm{e}$, $n_\mathrm{e}$, and $P_\mathrm{e}$ do not affect the
\lya\ output.  Sorting on these variables does not uniquely sort by \lya\
observables, and we speculate that these thermodynamic variables may be useful
to follow up as independent variables. 

\subsubsection{An Underlying Effect of Stellar Mass?} \label{sect:finds:mass}

After discussing many variables tracing ISM conditions, the remaining question
is that of galaxy mass.  Is this the factor that governs everything else by
secondary correlation?  Figure~\ref{fig:global_lya} shows there is an
anti-correlation between \ewlya\ and stellar mass, and our heatmaps show this
relationship to have $\tau\approx -0.4$ with $p\approx 10^{-9}$.  Indeed
stellar mass is a vital variable to consider.  However the amplitude of
variation of \fesclya\ with mass is comparable to that of \ewlya, which implies
the variations in EW are driven more by conditions in the ISM and not through
those of the stars (\xiion\ or evolutionary phase). 

Similar effects are seen in relationships between \lya\ output and the FWHM of
the optical emission lines, which would trace the total mass of the system
under dynamically relaxed conditions.  Both \lya\ tracers do decrease with
increasing FWHM and the absorption lines become stronger (suggestive of a mass
sequence) but these trends are only marginally significant.  We conclude that
mass and intrinsic dynamics of the \hII\ regions are important in the emission
of the \lya\ radiation, but mainly by correlation with ISM-related properties.

\subsection{Kinematic Properties of Lyman alpha}

\subsubsection{Velocity Shift and Width: Moments 1 and 2} \label{sect:disc:kin:mom12}

In this section we discuss the influence of our independent variables on the
higher order moments of the \lya\ line: the velocity offset, width, and
skewness.  In this section we must be cautious, as these measurements may be
influenced by varying contributions of the red and blueshifted components,
instead of actual variations in any one of the two peaks.  We discuss these
possible effects following the same structure as above.

\paragraph{Evolutionary Stages of Star Formation} As indicated by the EW of
Balmer lines and modeled age, evolutionary stage has a mild effect on the
\emph{total} \lya\ kinematics.  However in terms of offsets (moment 1), neither
the blue nor red peaks are in themselves strongly affected by age estimators.
The reason for the strong correlation in the total line (measured over a
2500~\kms\ window) ($p\approx 10^{-7}$), is that the blue peak increases
systematically in contribution with respect to the red peak (\Lbluered\
increases strongly), which shifts the central moment closer to $\Delta v=0$.
We do not see any strong or significant trends of the velocity width with the
evolutionary stage variables (e.g. \Lmech, \Emech). 

\paragraph{Abundances of Dust and Metals} \lya\ kinematic properties appear
invariant with dust obscuration, but certainly not with metal abundance.  The
shift of moment 1 of the total line evolves in the same way as described above,
because of changing \Lbluered.  More interestingly, when contrasted with N2 and
S2, the individual components are both systematically shifted away from
line-centre in more metal rich systems: the amplitude of this effect runs from
200--300~\kms\ and is significant at $p\approx 10^{-4}$.  Trends of both
comparable amplitude and significance are also visible in second moment.  In
summary, more of the \lya\ emission is concentrated closer to line centre in
less metallic galaxies.

\paragraph{Ionization State of the Gas} Clear kinematic trends are visible, and
are strongest among the line ratios that have already been discussed as having
the most influence over the global \lya\ output (e.g.
Section~\ref{sect:finds:ionstate}).  Both the first and second moments strongly
anti-correlate with ionization parameter, as probed by O$_{32}$
(Figure~\ref{fig:corrLyaO32}), Ne3O2 and also Ar43 (online-only materials).

\paragraph{Thermodynamic Variables} $n_\mathrm{e}$ and $P_\mathrm{e}$ do not
separate galaxies into significantly different sub-groups by \lya\ kinematics.
Electron temperature, however, shows a remarkable trend of decreasing velocity
offset (factor of 2, $p\approx 10^{-5}$) and width (similar amplitude, $p\approx
10^{-3}$).  Even a posteriori it is difficult to explain this trend, but we
speculate it may be coupled to the decline in absorbing column density with
increasing temperature that we discuss in Section~\ref{sect:finds:thermodyn}.

\paragraph{An Underlying Mass Effect?}  The answer to this question is a
resounding yes.  Velocity offsets from line centre (moment 1) of the individual
peaks are most strongly correlated with mass. The dynamic range in offset is
about a factor of 2 (as above) but the $p$-value drops to $10^{-7}$ ($10^{-4}$)
for the red (blue) peak.  However, the bulk of this effect must be attributed
to radiative transfer effects, and cannot be explained by a larger fraction of
\lya\ being emitted at greater velocity offsets (see \citealt{Stark.2017} and
\citealt{Mason.2018boost} for further discussion).  While it is absolutely the
case that more massive galaxies have broader intrinsic optical lines, the trend
between the FWHM of the optical features and \lya\ moments is barely
statistically significant.

\medskip 
As with the global \lya\ output, tracers of metallicity and ionization
parameter are best able to identify differences in \lya\ offset and width.  In
contrast, the evolutionary tracers (ages and Balmer EWs) appear unable to sort
galaxies this way.  Both offsets and \lya\ width are linked to radiative
transfer effects, where a higher column density implies the need for larger
frequency excursions at scattering: the effects concerning O/H and log~$U$
could therefore be the result of a harder ionizing spectrum that ionizes more
of the interstellar medium and reduces the overall column density and residual
\hI\ fraction.  

In this context, it is interesting that we do not see the kinematic
measurements evolve during the starburst event, or on timescales that can be
probed with UV-optical spectroscopy.  This is remarkable, since we do see
strong evolution in the EW and first moments in the LIS absorption lines over
these timescales.  We suspect that the critical quantity in shaping the \lya\
is the column density of gas close to zero velocity, which evolves much more
slowly than either the EW or centroid velocity of the absorption.  The fact
there is an overall trend with total galaxy mass shows that the line profile
will evolve on longer timescales as galaxies assemble, and this may also
explain some of the above trends as secondary correlations.  For example, less
metallic galaxies are also less massive on average
\citep[e.g.][]{Tremonti.2004}, which results in narrower intrinsic \lya\
profiles as observed here; dust is also built up in starburst events, mixed,
and accelerated to velocities where it may have varying effects on the line
profile.

\subsubsection{Skewness: Moment 3}

The skewness of the total \lya\ emission line was first derived (in slightly
different forms) to distinguish between \lya-emitting galaxies and other
single-line emitters in galaxy surveys
\citep[e.g.][]{Rhoads.2003,Shimasaku.2006,Kashikawa.2006}.   It has since been
studied in more detail in both galaxies \citep{U.2015,Childs.2018} and extended
\lya\ halos \citep{Herenz.2020}, and put forward as a diagnostic of scattering
scenarios in different geometries \citep[e.g.][]{Remolina-Gutierrez.2019}.
Clearly the \lya\ skewness must be connected to resonance scattering and
modulated by kinematic effects that preferentially reshape one side.

We find the skewness of total \lya\ line (measured over all velocities) to be
most strongly connected with variables concerning stellar evolutionary state,
ionization state, and nebular abundance (all with $p\approx 10^{-4}$), with a
weaker but still significant effect seen in total mass ($p\approx 0.01$).
However, as discussed at the top of Section~\ref{sect:disc:kin:mom12}, this
effect is entirely one in which skewness become increasingly negative as the
blue peak become systematically stronger.  It does not relate to the earlier
proposition of skewness in individual peaks that identify LAEs
\citep[e.g.][]{Kashikawa.2006} or investigate outflows \citep[][]{U.2015}. 

We proceed to investigate the skewness of the red and blue components
independently.  In the blue peak we find significant relations (defined
conventionally as $p<0.05$ between skewness and Ar43, $\Delta$[\sII],
$T_\mathrm{e}$, $n_\mathrm{e}$ (but not their product, $P_\mathrm{e}$) and
stellar mass.  The lack of some correlations may be the result of relatively
poor signal-to-noise in many of the blue peaks, and the lack of correlations
with O$_{32}$ and Ne3O2 cast doubt upon the Ar43 relation.  Indeed the red peak
shows stronger relations with skewness, which increases (becomes more extended
towards high velocities) at low metallicity, high ionization parameter,
increasing deficit of [\sII] and higher temperature.  Most of this appears
connected to ages derived over longer timescales, as evidenced by an
anti-correlation with total age and mass ($p\approx 10^{-3}$).

\subsection{Fluorescent Emission Lines and Circumgalactic
Gas}\label{sect:find:fluor}

A prominent feature in some of the stacked UV spectra is the presence of the
fluorescent emission lines arising from the split ground state of the \cII\ and
\siII\ transitions.  These lines have been discussed for some time in
connection with the circumgalactic gas content of galaxies: e.g. in low-$z$
\lya-emitting starburst KISSR\,242 \citep[a.k.a. LARS\,07;][]{France.2010} and
at $z\approx 3$ in the stacks of Lyman break galaxies \citep{Shapley.2003}.
They have also been used to estimate the 3-dimensional configuration of galaxy
outflows in unresolved observations \citep{Scarlata.2015,Carr.2021}.  The
overwhelming majority of this line emission is excited by absorption in the
$^2P^0_{1/2} \rightarrow ^2D_{1/2}$ (the resonance transition to the lower
ground state) that then decays instead to the $^2P^0_{3/2}$, releasing a photon
red-wards of the resonance.  The key point is that gas emitting a fluorescent
line must have absorbed in a resonance transition. 

Our stacked spectra show clear fluorescent lines in both the \cII* and \siII*
species, with typical EWs of 0.2--0.3~\AA.  The \siII* line (whose offset in
velocity space is $\approx 1250$~\kms\ from resonance) is well centered at the
systemic velocity (determined by optical nebular lines).  The peak of the \cII*
line appears redshifted, but this is probably a result of blending with the
absorption, as the intrinsic velocity difference is only 250~\kms.  A key
finding of this work is that, while the absorption EWs may vary by factors of
4--5 (e.g. with Ne3O2) the EW of the fluorescent line is invariant.  

Aperture size will of course be important to capture all the fluorescent
emission, but it is hard to envisage a scenario in which its EW does not
correlate with the resonantly scattered \lya\ line -- we tested this explicitly
with both \lya\ luminosity and EW as the independent variable.  Our
interpretation is that the larger amounts of absorbing gas are actually due to
a more extended halo -- in this scenario the line-of-sight distance grows to
provide a greater column density and deeper absorption.  Since the galaxies
with strongest absorption are also found to be more evolved, this would also
support the scenario in which wind envelopes are built up successively with
time \citep{Hayes.2023}.  At larger foreground distances, both scattering into
the line of sight (`resonant infilling') and fluorescent emission towards the
observer become successively less likely for geometric reasons.  Thus the
fluorescent line is dominated by emission from close to the star forming
regions and does not vary with total column density.

\section{Quantifying the scatter: multi-dimensional analysis of the relations}\label{sect:multipar}

\begin{figure*}
\centering
\includegraphics[width=\textwidth]{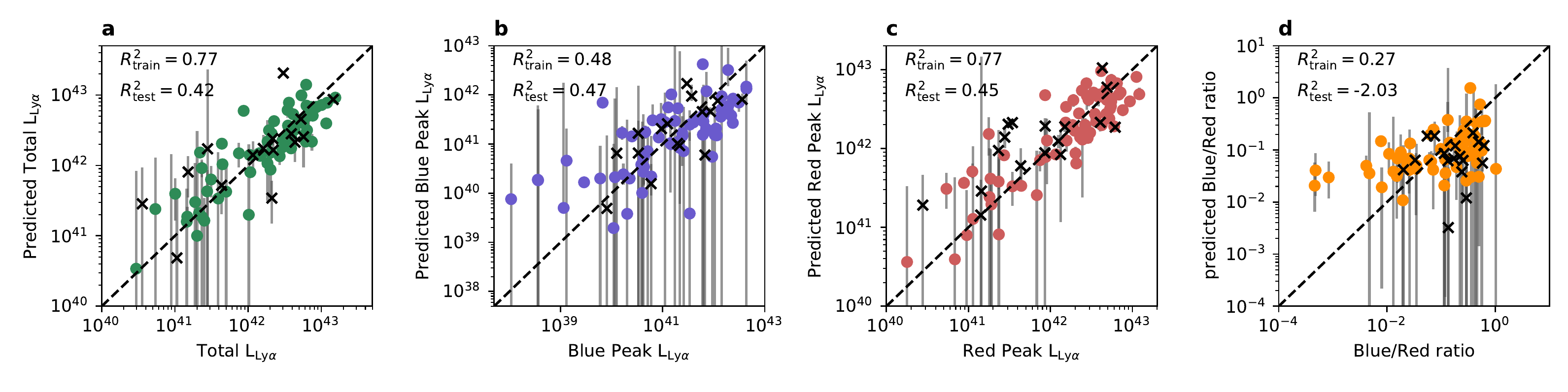}
\caption{Predictions of the total, blue, and red luminosities as well as the
ratio of the blue to red luminosity. The prediction is on the y-axis and the
true measured value on x. Colored points indicate the training sample and black
xs mark the test sample. $R^2$ values denoted in each panel show the
coefficient of determination for the training and testing sample respectively.}
\label{fig:multivar_lum}
\end{figure*}

\begin{figure*}
\centering
\includegraphics[width=\textwidth]{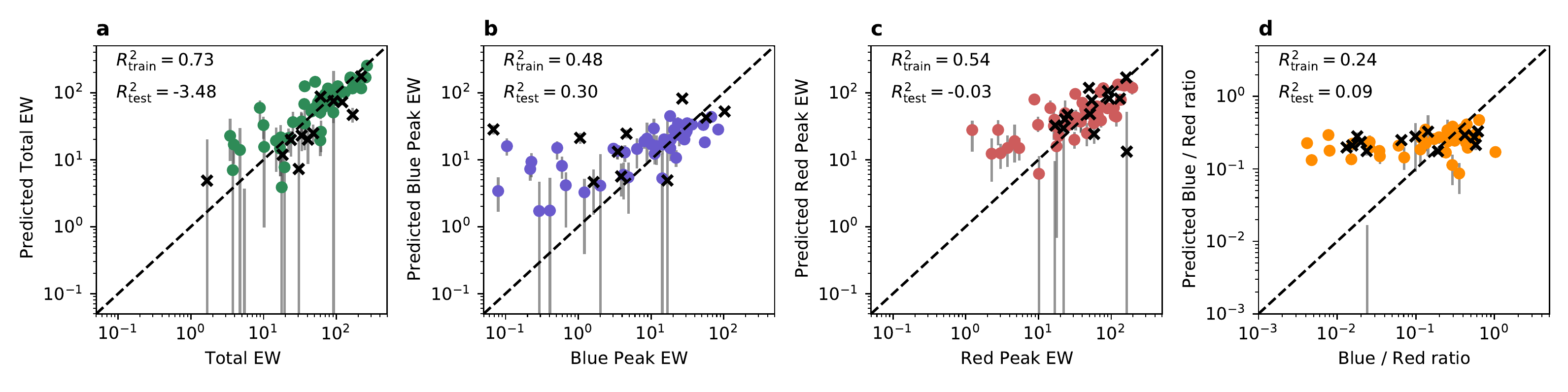}
\caption{Predictions of the total, blue, and red EWs as well as the blue over
red ratio. For details see caption of Figure\,\ref{fig:multivar_lum}}
\label{fig:multivar_EW}
\end{figure*}

So far we have presented a large number of correlations between optical and UV
line properties and \lya---some of which are unprecedentedly strong. In this
Section we investigate whether we can use multi-parametric analysis to provide
even stronger predictive relations for \lya. Similar analysis was done in
\citet{runnholm.2020}, with the aim of predicting global \lya\ luminosities
measured in narrowband imaging. With the quantities derived from the LASD we
can expand upon this and test whether predictions can be made for the blue and
red components of \lya\ separately. 

We start out by subdividing both our \lya\ quantities and our predictors into
two groups: luminosities and ratios. For the first group we want to predict the
total, blue and red \lya\ luminosities as well as the blue over red ratio using
optical line luminosities. We selected the line luminosities to use as
predictors by first filtering on signal-to-noise, requiring a median SNR of 10
across the sample of galaxies. We then curated this line list by  manually
removing lines with redundant information, for instance by including the sum of
the [\oII]$\lambda\lambda 3726,3729$~\AA\ doublet instead of the individual
lines, leaving us with a total of nine lines.

For the second group we aim to predict the \lya\ EW, again divided into total,
blue, and red components. For this model we select a set of line ratios, or
quantities derived from such based on each quasi-group presented in
Section~\ref{sect:indvar}. The parameters we consider  are: \hbeta\ EW
indicating stellar evolutionary state, \halpha/\hbeta\ ratio indicating dust,
N2 index as a metallicity tracer, [\neIII]/[\oII] as a marker of ionization
state, [\sII] deficit as an indirect indicator of the neutral gas column, and
the temperature and pressure measured from [\oIII] and [\sII], respectively. 

The model we use in both cases is a simple multivariate linear regression. We
choose this model because it has a low number of free parameters and maintains
a high level of interpretability. For the luminosity model we use a
standardization that includes first taking the logarithm of the variables, and
then performing a more ordinary standardization -- subtracting the means and
dividing by the standard deviations, often referred to as z-score
standardization. This ensures that all variables are comparable but it should
be noted that this means that the model in this case is a power law rather than
linear. For the prediction of ratios we only use a z-score standardization.
Standardizations and model fitting was done using the \texttt{sklearn} python
package.

\begin{figure*}
\centering
\includegraphics[width=\textwidth]{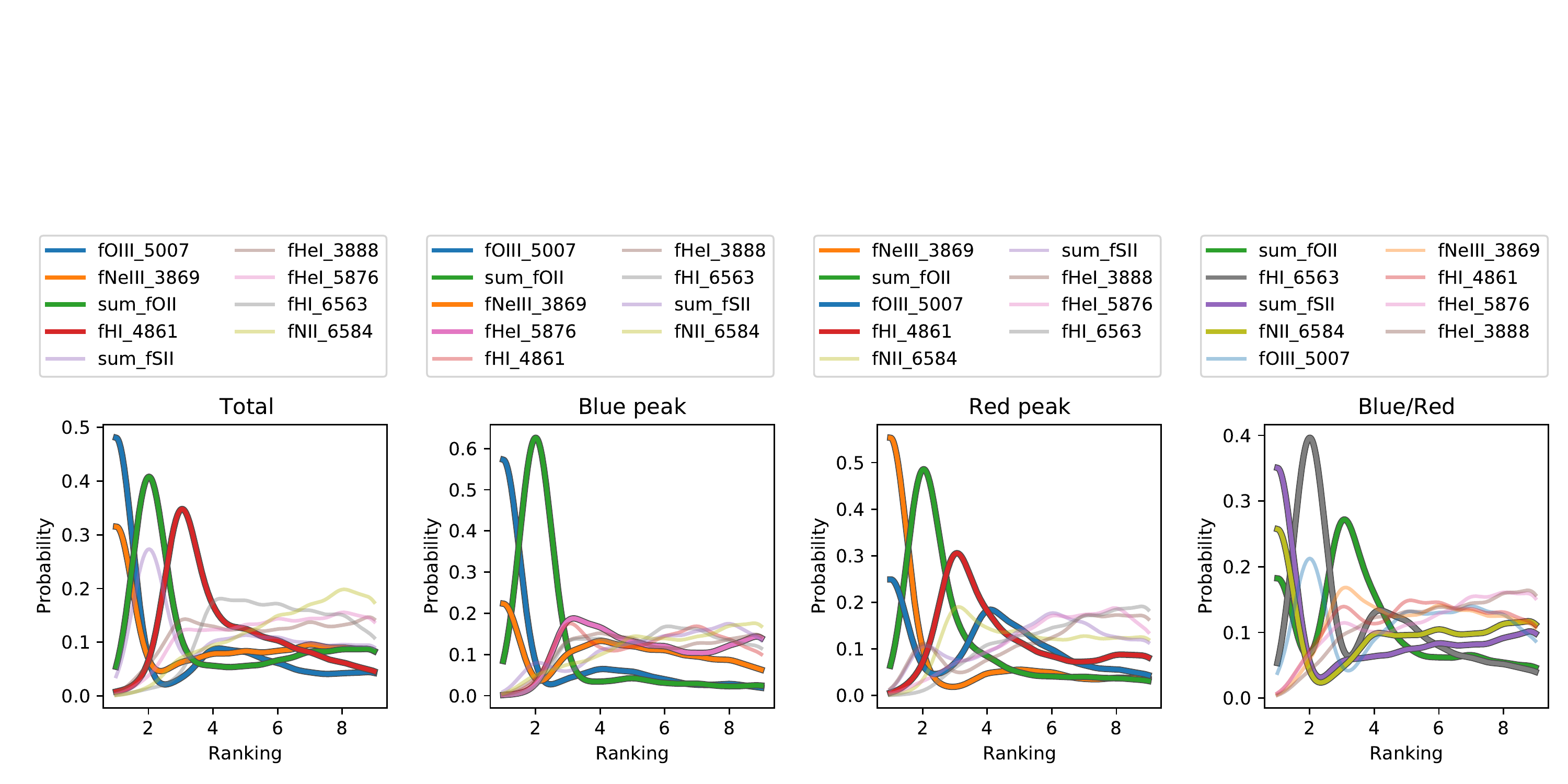}
\caption{Variable importance for the luminosity predictions. On the x-axis is
the variable ranking with 1 being the most important variable and 10 the least
important. Each curve shows the distribution of rankings for one variable over
the set of 2000 Monte Carlo iterations converted to a probability density
function by a kernel density estimator with a Gaussian kernel of width $0.5$.
The 4 most variables with the lowest mean rank are highlighted with thicker
lines.}
\label{fig:multivar_lum_rank}
\end{figure*}

\begin{figure*}
\centering
\includegraphics[width=\textwidth]{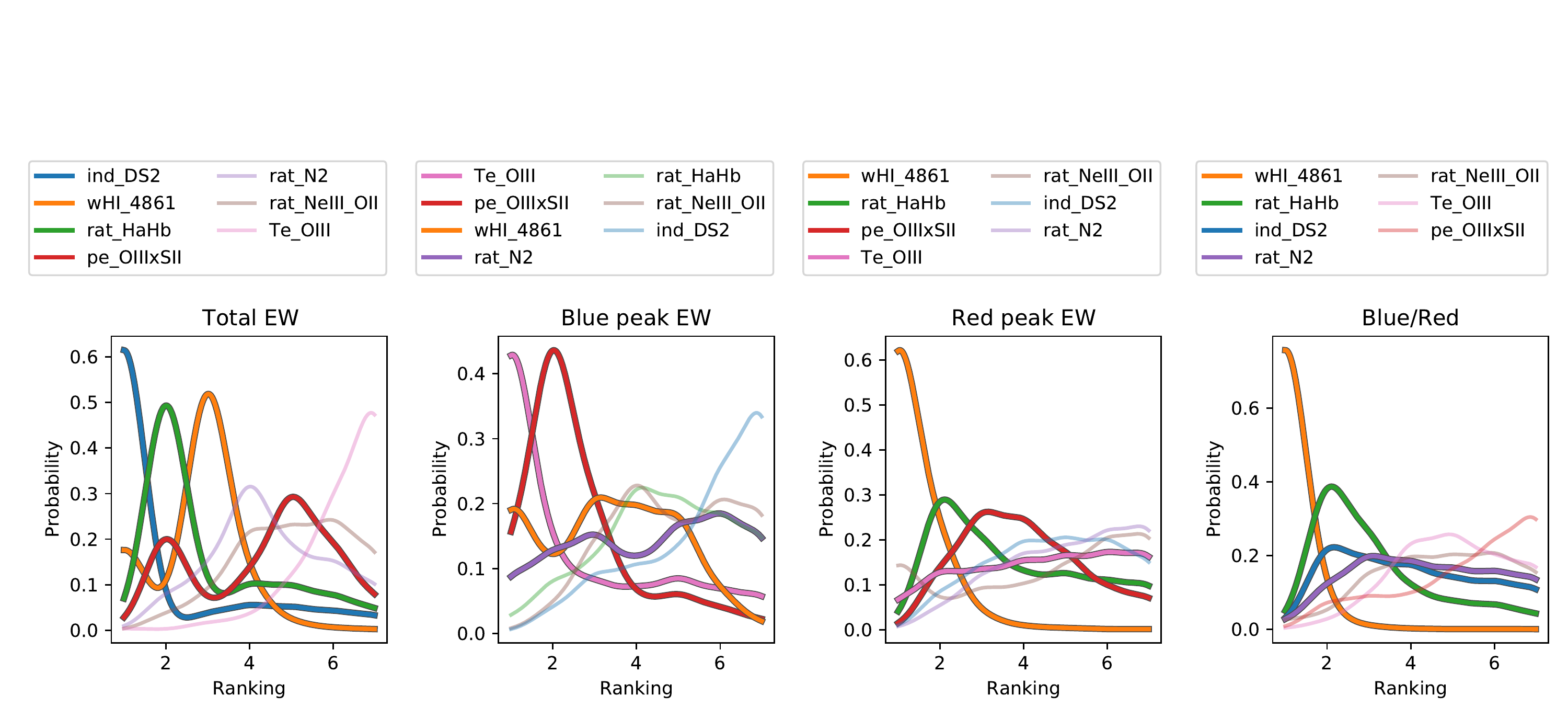}
\caption{Variable importances for the equivalent width predictions. For details
see caption of Figure\,\ref{fig:multivar_lum_rank}.}
\label{fig:multivar_EW_rank}
\end{figure*}

The resulting luminosity predictions are shown in
Figure\,\ref{fig:multivar_lum} and the EW predictions in
Figure\,\ref{fig:multivar_EW}. Colored points indicate the training part of the
sample and black crosses show the testing sample. We use a 20~\% test train
split for training the models in this case. The errorbars are estimated by
simple Monte Carlo simulation, where we resample all the predictors based on
their error estimates and refit the model 2000 times.

Attending first to panel \textbf{a} of Figure\,\ref{fig:multivar_lum} the
models performs well with the a coefficient of determination (R$^2$) of 0.8
meaning that it explains $\sim 80$\% of the variance in the data. The
performance of the model predicting the red peak (panel \textbf{c}) is very
similar which is expected since the red peak represents a majority of the \lya\
line flux. On the other hand the model predicting the blue peak performs worse,
explaining around 50\% of the variance---most likely due to the lower signal to
noise in the blue peak compared to the red. The last panel (\textbf{d}) shows
the model for the blue over red ratio. It is immediately apparent that the
performance of this model is degraded compared to the others with $R^2$ around
0.2. This appears surprising given the decent performance of the red and blue
separate models. One contributing factor is most likely that the dynamic range
of the luminosities is significantly larger,  which means that it is relatively
easy to get the correct order of magnitude of the prediction, but when that
large scale variation is removed, such as in a ratio, only the smaller scale
variation remains which is harder to predict.

Figure\,\ref{fig:multivar_EW} shows the results of our predictions of the \lya\
EWs.  The $R^2$ scores are lower for predicting the EW than the luminosity,
which is expected because absolute values (e.g. luminosity) no longer enter
when predicting the ratio of two quantities. However, models can still predict
more than 50\% of the variance. In this case we do not see a marked difference
in performance between the total, blue and red models but the blue/red ratio
model remains significantly worse. 

We use a train-test-split to quantify the performance of the models when
applied to galaxies that were not included in the fitting. The $R^2$ scores for
the test samples are noted in the corresponding panels
Figures~\ref{fig:multivar_lum} and \ref{fig:multivar_EW}. We note that they are
in general, as expected, lower than the $R^2_\mathrm{train}$ scores, in some
cases considerably so. This is an indication that the full models may be
somewhat overfitting the data, however it is difficult to use the specific
$R^2_\mathrm{test}$ score to evaluate to what degree, since the score is quite
sensitive to the exact selection of train and test samples.

In order to reduce overfitting we need to reduce the number of included
variables in the model. We can evaluate which variables to include and the
effect on both train and test performance using variable selection, in this
case forward selection. The method is quite simple. First we fit a 1-variable
model using all variables individually and check, using the $R^2$ score, which
is the best performing. This variable is selected as the most important. Then
we fit a 2-variable model using the one that was selected and each of the
remaining individually, with the best performing selected as the second most
important variable. This is repeated for the whole set of predictors to give a
complete set of rankings.  We do this for  each iteration of the Monte Carlo
simulation which gives us a distribution of rankings for each variable which we
then smooth with a kernel density estimator with a kernel size of 0.5. The
resulting probability density functions are shown in
Figure\,\ref{fig:multivar_lum_rank} and Figure\,\ref{fig:multivar_EW_rank} for
the luminosity and equivalent width models respectively.  The curves are
ordered by the mean ranking of the variable with lowest rank, i.e. most
important, first.

Attending first to the first panel of the luminosity variable selection, some
patterns appear from the what at first seems a noisy plot. The two most
important variables are the [\oIII]$_{5007}$  and [\neIII]$_{3869}$ lines, both
clearly peaking at first rank. Both of these lines trace the high ionization
gas. The second spot is shared by the [\oII] and [\sII] sums, both tracing low
ionization gas. The first two variables then clearly correspond to ionization
state tracers, c.f. the O$_{32}$, Ne3O2 or [\sII]/[\oIII] ratios. After these
ionization tracers \hbeta\ has the highest weight.  Since we are predicting a
luminosity our interpretation of this is that \hbeta\ acts as a flux
normalization quantity. Additionally its dust attenuation is, most likely,
closer to that of \lya\ than for example \halpha, due to the bluer wavelength.

The same patterns of importance are also seen in the predictions for the blue
and red peak, with the only major difference being a reduced importance of
\hbeta. The biggest difference is notable in the prediction of the blue over
red ratio. For this prediction lower ionization lines together with \halpha\
instead dominate. It is important not to over-interpret this since the
prediction can in this case only explain about 30\% of the variance in the
sample.

Looking now to the equivalent width result, we find that there is more variance
in the rankings across the different models. This is most likely due to the
poorer performance of the models overall, but nevertheless some trends are
clear. For instance the \hbeta\ equivalent width maintains a high degree of
importance in all relations---in particular the red peak and blue over ratio
models.  This is in line with our expectations since a high \hbeta\ EW should
correlate with a high production of \lya\ photons as well as the evolutionary
stage of the stars, and therefore potentially the \lya\ EW. The \halpha/\hbeta\
ratio, which encodes the dust extinction of the system, is also consistently
ranked well. Again this is expected since dust is the primary destruction
mechanism of \lya, however, what is less expected is that this variable seems
to hold very little importance for the equivalent width of the blue peak. In
fact the blue peak models seem dominated by thermodynamic variables, such as
the temperature and pressure.  While these variables describe the state of the
gas and as such should impact \lya\ transfer, we have found them to be only
weakly correlated with \lya\ properties throughout this work which makes this
result surprising.  Further, more complete multi-parametric analysis of these
trends will be the topic of a forthcoming dedicated paper.

\section{Summary, Conclusions and Outlook} \label{sect:conclusions}

In this paper we have performed an extensive and complete analysis of the
medium resolution ultraviolet spectral data of local starburst galaxies,
obtained with the Cosmic Origin Spectrograph on the Hubble Space Telescope.  We
have modeled the stellar continuum using population synthesis models to obtain
many properties of the stellar population, including mass, metallicity, age,
and inferred secondary quantities of ionizing photon production rates and
mechanical energy returns.  We have also measured the properties of over 40
optical emission lines in aperture-matched spectra, and measured quantities
related to metal abundance, dust reddening, ionization state, and thermodynamic
variables.  

We show this full sample lies along standard loci in diagnostic diagrams, such
as excitation and metallicity sequences (the `BPT diagrams'), the mass
metallicity relation, and SFR-stellar mass diagram (the `star-forming main
sequence').  There is a strong correlation between stellar metallicity and
abundance estimates from the nebular gas, and also between starburst age and
the equivalent width of \hbeta\ -- in concert these show that our modeling is
able to break the well-known age-metallicity degeneracy, and the quantities we
refer to as `age' meaningfully represent the evolutionary stage of the
starburst episode.  We conclude that our inferred stellar parameters are
accurate. 

In the UV spectra we have measured a large number of photometric and kinematic
properties of the spectrally resolved \lya\ emission line; we then use the
properties derived from the nebular lines and stellar modeling to investigate
how the \lya\ output and spectral profile depend on physical conditions and how
they may be predicted from other data.  We back up some of these findings using
estimates of the cool gas covering and kinematics, which are also derived from
the UV spectra.

Our main conclusions are: 

\begin{itemize}
\item{Total \lya\ output is strongest among young galaxies, when the most
massive stars remain on the main sequence.  Evidence for this is mainly
provided by modeled age, and the equivalent width of Balmer lines.  This is
natural since a young stellar population is necessary to produce the maximum
\lya\ emission.}
\item{Stellar age cannot uniquely explain the \lya\ output.  \lya\ EW
correlates most strongly with ionization parameter, as inferred from ratios of
strong forbidden lines [\oIII]/[\oII] and [\neIII]/[\oII], and higher nebular
EWs are also found among more highly ionized ISM conditions.}
\item{The effects of age and ionization conditions do not only influence the
\lya\ production.  Relations of equivalent amplitude and higher significance
are found when contrasting these quantities with the \lya\ escape fraction,
which traces only the output and transfer of \lya\ photons.  These properties
have a significant impact on the large scale ISM.}
\item{Correlations with dust content are significant, but are insufficient to
explain the \lya\ output.  Since ionization parameter is the strongest
correlating variable, we expect that the main factor in influencing \lya\
emission at early times, and therefore the observed EW, is the ionization
condition. The \hI\ column density is reduced by the extreme ionizing
conditions and the \lya\ transfer is eased.}
\item{This conclusion is further supported by the resolved \lya\ line profile,
which shows more blueshifted emission at early times and with higher ionization
parameters.  Most of the relations described above hold for the blue and red
components of the line, but relations for the blueshifted part alone exhibit a
larger dynamic range.  } 
\item{By stacking the ultraviolet continuum spectra and examining the low
ionization absorption lines, we see that all these phenomena are directly
associated with the column of absorbing gas.  This decreases towards younger
ages and higher ionization parameters, reduces the \lya\ scattering, and leads
to higher escape fractions.  Moreover, the column of gas at zero velocity is
lowest in the galaxies with highest contributions of blueshifted \lya,
indicating that the removal of this gas by photoionization at early times is
responsible for the blueshifted emission.}
\item{Simple multi-parametric analysis can predict up to 80\% of the variance
in the \lya\ luminosity, performing equally well for the total and red peak.
Similar models can predict approximately 50\% of the variance in the EWs.}
\item{We use variable selection methods to quantify which variables carry the
most weight in the relations. We find that the luminosity relations appear
dominated by ionization tracers such as [\oIII] and [\neIII] together with low
ionization lines like [\oII]. This is in line with the correlation analysis
presented earlier in this work. We also find that \hbeta\ provides a better
absolute value calibration than \halpha\ most likely due to the effect of the
bluer wavelength on the dust attenuation.  In this multi-parametric experiment
we are unable to predict the \Lbluered\ ratio.}
\item {The variable selection results for the EW relations is somewhat less
clear cut but does seem to give preference to dust tracers, in this case
\halpha/\hbeta\ and to the \hbeta\ EW.  Improving these multi-parametric
estimates will be the subject of future work.}
\item{We have no uniform estimates of the inclination nor morphology of our
galaxies, and neither can we estimate the influence of viewing angle on the
\lya\ escape. These effects would cause our observations to miss \lya\ in some
cases, and must certainly contribute scatter to some of the relations, hiding
some of the weaker trends.}
\end{itemize}

\section*{Acknowledgements}
M.H. is Fellow of the Knut and Alice Wallenberg Foundation.  M.G. thanks the
Max Planck Society for support through the Max Planck Research Group.

\section*{Data Availability}

All data are available in the Barbara A. Mikulski Archive for Space Telescopes
(MAST) at
\href{https://archive.stsci.edu/hst/search.php}{https://archive.stsci.edu/hst/search.php}

\bibliographystyle{mnras}
\input{main.bbl}

\clearpage

\appendix
\section{Galaxy Coordinates and Redshifts}

\onecolumn
\centering
\begin{longtable}{ccccccccc} 
\caption{Galaxy names and properties}\\
\hline
Name              & Right Ascension & Declination & Redshift & $M_\mathrm{UV}$ & SFR(\halpha) & \ebv\ & 12+log(O/H) & log($L_{\mathrm{Ly}\alpha}$)\\
                  & [$^\circ$] J2000 & [$^\circ$] J2000  &   &                 & [\msunyr]    & [mag] &             & \ergsec     \\
\hline
  SDSSJ0055-0021 & $13.86446$ & $-0.36349$ & $0.16743$ & $-20.33$ & $35.01$ & $0.227$ & $8.19$ & $41.45$ \\
  SDSSJ0150+1308 & $27.61838$ & $13.14956$ & $0.14674$ & $-20.54$ & $18.61$ & $0.188$ & $8.30$ & $41.37$ \\
  SDSSJ0213+1259 & $33.45221$ & $12.99761$ & $0.21897$ & $-20.36$ & $34.47$ & $1.024$ & $8.49$ & $41.70$ \\
  SDSSJ0808+3948 & $122.18450$ & $39.81459$ & $0.09125$ & $-19.97$ & $5.91$ & $0.326$ & $8.53$ & $42.01$ \\
  SDSSJ0921+4509 & $140.49746$ & $45.15344$ & $0.23500$ & $-21.02$ & $29.59$ & $0.416$ & $8.48$ & $41.49$ \\
  SDSSJ0926+4427 & $141.50183$ & $44.46015$ & $0.18068$ & $-20.84$ & $21.79$ & $0.096$ & $8.02$ & $42.78$ \\
  SDSSJ0938+5428 & $144.55621$ & $54.47364$ & $0.10207$ & $-20.43$ & $21.47$ & $0.174$ & $8.16$ & $41.34$ \\
  SDSSJ2103-0728 & $315.99475$ & $-7.46736$ & $0.13688$ & $-20.32$ & $16.69$ & $0.108$ & $8.49$ & $42.21$ \\
  SDSSJ0021+0052 & $5.25429$ & $0.88003$ & $0.09835$ & $-20.90$ & $20.11$ & $0.044$ & $8.12$ & $42.59$ \\
  SDSSJ1025+3622 & $156.45158$ & $36.38289$ & $0.12650$ & $-20.20$ & $16.07$ & $0.136$ & $8.09$ & $42.31$ \\
  SDSSJ1112+5503 & $168.18354$ & $55.06306$ & $0.13164$ & $-99.00$ & $27.33$ & $0.208$ & $8.37$ & $41.87$ \\
  SDSSJ1144+4012 & $176.09283$ & $40.20589$ & $0.12694$ & $-20.46$ & $11.52$ & $0.322$ & $8.30$ & $40.73$ \\
  SDSSJ1414+0540 & $213.72554$ & $5.67989$ & $0.08186$ & $-20.26$ & $10.02$ & $0.286$ & $8.21$ & $40.55$ \\
  SDSSJ1416+1223 & $214.05363$ & $12.39456$ & $0.12315$ & $-20.60$ & $24.02$ & $0.260$ & $8.34$ & $41.18$ \\
  SDSSJ1428+1653 & $217.23504$ & $16.89426$ & $0.18167$ & $-21.29$ & $27.71$ & $0.169$ & $8.24$ & $42.46$ \\
  SDSSJ1429+0643 & $217.44583$ & $6.72637$ & $0.17353$ & $-20.92$ & $42.88$ & $0.048$ & $8.08$ & $42.73$ \\
  SDSSJ1521+0759 & $230.42258$ & $7.98937$ & $0.09426$ & $-20.23$ & $8.53$ & $0.187$ & $8.20$ & $41.64$ \\
  SDSSJ1612+0817 & $243.18967$ & $8.28361$ & $0.14914$ & $-20.84$ & $38.59$ & $0.303$ & $8.35$ & $42.31$ \\
     GP0303-0759 & $45.83921$ & $-7.98978$ & $0.16485$ & $-20.56$ & $13.48$ & $-0.041$ & $8.02$ & $41.94$ \\
     GP0911+1831 & $137.80558$ & $18.51894$ & $0.26220$ & $-20.99$ & $32.52$ & $0.159$ & $8.20$ & $42.81$ \\
     GP1054+5238 & $163.37833$ & $52.63136$ & $0.25263$ & $-21.31$ & $38.74$ & $0.078$ & $8.12$ & $42.54$ \\
     GP1133+6514 & $173.26583$ & $65.22817$ & $0.24140$ & $-20.79$ & $8.72$ & $0.022$ & $8.06$ & $42.58$ \\
     GP1137+3524 & $174.34225$ & $35.40742$ & $0.19437$ & $-20.86$ & $27.82$ & $0.047$ & $8.10$ & $42.62$ \\
     GP1219+1526 & $184.76658$ & $15.43569$ & $0.19563$ & $-20.78$ & $19.65$ & $-0.029$ & $8.00$ & $43.19$ \\
     GP1244+0216 & $191.09737$ & $2.26122$ & $0.23943$ & $-20.66$ & $43.90$ & $0.066$ & $8.07$ & $42.56$ \\
     GP1249+1234 & $192.14429$ & $12.56747$ & $0.26339$ & $-20.91$ & $24.78$ & $0.080$ & $8.05$ & $43.07$ \\
     GP1424+4217 & $216.02383$ & $42.27953$ & $0.18477$ & $-20.79$ & $30.47$ & $0.064$ & $8.05$ & $42.94$ \\
      J0925+1403 & $141.38487$ & $14.05363$ & $0.30123$ & $-20.47$ & $26.92$ & $0.075$ & $8.01$ & $42.83$ \\
      J1152+3400 & $178.02033$ & $34.01386$ & $0.34193$ & $-20.77$ & $21.29$ & $0.001$ & $8.02$ & $43.00$ \\
      J1333+6246 & $203.26650$ & $62.76772$ & $0.31812$ & $-20.52$ & $-99.00$ & $-99.000$ & $8.09$ & $42.75$ \\
      J1442-0209 & $220.63079$ & $-2.16445$ & $0.29368$ & $-21.11$ & $29.37$ & $0.104$ & $7.93$ & $43.16$ \\
      J1503+3644 & $225.92846$ & $36.74743$ & $0.35569$ & $-20.43$ & $23.11$ & $0.068$ & $8.04$ & $42.86$ \\
      J0901+2119 & $135.44004$ & $21.32438$ & $0.29931$ & $-19.06$ & $15.92$ & $0.072$ & $7.99$ & $42.48$ \\
      J1011+1947 & $152.90950$ & $19.78915$ & $0.33217$ & $-19.33$ & $33.73$ & $0.302$ & $7.92$ & $42.64$ \\
      J1154+2443 & $178.70354$ & $24.72584$ & $0.36899$ & $-19.50$ & $12.74$ & $0.135$ & $7.92$ & $42.87$ \\
      J1243+4646 & $190.75262$ & $46.78067$ & $0.43171$ & $-20.67$ & $23.77$ & $0.098$ & $7.70$ & $43.08$ \\
      J1248+4259 & $192.04367$ & $42.99822$ & $0.36302$ & $-99.00$ & $25.27$ & $0.025$ & $7.93$ & $42.79$ \\
      J1256+4509 & $194.18396$ & $45.15472$ & $0.35320$ & $-19.85$ & $11.24$ & $0.076$ & $7.93$ & $42.54$ \\
      J0159+0751 & $29.96979$ & $7.86356$ & $0.06106$ & $-16.58$ & $1.99$ & $0.258$ & $7.79$ & $41.59$ \\
      J1205+4551 & $181.26479$ & $45.86415$ & $0.06539$ & $-16.41$ & $2.91$ & $0.131$ & $8.01$ & $41.65$ \\
      J1242+4851 & $190.61025$ & $48.86602$ & $0.06216$ & $-15.81$ & $0.70$ & $0.009$ & $7.69$ & $41.44$ \\
  SDSSJ0815+2156 & $123.96667$ & $21.93990$ & $0.14095$ & $-19.25$ & $7.04$ & $0.040$ & $7.96$ & $42.34$ \\
      J0240-0828 & $40.21746$ & $-8.47428$ & $0.08224$ & $-18.20$ & $8.63$ & $0.067$ & $7.99$ & $42.32$ \\
      J0851+5840 & $132.81521$ & $58.68194$ & $0.09190$ & $-18.35$ & $9.21$ & $0.108$ & $7.97$ & $41.64$ \\
      J1200+2719 & $180.06867$ & $27.33306$ & $0.08188$ & $-18.83$ & $8.01$ & $0.123$ & $7.95$ & $42.43$ \\
      J1226+0415 & $186.54954$ & $4.26002$ & $0.09422$ & $-18.06$ & $10.24$ & $0.168$ & $8.00$ & $42.02$ \\
      J1311-0038 & $197.88033$ & $-0.64566$ & $0.08106$ & $-18.08$ & $5.25$ & $0.184$ & $8.03$ & $42.03$ \\
     GP0822+2241 & $125.69859$ & $22.69558$ & $0.21622$ & $-20.12$ & $52.36$ & $0.198$ & $8.06$ & $42.35$ \\
     GP0917+3152 & $139.26052$ & $31.87238$ & $0.30038$ & $-20.96$ & $25.55$ & $0.176$ & $8.22$ & $42.59$ \\
     GP1009+2916 & $152.32914$ & $29.27264$ & $0.22192$ & $-19.75$ & $7.53$ & $0.069$ & $8.01$ & $42.34$ \\
     GP1122+6154 & $170.58222$ & $61.91263$ & $0.20456$ & $-19.24$ & $9.17$ & $0.150$ & $8.06$ & $42.26$ \\
     GP1440+4619 & $220.04142$ & $46.32693$ & $0.30078$ & $-21.43$ & $47.94$ & $0.137$ & $8.13$ & $42.81$ \\
     GP1454+4528 & $223.64825$ & $45.48231$ & $0.26852$ & $-20.31$ & $27.91$ & $0.181$ & $8.17$ & $42.27$ \\
     GP1514+3852 & $228.53598$ & $38.86871$ & $0.33263$ & $-20.68$ & $9.38$ & $-0.275$ & $8.14$ & $42.76$ \\
     GP1559+0841 & $239.85824$ & $8.68865$ & $0.29719$ & $-19.80$ & $-99.00$ & $-99.000$ & $8.18$ & $42.57$ \\
     GP2237+1336 & $339.39608$ & $13.61306$ & $0.29350$ & $-21.15$ & $41.17$ & $0.136$ & $8.10$ & $42.29$ \\
  GALEX1417+5228 & $214.43094$ & $52.46828$ & $0.20813$ & $-18.28$ & $1.77$ & $-0.083$ & $7.68$ & $42.06$ \\
      J0232-0426 & $38.06704$ & $-4.44076$ & $0.45237$ & $-99.00$ & $9.52$ & $0.101$ & $7.96$ & $42.58$ \\
      J0919+4906 & $139.98241$ & $49.10243$ & $0.40515$ & $-19.28$ & $6.24$ & $0.041$ & $7.99$ & $42.75$ \\
      J1046+5827 & $161.50825$ & $58.46582$ & $0.39675$ & $-20.40$ & $8.93$ & $-0.061$ & $7.93$ & $42.59$ \\
      J1121+3806 & $170.32591$ & $38.11189$ & $0.31787$ & $-99.00$ & $4.80$ & $-0.017$ & $7.96$ & $42.47$ \\
      J1127+4610 & $171.83750$ & $46.17847$ & $0.32221$ & $-99.00$ & $4.45$ & $-0.139$ & $7.88$ & $42.15$ \\
      J1233+4959 & $188.37825$ & $49.99707$ & $0.42199$ & $-99.00$ & $10.12$ & $-0.050$ & $8.03$ & $42.72$ \\
      J1349+5631 & $207.47958$ & $56.51969$ & $0.36361$ & $-99.00$ & $6.57$ & $0.157$ & $8.03$ & $42.32$ \\
      J1355+1457 & $208.97275$ & $14.95041$ & $0.36513$ & $-99.00$ & $11.92$ & $0.060$ & $7.92$ & $42.56$ \\
      J1455+6107 & $223.99820$ & $61.12214$ & $0.36786$ & $-20.05$ & $7.17$ & $-0.022$ & $7.99$ & $42.57$ \\
  J010534+234960 & $16.39057$ & $23.83325$ & $0.33807$ & $-19.71$ & $22.32$ & $0.121$ & $8.09$ & $42.69$ \\
  J015208-043117 & $28.03331$ & $-4.52145$ & $0.38359$ & $-20.23$ & $24.67$ & $0.168$ & $8.15$ & $42.79$ \\
  J020819-040136 & $32.07877$ & $-4.02679$ & $0.38445$ & $-20.56$ & $16.32$ & $-0.130$ & $8.02$ & $42.92$ \\
  J110359+483456 & $165.99582$ & $48.58219$ & $0.41795$ & $-20.25$ & $24.49$ & $0.177$ & $8.07$ & $42.64$ \\
  J110506+594741 & $166.27636$ & $59.79484$ & $0.40540$ & $-19.72$ & $17.02$ & $0.155$ & $8.06$ & $42.80$ \\
  J121948+481411 & $184.94943$ & $48.23625$ & $0.42022$ & $-20.42$ & $13.69$ & $0.055$ & $8.10$ & $42.63$ \\
  J124619+444902 & $191.58118$ & $44.81730$ & $0.32217$ & $-20.75$ & $40.72$ & $0.150$ & $8.07$ & $42.88$ \\
  J142535+524902 & $216.39628$ & $52.81728$ & $0.38695$ & $-19.74$ & $17.46$ & $0.156$ & $7.97$ & $42.49$ \\
\hline
%
\multicolumn{9}{l}{Names are according to MAST archive labeling.}\\
\multicolumn{9}{l}{$M_\mathrm{UV}$ is corrected for Milky Way extinction.}\\
\multicolumn{9}{l}{SFR(\halpha) is corrected for internal attenuation and uses the \citet{Kennicutt.2012} calibration}\\
\multicolumn{9}{l}{Oxygen abundance is derived from the O3N2 index using \citet{Marino.2013}.}\\
\multicolumn{9}{l}{\llya\ is derived from the \emph{Lyman alpha Spectral Database} \citep[LASD][]{Runnholm.2021}.}\\
\label{tab:galaxies} 
\end{longtable}

\bsp    
\label{lastpage}

\end{document}